\documentclass[%
 reprint,
 amsmath,amssymb,
 aps,
]{revtex4-2}

\usepackage{amsthm}
\usepackage{graphicx}
\usepackage{dcolumn}
\usepackage{bm}
\usepackage{mathtools}

\usepackage{changes}
\definecolor{darkviolet}{rgb}{0.58, 0.0, 0.83}
\definecolor{trueblue}{rgb}{0.1, 0.35, 0.91}
\definechangesauthor[name = {Lock}, color = darkviolet]{LY}
\definechangesauthor[name = {Rafael}, color = trueblue]{RF}

\newtheorem{definition}{Definition}

\begin{document}

\preprint{APS/123-QED}

\title{Thermoinformational State Construction: \\Generative Energies, Entropies, and H-Theorem Consistency}


\author{George-Rafael Domenikos}
 \email{corresponding author: georgios.rd@ntu.edu.sg}
\affiliation{EMPOWER Centre, Nanyang Technological University, 11 Mandalay Road, 308232, Singapore
}%


\author{Lock Yue Chew}
\affiliation{School of Mathematical and Physical Sciences, Nanyang Technological University, 21 Nanyang Link, 637371, Singapore}%

\author{Victoria Leong}
\affiliation{School of Social Sciences, Nanyang Technological University, 21 Nanyang Link, 637371, Singapore}%

\date{\today}

\begin{abstract}
We introduce a constructive framework for assigning thermodynamic structure to an arbitrary data system from its measured microstates. Starting from an empirical distribution over configurations, we first infer a data-driven energy function by fitting a Boltzmann-type model to the observed statistics, thereby defining an energy axis that is intrinsic to the system. We then push the empirical distribution onto this energy coordinate and pose an inverse maximum-entropy problem: we learn a strictly concave trace-form entropy functional whose maximizer, under a small set of constraints extracted from the data, reproduces the observed energy-space histogram. With energy and entropy defined in this coupled, system-specific manner, macroscopic variables such as internal energy, an entropy–energy relation $S(U)$, and a thermoinformational temperature $T^{-1}=\partial S/\partial U$ follow consistently along admissible families of states. We demonstrate the construction on canonical unimodal and multimodal examples, including a harmonic well (recovering the classical equilibrium limit up to gauge) and a bistable double-well where global-constraint MaxEnt surrogates can obscure barrier and coexistence structure. The resulting formulation provides a principled route from microstate data to thermodynamically consistent macroscopic descriptors, with an optimized entropy matched to the empirical system.
\end{abstract}

\maketitle


\section{Introduction}
\label{sec:introduction}

Thermodynamics provides a compact macroscopic description of complex systems with linkages to microstate statistics through statistical mechanics to state variables such as internal energy, entropy,
and temperature. In equilibrium statistical mechanics this link is mediated by an
energy function (Hamiltonian) and a variational principle: given a fixed energy and
a set of macroscopic constraints, the equilibrium distribution is obtained by
maximizing an entropy functional (classically Shannon/Boltzmann--Gibbs) subject to
those constraints \cite{Shannon1948,Jaynes1957,Presse2013}. This variational structure
is not merely aesthetic; it underpins thermodynamic consistency relations and, in
stochastic settings, allows trajectory-level definitions of heat, work, and entropy
production that refine the second law \cite{Seifert2008,Crooks1999,Parrondo2015}.

In modern empirical domains, however, we often observe microstates without an
underlying Hamiltonian. A broad literature therefore imports thermodynamic language
into data analysis by choosing an ``energy'' (often a negative log-density, a loss, or
a score) and then applying a fixed entropy form or a fixed exponential-family model class.
This perspective has proved useful in diverse contexts, from maximum-entropy
constructions in complex systems \cite{ParkNewman2004,Banavar2010,DeMartino2018}
to energy-based learning and probabilistic modeling in machine learning
\cite{Ackley1985,Hinton2002,LeCun2006EBL,Hyvarinen2005,WainwrightJordan2008}. Yet a basic
tension remains: if the energy representation is itself inferred from data, and the
entropy functional is assumed a priori, the resulting macroscopic quantities need not
be mutually compatible with a MaxEnt principle in the representation where the energy
is defined. In practice this can manifest as systematic mismatch in non-Gaussian or
multimodal settings, where global-constraint surrogates compress distinct basins,
barriers, or tails into a single effective mode. 

Here we propose a constructive remedy: rather than postulating either ingredient, we
\emph{construct both} (i) an intrinsic energy coordinate and (ii) a system-tailored
entropy functional directly from the observed microstate statistics. The procedure
proceeds in two coupled stages. First, we infer a \emph{generative energy}
$E_{\mathrm{gen}}(x)$ from data by fitting a structured energy family so that the
induced Boltzmann-type model locally matches the empirical distribution (up to an
affine gauge). Second, we push the empirical microstate distribution onto the learned
energy axis and solve an \emph{inverse} maximum-entropy problem: we learn a strictly
concave trace-form entropy generator for which the empirical energy histogram is the
unique MaxEnt solution under a small set of constraints extracted from the data
\cite{Jaynes1957,Presse2013}. Only after these two steps are completed do macroscopic
state variables become well-defined: internal energy as the mean of $E_{\mathrm{gen}}$,
an entropy--energy relation $S(U)$, and a thermoinformational temperature defined by
$T^{-1}=\partial S/\partial U$ along admissible state families.

We demonstrate the framework on canonical testbeds that progressively relax physical
assumptions. A harmonic single-well potential provides a calibration case in which
the learned energy recovers the expected quadratic landscape and temperature scaling,
showing that the construction contains the classical Boltzmann--Gibbs limit. A
bistable double-well potential provides a canonical multimodal setting where barrier
and coexistence structure are essential, and where fixed-form global surrogates can
obscure the underlying landscape. Finally, a mixture-of-Gaussians example treats the
problem as a pure data system with no assumed Hamiltonian, illustrating that the
method constructs an effective thermodynamic description directly from an empirical
distribution. Across these examples, the emphasis is not on introducing additional
free parameters, but on enforcing a variationally consistent micro-to-macro mapping
while selecting an entropy functional that is optimized for the observed system.

\section{Thermoinformational state construction:
Generative energy $\rightarrow$ energy distribution $\rightarrow$ generative entropy $\rightarrow T$}

We present a four--step construction that starts from observed microstates, learns a
data--driven energy function, pushes the empirical distribution onto this learned energy axis,
and then applies the generative entropy formalism to obtain thermodynamically consistent
state variables $(U,S,T)$. The presentation below is self--contained and consistent with the
definitions of generative energy and generative entropy described previously.

\subsection{Step 1: Generative energy from microstate statistics}

\subsubsection*{Microstate space and empirical distribution}

Let $\mathcal{X}$ denote the microstate space of the system and let
\[
  x \in \mathcal{X}
\]
denote a microstate (configuration). Repeated observations yield an empirical probability
distribution (or density) $p$ on $\mathcal{X}$:
\[
  p(x), \qquad x \in \mathcal{X},
\]
satisfying
\[
  p(x)\ge 0,
  \qquad
  \int_{\mathcal{X}} p(x)\,\mathrm{d}\mu(x)=1,
\]
where $\mu$ is a reference measure on $\mathcal{X}$ (counting measure if $\mathcal{X}$ is
discrete; Lebesgue measure if $\mathcal{X}\subseteq\mathbb{R}^d$).

\subsubsection*{Parametric energy family}

We posit a parametric family of energy functions
\[
  \mathcal{E}(\,\cdot\,;\theta):\mathcal{X}\to\mathbb{R},
  \qquad
  \theta\in\Theta\subseteq\mathbb{R}^K,
\]
taken to be linear in parameters:
\begin{equation}
  \mathcal{E}(x;\theta)
  \;=\;
  \sum_{k=1}^K \theta_k\,\phi_k(x),
  \label{eq:lin-energy}
\end{equation}
where $\phi_k:\mathcal{X}\to\mathbb{R}$ are fixed basis functions (features). The feature functions $\{\phi_k\}_{k=1}^K$ are user-chosen observables on microstates. In classical settings they are typically low-order moments or sufficient statistics (e.g.\ $x$, $x^2$, pairwise couplings, indicator functions of regions), but the framework allows any measurable, possibly non-linear $\phi_k(x)$ that encodes a macroscopic constraint of interest. Vector or tensor-valued observables can be accommodated by including their scalar components or invariants (e.g.\ norms) as separate features. When $\phi_k$ are monomials (or polynomials) in $x$, the constraints $\mathbb{E}_p[\phi_k(X)]$ correspond to moment constraints; more general $\phi_k$ correspond to non-moment macroscopic observables.

At this stage, the choice of feature family $\{\phi_k\}$ is the only modelling input and it can be user-dependent. This dependence is deliberate: in many empirical systems the histogram exhibits distinctive structure
(e.g.\ asymmetry, heavy tails, multi-modality), and a domain-informed analyst can select constraints that target that structure and yield the sharpest thermodynamic description.
In principle, an infinite (complete) feature expansion over the support would provide enough expressivity to represent arbitrary histogram geometry and thus recover an exact match in the limit; in practice, real applications require finite families and therefore introduce an explicit expressivity--parsimony trade-off. For user-agnostic deployment, one can replace hand-chosen constraints by automatic ``failsafe'' libraries that provide nested, systematically improvable spans---such as Fourier/orthogonal-polynomial bases on
one-dimensional or product domains, or Laplacian eigenfunction bases constructed directly on the discrete histogram support. In all cases, the residual mismatch between the induced Boltzmann surrogate and the empirical histogram
provides a quantitative diagnostic of whether the chosen feature family is sufficiently expressive. We describe these default constructions and their recommended regimes of use in the Supplementary Note S0. For the remainder of the main text, we proceed with an abstract feature family $\{\phi_k\}$, assumed to be sufficiently expressive to provide a good empirical fit for the distributions considered.

\subsubsection*{Boltzmann--type model induced by $\mathcal{E}$}

Given $\mathcal{E}(x;\theta)$, define the induced model distribution
\begin{equation}
  q_\theta(x)
  \;=\;
  \frac{1}{Z(\theta)}\exp\!\big(-\beta\,\mathcal{E}(x;\theta)\big),
  \qquad x\in\mathcal{X},
  \label{eq:qtheta}
\end{equation}
with partition function
\begin{equation}
  Z(\theta)
  =
  \int_{\mathcal{X}}\exp\!\big(-\beta\,\mathcal{E}(x;\theta)\big)\,\mathrm{d}\mu(x).
  \label{eq:Ztheta}
\end{equation}
For the purpose of defining the generative energy we set $\beta=1$ and absorb any global
scaling into $\theta$. Because the energy family is linear in parameters, the inverse-temperature $\beta$ is not
identifiable separately from the scale of $\theta$:
\[
q_{\theta,\beta}(x)\propto \exp\!\Big(-\beta\sum_{k}\theta_k\phi_k(x)\Big)=
\]

\[=\exp\!\Big(-\sum_{k}(\beta\theta_k)\phi_k(x)\Big)=q_{\tilde\theta,1}(x),
\quad \tilde\theta:=\beta\theta.\]
Hence we may fix $\beta=1$ when defining $E_{\rm gen}$, with any global energy scaling
reappearing as a temperature-unit rescaling in Step~4.

(A global scaling of $\mathcal{E}_{\mathrm{gen}}$ simply rescales
temperature units; cf.\ Step~4.)

\subsubsection*{$L^2$ discrepancy and generative energy}

We measure the discrepancy between the empirical distribution $p$ and the induced model
$q_\theta$ by the squared $L^2(\mu)$ distance
\begin{equation}
  J(\theta)
  \;\equiv\;
  \|p-q_\theta\|_{L^2(\mu)}^2
  =
  \int_{\mathcal{X}}\big(p(x)-q_\theta(x)\big)^2\,\mathrm{d}\mu(x).
  \label{eq:J-theta}
\end{equation}
In the discrete case $\mathcal{X}=\{x_1,\dots,x_M\}$,
\[
  J(\theta)
  =
  \sum_{i=1}^M\big(p(x_i)-q_\theta(x_i)\big)^2.
\]

Although \eqref{eq:qtheta} has the familiar Boltzmann form, we do \emph{not} assume the
data-generating distribution $p$ is thermal or belongs to an exponential family.  The
Boltzmann map $\,\mathcal{E}\mapsto q_\theta \propto e^{-\mathcal{E}}\,$ is used only as
a thermodynamically consistent way to convert an energy function—constrained to the
user-chosen basis/feature family $\{\phi_k\}$ (linear or non-linear)—into a normalised
density.  The optimisation of $J(\theta)$ therefore identifies an \emph{energy
representation} within this restricted family, rather than fitting a presupposed
probabilistic law to $p$.

We use an $L^2$ criterion (rather than KL divergence) because the goal is a stable
histogram-matching energy embedding, not likelihood maximisation under a fixed model
family.  In particular, $L^2$ is less sensitive to low-probability bins and to zeros in
empirical frequencies, and it yields a smooth objective with a least-squares structure
once discretised.

If the chosen feature family is expressive, the induced $q_{\theta^\star}$ can match $p$
arbitrarily well; if it is restrictive, the mismatch quantifies exactly which aspects of
$p$ are not representable by that energy family.

\begin{definition}[Generative energy]
    Let $p$ be the empirical distribution on $\mathcal{X}$ and let $q_\theta$ be defined by
    \eqref{eq:qtheta} with energy \eqref{eq:lin-energy}. Any minimiser $\theta^\star\in\Theta$ of
    \begin{equation}
      \theta^\star \in \arg\min_{\theta\in\Theta} J(\theta)
      \label{eq:theta-star}
    \end{equation}
    defines the \emph{generative energy}
    \begin{equation}
      \mathcal{E}_{\mathrm{gen}}(x)
      \;\equiv\;
      \mathcal{E}(x;\theta^\star)
      =
      \sum_{k=1}^K \theta_k^\star\,\phi_k(x).
      \label{eq:E-gen}
    \end{equation}
\end{definition}

\noindent Among all energies in the parametric family, $\mathcal{E}_{\mathrm{gen}}$ is the one
whose induced Boltzmann distribution is closest to $p$ in the $L^2(\mu)$ sense.

\medskip
\noindent\textbf{Remark (least--squares in surprisal space).}
Define the empirical surprisal $s(x)=-\log p(x)$ (on the support where $p(x)>0$). One may
alternatively fit $\mathcal{E}(x;\theta)$ to $s(x)$ via a weighted least--squares objective
\[
  J_{\mathrm{s}}(\theta)
  =
  \int_{\mathcal{X}}\big(\mathcal{E}(x;\theta)-s(x)\big)^2\,w(x)\,\mathrm{d}\mu(x),
  \qquad w(x)\ge 0,
\]
which is quadratic in $\theta$ for linear $\mathcal{E}$. We treat \eqref{eq:theta-star} as the
primary definition and $J_{\mathrm{s}}$ as a convenient implementation route.
In practice we use weights defined on the same discretisation as $\hat p$.
A stable default is $w(x)=\hat p(x)$ (mass-weighted least squares), which prioritises
the high-probability region and avoids tail overfitting; when tail fidelity is important
we use $w(x)\propto (\hat p(x)+\varepsilon)^{-\alpha}$ with $\alpha\in[0,1/2]$.
All results in this paper use $w(x)=\hat p(x)$ unless stated otherwise.

\subsection{Step 2: Energy random variable and its distribution}

Given $\mathcal{E}_{\mathrm{gen}}$ and the empirical microstate distribution $p$, define the
energy random variable
\[
  E=\mathcal{E}_{\mathrm{gen}}(X),
\]
where $X$ is a random microstate with law $p$:
\[
  \mathbb{P}(X\in A)=\int_A p(x)\,\mathrm{d}\mu(x),
  \qquad A\subseteq\mathcal{X}.
\]

\subsubsection*{Pushforward to energy space}

The probability law of $E$ is the pushforward of $p$ under $\mathcal{E}_{\mathrm{gen}}$.
For any Borel set $B\subseteq\mathbb{R}$,
\[
  \mathbb{P}(E\in B)
  =
  \int_{\{x\in\mathcal{X}:\,\mathcal{E}_{\mathrm{gen}}(x)\in B\}}
    p(x)\,\mathrm{d}\mu(x).
\]
If $E$ admits a density $p_E$ with respect to Lebesgue measure on $\mathbb{R}$, then
\begin{equation}
  p_E(e)
  =
  \int_{\mathcal{X}}
    \delta\!\big(e-\mathcal{E}_{\mathrm{gen}}(x)\big)\,p(x)\,\mathrm{d}\mu(x),
  \qquad e\in\mathbb{R}.
  \label{eq:pE-general}
\end{equation}
In the discrete microstate case $\mathcal{X}=\{x_i\}_{i=1}^M$ with energies
$E_i=\mathcal{E}_{\mathrm{gen}}(x_i)$, this reduces to
\begin{equation}
  p_E(e)=\sum_{i=1}^M p(x_i)\,\delta(e-E_i).
  \label{eq:pE-discrete}
\end{equation}

\subsubsection*{Discretisation (energy histogram)}

Since Step~3 is posed on a finite simplex, we discretise energy space into bins
$\{I_j\}_{j=1}^n$ with representative centres $c_j$ and define the energy histogram
\begin{equation}
  p_{E,j}
  \;\equiv\;
  p_E(I_j)
  =
  \mathbb{P}(E\in I_j)
  =
  \sum_{i:\,E_i\in I_j} p(x_i),
  \qquad
  \sum_{j=1}^n p_{E,j}=1.
  \label{eq:pE-binned}
\end{equation}
We write $p_E=(p_{E,1},\dots,p_{E,n})\in\Delta_n$ for the resulting discrete energy--space
representation.

\subsection{Step 3: Generative entropy via inverse maximum entropy}
\label{subsec:gen-entropy-general}

We construct an entropy functional directly from the empirical histogram, without assuming
\emph{a priori} membership in a particular family (Shannon, Tsallis, R\'enyi, Fermi--Dirac,
etc.). The guiding principle is variational: we seek an entropy $H_{\mathrm{gen}}$ whose
unique maximiser, under appropriately chosen linear constraints, is the empirical histogram
itself. This is achieved by learning an entropy generator from data through two
finite--dimensional quadratic programmes (QP): a \emph{core} QP enforcing global constraints,
and a \emph{shape} QP enforcing local structure.

Throughout this step, $p=(p_1,\dots,p_n)$ denotes a discrete probability vector on a finite
support $\mathcal{C}=\{c_1,\dots,c_n\}$, and in our application we take\textbf{ \textbf{$p=p_E$} }from
\eqref{eq:pE-binned}.
In our application, $C=\{c_j\}_{j=1}^n$ denotes the discretised energy support
(i.e. the energy-bin representatives/centres used in Eq.~(9)), and $p=p_E$ is the
corresponding energy histogram on that support.

\subsubsection*{Trace--form entropies and generator}

We restrict attention to trace--form entropies
\begin{equation}
  H_G(p)
  \;=\;
  \sum_{j=1}^n G(p_j),
  \qquad
  G\in C^2(0,1),
  \qquad
  G''(u)<0,
  \label{eq:trace}
\end{equation}
where strict concavity ensures that $H_G$ is concave on the simplex and yields unique
maximisers under linear constraints. The function $G:[0,1]\to\mathbb{R}$ is the
\emph{entropy generator}.

\subsubsection*{Inverse maximum--entropy condition}

Let $\{f_k(c_j)\}_{k=0}^m$ be a collection of constraint functions on $\mathcal{C}$
(e.g.\ low--order moments for the ``core'', and local masks for the ``shape'').
We impose expectation (moment-matching) constraints of the form
\begin{equation}
\label{eq:constraints}
  \sum_{j=1}^n q_j f_k(c_j)
  =
  \sum_{j=1}^n p_j f_k(c_j),
  \qquad k=0,\dots,m,
\end{equation}
where the feature functions $f_k:\mathcal{C}\to\mathbb{R}$ are user-chosen and may be
nonlinear in $c_j$ (e.g.\ moments, indicator masks, or other transforms).  The constraints
are linear in the decision variable $q$.

We require that $p$ is the maximiser of $H_G$ under these constraints:
\begin{equation}
  p\in
  \arg\max_{q\in\Delta_n} H_G(q)
  \quad \text{subject to \eqref{eq:constraints}}.
  \label{eq:ME-problem}
\end{equation}
The KKT stationarity conditions at $q=p$ yield
\begin{equation}
  G'(p_j)=\sum_{k=0}^m \lambda_k f_k(c_j),
  \qquad j=1,\dots,n,
  \label{eq:Gprime-eq}
\end{equation}
for some multipliers $\lambda=(\lambda_0,\dots,\lambda_m)$. Thus $G'(p)$ must lie in the
span of the constraint features evaluated on $\mathcal{C}$.

\subsubsection*{Linear model for $G'$}

We represent the unknown derivative $G'$ as a linear combination of basis functions
$\{\psi_\ell(u)\}_{\ell=0}^L$ on $[0,1]$:
\begin{equation}
  G'(u)=\sum_{\ell=0}^L \alpha_\ell \psi_\ell(u),
  \label{eq:Gprime-basis}
\end{equation}
with coefficients $\alpha\in\mathbb{R}^{L+1}$. Evaluating at $u=p_j$ gives
\[
  G'(p_j)=\sum_{\ell=0}^L \alpha_\ell \psi_\ell(p_j)=(A\alpha)_j,
\]
where $A\in\mathbb{R}^{n\times(L+1)}$ has entries $A_{j\ell}=\psi_\ell(p_j)$. Let
$F\in\mathbb{R}^{n\times(m+1)}$ denote the constraint matrix with entries
$F_{jk}=f_k(c_j)$. Then \eqref{eq:Gprime-eq} becomes
\begin{equation}
  A\alpha = F\lambda.
  \label{eq:Aalpha=Flambda}
\end{equation}

\subsubsection*{Core QP (global constraints)}

We select $(\alpha,\lambda)$ by a regularised least--squares fit:
\begin{equation}
  \min_{\alpha,\lambda}\;
  \|A\alpha - F\lambda\|_2^2 + \gamma\|\alpha\|_2^2
  \quad\text{subject to concavity constraints},
  \label{eq:QP-core}
\end{equation}
where $\gamma>0$ stabilises the solution and selects a minimal--norm generator derivative.

Concavity is enforced discretely via
\begin{equation}
  G''(p_j)=\sum_{\ell=0}^L \alpha_\ell \psi_\ell'(p_j)\le -\varepsilon,
  \qquad j=1,\dots,n,
  \label{eq:concavity}
\end{equation}
with $\varepsilon>0$ small. Strict concavity prevents flat regions that can otherwise yield
non--unique maximum--entropy solutions.

Let $(\alpha^\star,\lambda^\star)$ denote a solution. Define
\[
  G'_{\mathrm{core}}(u)=\sum_{\ell=0}^L \alpha^\star_\ell \psi_\ell(u),
  \qquad
  G_{\mathrm{core}}(u)=\int_0^u G'_{\mathrm{core}}(t)\,\mathrm{d}t,
\]
with $G_{\mathrm{core}}(0)=0$.

\subsubsection*{Shape QP (local structure)}

Global constraints alone are typically insufficient to encode local structure (heavy tails,
multimodality, pronounced skew, barrier--like features in energy space \footnote{Phenomena such as multimodality can, in principle, be represented either in the core (by including mode-resolving features in $F$) or in the shape term (via localized constraints). In practice we default to capturing such local structure in the shape correction, which keeps the core feature family low-dimensional and simplifies concavity enforcement, unless a specific application motivates encoding it directly in the core.}). To capture such effects, introduce additional localised constraint functions $\{g_r(c_j)\}_{r=1}^R$ (e.g. indicator masks for tails or side lobes). Let $M\in\mathbb{R}^{n\times R}$ be the corresponding constraint matrix $M_{jr}=g_r(c_j)$.

Although both stages use linear constraint matrices, the difference is structural:
the core stage uses a low-dimensional \emph{global} feature family (smooth functions of $c_j$),
whereas the shape stage uses \emph{localised masks} that are (by construction) orthogonal to the
core span so that they can adjust only local deviations without altering the global fit.
Concretely, we enforce $F^\top M = 0$ after Gram--Schmidt orthogonalisation (or equivalently
impose $(\Pi_{\mathrm{col}(F)} M)\eta=0$ as linear equalities), so the shape correction lies in
$\mathrm{col}(F)^\perp$.

We learn a \emph{shape} generator derivative by solving an analogous QP
\begin{equation}
  \min_{\alpha^{(S)},\eta}\;
  \|A\alpha^{(S)} - M\eta\|_2^2 + \gamma_S\|\alpha^{(S)}\|_2^2
  \label{eq:QP-shape}
\end{equation}
subject to concavity constraints, 
again enforcing \eqref{eq:concavity} (with $\alpha$ replaced by $\alpha^{(S)}$). In practice,
we additionally ensure that the shape correction does not alter the global constraints,
either by orthogonalising the shape features with respect to the core feature span or by
adding linear equalities that remove any component of the shape fit lying in
$\mathrm{span}(F)$.

Let $\alpha^{(S)\star}$ denote the resulting coefficients and define
\[
  G'_{\mathrm{shape}}(u)=\sum_{\ell=0}^L \alpha^{(S)\star}_\ell \psi_\ell(u),
  \qquad
  G_{\mathrm{shape}}(u)=\int_0^u G'_{\mathrm{shape}}(t)\,\mathrm{d}t,
\]
with $G_{\mathrm{shape}}(0)=0$.

\subsubsection*{Full generative generator and entropy}

The full learned generator is the sum
\begin{equation}
  G_{\mathrm{gen}}(u)=G_{\mathrm{core}}(u)+G_{\mathrm{shape}}(u),
  \qquad 0\le u\le 1,
  \label{eq:Ggen}
\end{equation}
which remains strictly concave. The \emph{generative entropy} of a histogram $p\in\Delta_n$
is then
\begin{equation}
  H_{\mathrm{gen}}(p)=\sum_{j=1}^n G_{\mathrm{gen}}(p_j).
  \label{eq:Hgen}
\end{equation}

By construction, $H_{\mathrm{gen}}$ is a data--driven, system--specific entropy: it is the
(trace--form) functional whose maximum--entropy principle, under the learned constraints,
selects the empirical histogram. Classical entropies are recovered only in regimes where
their generators satisfy the same inverse optimality conditions.

\subsection{Step 4: Internal energy, entropy--energy relation, and temperature}

\subsubsection*{Internal energy}

Given $\mathcal{E}_{\mathrm{gen}}$ and $p$, define the internal energy as the expectation
\begin{equation}
  U
  \;\equiv\;
  \mathbb{E}_p\!\big[\mathcal{E}_{\mathrm{gen}}(X)\big]
  =
  \int_{\mathcal{X}} \mathcal{E}_{\mathrm{gen}}(x)\,p(x)\,\mathrm{d}\mu(x).
  \label{eq:U-micro}
\end{equation}
Equivalently, in energy space (continuous density or discrete histogram),
\begin{equation}
  U
  =
  \mathbb{E}_{p_E}[E]
  =
  \int e\,p_E(e)\,\mathrm{d}e
  \quad\text{or}\quad
  U=\sum_{j=1}^n c_j\,p_{E,j}.
  \label{eq:U-energy}
\end{equation}

\subsubsection*{$S(U)$ along an admissible state family}

Consider a one--parameter family of admissible states $\alpha\mapsto p_\alpha(x)$, with
corresponding generative energies $\mathcal{E}_{\mathrm{gen},\alpha}$ and energy histograms
$p_{E,\alpha}\in\Delta_n$ obtained by Step~2. For each $\alpha$ define
\[
  U(\alpha)=\int_{\mathcal{X}} \mathcal{E}_{\mathrm{gen},\alpha}(x)\,p_\alpha(x)\,\mathrm{d}\mu(x),
  \qquad
  S(\alpha)=H_{\mathrm{gen}}\!\big(p_{E,\alpha}\big).
\]
If $U(\alpha)$ is monotone on the domain of interest, we may invert $\alpha\mapsto U(\alpha)$
and view entropy as a function of internal energy along this family: $S=S(U)$.

\subsubsection*{Temperature from the entropy--energy slope}

We define (thermoinformational) temperature by the slope of $S(U)$ along the chosen family:
\begin{equation}
  \frac{1}{T}
  =
  \left.\frac{\partial S}{\partial U}\right|_{\text{along the family}}
  =
  \frac{\dfrac{\mathrm{d}}{\mathrm{d}\alpha} S(\alpha)}
       {\dfrac{\mathrm{d}}{\mathrm{d}\alpha} U(\alpha)}.
  \label{eq:T-def}
\end{equation}
Writing $p_{E,\alpha}=(p_{E,\alpha,1},\dots,p_{E,\alpha,n})$, we have in discrete form
\[
  \frac{\mathrm{d}}{\mathrm{d}\alpha}S(\alpha)
  =
  \sum_{j=1}^n G_{\mathrm{gen}}'\!\big(p_{E,\alpha,j}\big)\,
  \frac{\mathrm{d}}{\mathrm{d}\alpha}p_{E,\alpha,j},
\]
and similarly $U(\alpha)=\sum_{j=1}^n c_j\,p_{E,\alpha,j}$ yields
\[
  \frac{\mathrm{d}}{\mathrm{d}\alpha}U(\alpha)
  =
  \sum_{j=1}^n c_j\,\frac{\mathrm{d}}{\mathrm{d}\alpha}p_{E,\alpha,j}
  \;+\;
  \text{($\alpha$--dependent terms $\mathcal{E}_{\mathrm{gen},\alpha}$)*}.
\]
  \\
  \text{* when applicable}

If the learned energy map changes along the family, the bin representatives $c_j$ (energy levels)
also depend on $\alpha$, and
\[
U(\alpha)=\sum_{j=1}^n c_j(\alpha)\,p_{E,\alpha,j} \quad\Rightarrow
\]
\[\Rightarrow\quad
\frac{dU}{d\alpha}=\sum_{j} c_j(\alpha)\frac{d}{d\alpha}p_{E,\alpha,j}
+\sum_{j} p_{E,\alpha,j}\frac{d}{d\alpha}c_j(\alpha).
\]
The first term is redistribution at fixed levels; the second is motion of the levels
(work-like contribution in Supplementary Note~S2.1).

For a one-parameter family of states $p_\alpha$ and learned energies
$E_{\mathrm{gen},\alpha}$, the internal-energy derivative splits as
\[
\frac{dU}{d\alpha}
=\]

\[=\int E_{\mathrm{gen},\alpha}(x)\,\partial_\alpha p_\alpha(x)\,d\mu(x)
+
\int \big(\partial_\alpha E_{\mathrm{gen},\alpha}(x)\big)\,p_\alpha(x)\,d\mu(x)\]
\[
\frac{dU}{d\alpha}\equiv \delta Q(\alpha)+\delta W(\alpha).
\]
The first term corresponds to redistribution on a fixed energy landscape (heat-like
contribution), whereas the second captures deformation of the learned energy landscape
(work-like contribution).

The ratio \eqref{eq:T-def} then defines $T$ for that state family.

\paragraph{Operational validation of temperature via thermal contact.}
To verify that the inferred temperature has an operational meaning beyond the formal identity
$\beta(U)=\partial S/\partial U$, we perform a \emph{thermal-contact} test on two synthetic data systems
(two Gaussian mixtures with $\mu=2$ and $\mu=3$), using a shared learned energy axis but
system-specific inferred entropies.
For a fixed total energy $U_{\mathrm{tot}}$ (chosen in the interior of the common admissible range),
the predicted equilibrium split is the maximiser of
$S_A(U_A)+S_B(U_{\mathrm{tot}}-U_A)$, which implies $\beta_A(U_A^\star)=\beta_B(U_B^\star)$.
In this experiment we obtain $U_A^\star=3.289$, $U_B^\star=3.360$ and
$\lvert \beta_A(U_A^\star)-\beta_B(U_B^\star)\rvert=3.21\times 10^{-3}$
(Fig.~S5; details in Supplementary Note~S10).

\medskip

This construction is intentionally \emph{system-specific}: the objective is to obtain
the most informative energy and entropy representation of a given empirical data system,
rather than to impose a fixed entropy form across disparate systems.
Because the entropy functional is inferred from the empirical macrostate geometry,
the resulting thermodynamic variables $(U,S,T,F)$ are then defined \emph{consistently}
via the associated MaxEnt and Legendre structure on that same system.
Falsifiability is assessed by a within-system holdout test in which the entropy functional
is learned on a training subset and then held fixed while predicting held-out distributions
from held-out macrostates (Supplementary Note~S3.1).

\noindent\textbf{Summary.}
The construction proceeds as:
\begin{enumerate}
  \item infer $\mathcal{E}_{\mathrm{gen}}$ by minimising an $L^2$ discrepancy between $p$ and a
  Boltzmann--type model $q_\theta$;
  \item push $p$ forward under $\mathcal{E}_{\mathrm{gen}}$ to obtain an energy--space histogram $p_E$;
  \item construct $H_{\mathrm{gen}}$ by inverse maximum entropy (core + shape QPs) and set $S=H_{\mathrm{gen}}(p_E)$;
  \item define $U=\mathbb{E}_p[\mathcal{E}_{\mathrm{gen}}(X)]$ and $T^{-1}=\mathrm{d}S/\mathrm{d}U$ along admissible families.
\end{enumerate}
This yields a thermodynamically consistent description $(U,S,T)$ grounded in both
data--driven energy and data--driven entropy.

\section{Results}

\paragraph*{Role of the Hamiltonian across the examples.}
The three validation examples are ordered from (i) systems with a known mechanistic Hamiltonian,
to (ii) systems where a Hamiltonian exists in principle but is treated as unknown and inferred
from samples, to (iii) non-mechanistic multimodal distributions where no \emph{a priori} physical
Hamiltonian is available. In case (i) (harmonic/single well), the canonical Boltzmann--Gibbs form
is known and Stage-I on the energy axis is therefore optional and mainly serves as a sanity check.
In case (ii) (bistable double well), the Hamiltonian $V(x)$ exists but we deliberately perform a
Hamiltonian-free recovery from data; Stage-I is useful because the empirical pushforward
$\hat p_E$ is sensitive to binning and barrier-region sparsity, and the MaxEnt projection yields a
strictly positive, constraint-consistent energy-state for stable computation of $(S,U,T)$.
In case (iii) (Gaussian mixtures), there is no physically dictated Hamiltonian; the generative
energy and inverse-MaxEnt steps provide the only principled route to a thermodynamically
consistent state-space representation.

\subsection{Calibration on a harmonic single--well potential}
\label{subsec:harmonic}

We first calibrate the generative thermodynamic construction on a simple
equilibrium system where the usual Boltzmann--Gibbs picture is known to be
exact.  Consider a one–dimensional overdamped particle in a harmonic
potential,
\begin{equation}
  V(x) = \tfrac{1}{2} k x^2 ,
\end{equation}
coupled to a thermal bath at temperature $T$.  The stationary distribution
over microstates is Gaussian,
\begin{equation}
  p(x) = \sqrt{\frac{\beta k}{2\pi}}
  \exp\!\bigl[-\tfrac{1}{2}\beta k x^2\bigr],
  \qquad \beta = (k_B T)^{-1},
\end{equation}
with variance $\langle x^2\rangle = k_B T/k$ and internal energy
$V_\mathrm{phys} = \tfrac{1}{2} k_B T$.

Here we use the known Hamiltonian as a ground-truth baseline; applying the energy-space Stage-I
projection would recover the same Boltzmann--Gibbs form up to discretisation error, so we omit it. In this unimodal equilibrium regime the generative construction should reduce
to the standard description: the learned generative energy should be quadratic
in $x$, and the resulting entropy and temperature should coincide with the
usual thermodynamic ones.  To verify this, we drew $N=2\times 10^5$ samples
from the exact Gaussian distribution for several temperatures
$T\in\{0.5,1.0,1.5,2.0\}$ (with $k_B=k=1$).  From the empirical histogram
$\hat p(x)$ we performed the generator step restricted to a quadratic energy
family,
\begin{equation}
  E_{\mathrm{gen}}(x;\omega) = \omega_0 + \omega_2 x^2 ,
\end{equation}
by fitting $E_{\mathrm{gen}}$ to $-\log \hat p(x)$ in least–squares sense
(see Supplemental Material for numerical details).  In the infinite–data
limit the exact relation $-\log p(x) = \mathrm{const} + \tfrac{1}{2}\beta k
x^2$ implies that the optimal generator within this family is quadratic with
coefficient $\omega_2 = \tfrac{1}{2}\beta k$.

\begin{figure*}[t]
  \centering
  \includegraphics[width=0.4\textwidth]{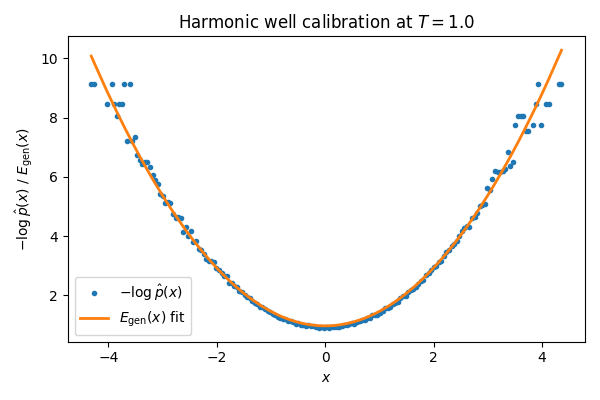}
  \includegraphics[width=0.4\textwidth]{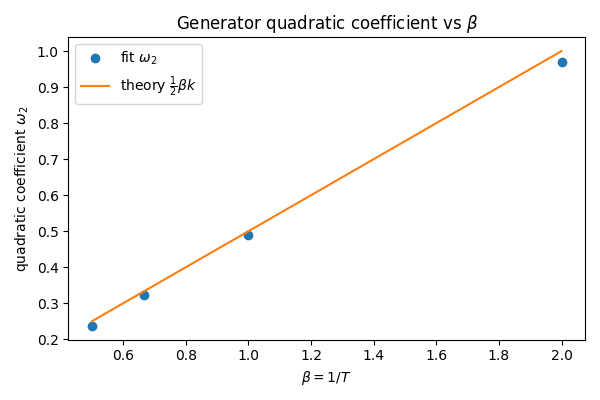}
  \caption{\textbf{Harmonic-well calibration of the generative energy.}
\textbf{(a)} Empirical surprisal $-\log\hat p(x)$ (blue) obtained from samples of the exact Gaussian equilibrium distribution in a quadratic potential at $T=1$, shown together with the fitted generative energy $E_{\mathrm{gen}}(x)$ (orange). The fit recovers the expected quadratic landscape (up to an additive constant) across the sampled support.
\textbf{(b)} Fitted quadratic coefficient $\omega_2$ of $E_{\mathrm{gen}}(x)\approx \omega_2 x^2+\omega_0$ versus inverse temperature $\beta=1/T$ over the calibration sweep. The fitted coefficients follow the theoretical prediction $\omega_2=\tfrac{1}{2}\beta k$ (line), confirming the correct temperature scaling and the expected affine gauge freedom in the learned energy.}
  \label{fig:harmonic_calibration}
\end{figure*}

Figure~\ref{fig:harmonic_calibration}(a) shows the result of this procedure
for $T=1$.  The fitted $E_{\mathrm{gen}}(x)$ is visually indistinguishable
from $-\log \hat p(x)$ across the support of the distribution, confirming
that the generator recovers the expected quadratic energy landscape from
finite data.  Panel~\ref{fig:harmonic_calibration}(b) summarizes the fitted
quadratic coefficient $\omega_2$ as a function of $\beta=1/T$ for all
temperatures.  The points lie on the theoretical line
$\omega_2 = \tfrac{1}{2}\beta k$ to within a few percent, with typical
relative errors below $5\%$.  This calibration demonstrates that, for a
simple harmonic single–well, the generative energy coincides with the usual
Hamiltonian up to the expected affine gauge freedom, and the framework
contains classical Boltzmann--Gibbs thermodynamics as a special unimodal
limit.

\subsection{Bistable double–well potential}

To test the framework on a genuinely multimodal system we consider a
one–dimensional particle in a bistable quartic potential,
\begin{equation}
  V(x) = a x^{4} - b x^{2}, \qquad a>0,\,b>0,
\end{equation}
sampled in thermal equilibrium at inverse temperatures
$\beta \in \{0.5,1,2\}$. For each $\beta$ we generate $N=10^5$ samples
from the canonical density
$p_\beta(x)\propto \exp[-\beta V(x)]$ using Metropolis–Hastings
dynamics. The resulting stationary distribution is strongly bimodal,
with mass concentrated near the two minima of $V(x)$.
Although $V(x)$ is known in this synthetic test, we treat the Hamiltonian as unknown and infer the energy representation from samples; Stage-I is used here to stabilise the energy-space state by projecting the empirical pushforward onto the MaxEnt manifold induced by the learned generator.

\begin{figure*}[t]
  \centering
  \includegraphics[width=0.32\textwidth]{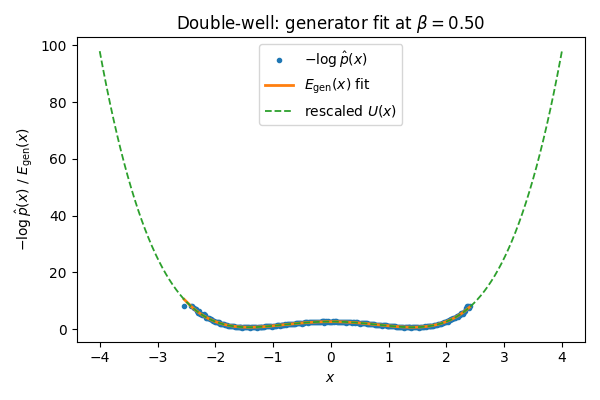}
  \includegraphics[width=0.32\textwidth]{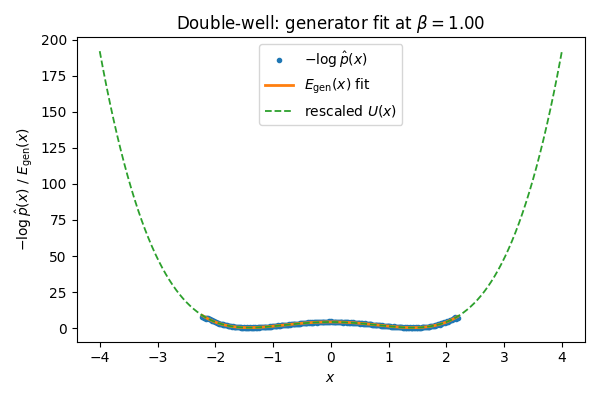}
  \includegraphics[width=0.32\textwidth]{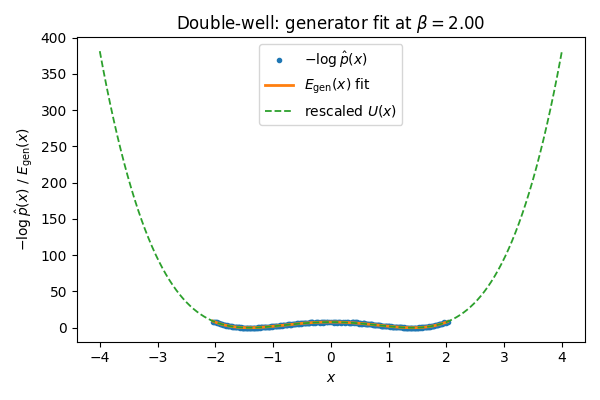}
  \caption{Double–well generator calibration.
  For each inverse temperature $\beta\in\{0.5,1,2\}$ we show the
  empirical curve $-\log \hat p(x)$ (blue points), the fitted generator
  $E_{\mathrm{gen}}(x)$ (orange line), and the true quartic potential
  $V(x)$ after an optimal affine rescaling (green dashed line).
  The generative energy reconstructs the bistable landscape across the
  full range of $x$, including both minima and the barrier region.}
  \label{fig:dw_generator}
\end{figure*}

\paragraph*{Generator recovery from $-\log \hat p(x)$.}
From each trajectory we estimate the empirical density $\hat p(x)$ by
kernel density estimation and then fit a quartic generator
\begin{equation}
  E_{\mathrm{gen}}(x) = w_0 + w_2 x^2 + w_4 x^4
\end{equation}
by least squares to the curve $-\log \hat p(x)$ over a central window
containing both wells. Figure~\ref{fig:dw_generator} shows the fit for
all three inverse temperatures. In each case the recovered generator
tracks $-\log \hat p(x)$ closely across the entire bimodal region, and
after an affine rescaling it is almost indistinguishable from the true
potential $V(x)$. This confirms that the generative–energy step recovers
the correct energy landscape, even in the presence of two stable
basins separated by a barrier.

\paragraph*{Energy--space Stage-I (trace-form) reconstruction and internal energy.}
Given $E_{\mathrm{gen}}$, we push the empirical distribution onto
energy space and discretize the resulting energy histogram
$\hat p_E$ on a fixed grid $\{E_j\}$.
On this energy axis we perform the Stage-I (core) trace-form construction:
we learn a strictly concave generator derivative $G'(\cdot)$ whose
values on the empirical histogram bins satisfy a low-dimensional
\emph{slope-in-span} condition,
$G'(\hat p_{E,j}) \approx (F\theta)_j$ for a chosen feature family $F(E)$,
with concavity enforced by monotonicity of $G'$ in probability.
The corresponding MaxEnt distribution on the same grid is then obtained
in KKT-closed form as
$p_{E,j}^\star = (G')^{-1}((F\theta)_j)$, followed by normalization
(see SM for numerical details and diagnostics).
We define the generative internal energy from this reconstructed
distribution as
\begin{equation}
  U_{\mathrm{gen}} = \sum_j E_j\,p_E^\star(E_j)\,\Delta E .
\end{equation}

\begin{figure}[t]
  \centering
  \includegraphics[width=0.45\textwidth]{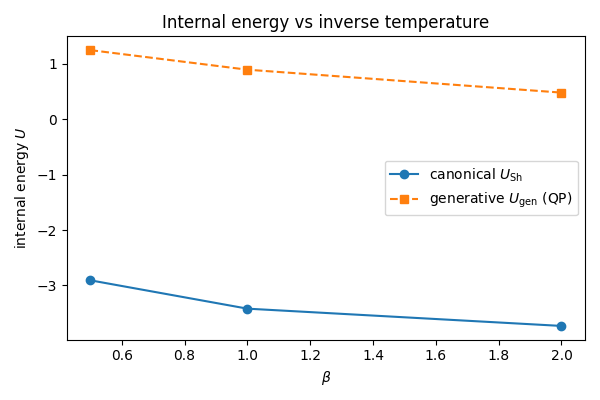}
  \caption{Internal energy as a function of inverse temperature for the
  double–well system. Blue: canonical Shannon internal energy
  $U_{\mathrm{Sh}}(\beta)=\mathbb{E}_{p_\beta}[V(x)]$ from the known
  potential. Orange: generative internal energy
  $U_{\mathrm{gen}}(\beta)$ computed from the QP energy distribution
  $p_E^\star$. Both energies vary monotonically with $\beta$ but live on
  different scales: $U_{\mathrm{Sh}}$ is negative and becomes more
  negative as the ensemble cools, whereas $U_{\mathrm{gen}}$ is
  positive and moves towards lower values as probability mass
  concentrates in the lowest generative–energy band.}
  \label{fig:dw_energy_vs_beta}
\end{figure}

Across the three inverse temperatures we obtain
$U_{\mathrm{gen}}\approx \{1.25, 0.90, 0.48\}$, which decrease
monotonically with $\beta$ as expected for a cooling bistable system
(Fig.~\ref{fig:dw_energy_vs_beta}, orange curve).

For comparison we also compute the conventional canonical internal
energy
$U_{\mathrm{Sh}}(\beta)=\mathbb{E}_{p_\beta}[V(x)]$, which is negative
because the minima of $V(x)$ lie below zero. As $\beta$ increases, the
canonical energy $U_{\mathrm{Sh}}$ becomes more negative, whereas the
generative internal energy $U_{\mathrm{gen}}$ moves towards lower
positive values. 
branches of the same underlying energy landscape: $U_{\mathrm{Sh}}$
measures how deeply the canonical ensemble sits in the wells of the
physical potential $V(x)$, while $U_{\mathrm{gen}}$ measures how far
into the low–energy band of the data–driven generator the system has
progressed. This illustrates that, even for the same microscopic
potential, Shannon–based and generative thermodynamics can assign
different, complementary internal–energy scales.
We do \emph{not} expect $U_{\rm Sh}$ and $U_{\rm gen}$ to coincide numerically, even for the same
microscopic system, because they are defined on different representations and gauges.
$U_{\rm Sh}(\beta)=\mathbb E_{p_\beta}[V(x)]$ uses the physical potential with its chosen zero,
whereas $U_{\rm gen}$ is the mean of the learned energy coordinate (defined only up to affine
gauge) computed under the MaxEnt-consistent energy-space reconstruction $p_E^\star$.
The invariant comparison is therefore the \emph{ordering/trend} with $\beta$ and the
consistency relations (e.g. $T^{-1}=dS/dU$), not equality of absolute values.

In the harmonic case, the energy distribution can be obtained in closed form (and is smooth and strictly positive on its support), so applying the energy-space Stage-I
MaxEnt projection is not necessary for defining the state variables and would be redundant. We therefore use the harmonic system primarily as a ground-truth sanity check. 
In contrast, for the bistable potential the empirical pushforward histogram $\hat p\_E$ is highly sensitive to finite-sample and binning effects (notably near the barrier region), and may contain empty or near-empty bins. Because the thermodynamic construction uses state functions and derivatives such as $T^{-1}=dS/dU$, these artefacts can propagate and obscure the intended structural comparisons. We therefore compute $(U,S,T)$ on the Stage-I energy-space MaxEnt state $p_E^\star$, i.e. the unique distribution that
maximizes the learned trace-form entropy under the chosen constraint family. This projection yields a strictly positive, constraint-satisfying macrostate for which the KKT
relations hold exactly, ensuring that the subsequent thermodynamic relations are evaluated on the same MaxEnt manifold induced by the learned generator, rather than on a noisy finite-sample histogram $\hat p_E$. More generally, Stage-I on energy space is optional when $\hat p_E$ is already
smooth/positive and derivatives are stable, but it is recommended for multimodal systems where $\hat p_E$ is noisy and thermodynamic derivatives are sensitive.

\subsection*{Mixture-of-Gaussians “data system”: discovering structure without a known Hamiltonian}

As a third example we consider a purely data–driven setting where no
physical Hamiltonian is specified. In this example there is no \emph{a priori} mechanistic Hamiltonian that would dictate a canonical
Boltzmann--Gibbs form; the energy and entropy are therefore inferred directly from the observed
distribution via the generative construction. We draw samples from a
one–dimensional, symmetric mixture of Gaussians
\[
  p(x)
  = \tfrac{1}{2}\,\mathcal{N}(-\mu,1) + \tfrac{1}{2}\,\mathcal{N}(+\mu,1),
\]
with mixture separation $\mu\in\{1,2,3\}$. In contrast to the double–well
system, here the mixture is treated as an arbitrary “data system”:
the only inputs to the construction are the observed microstates
$x_n$ and their empirical density $\hat{p}(x)$; there is no notion of a
true potential $V(x)$.

\begin{figure*}[t]
  \centering
  \includegraphics[width=0.98\textwidth]{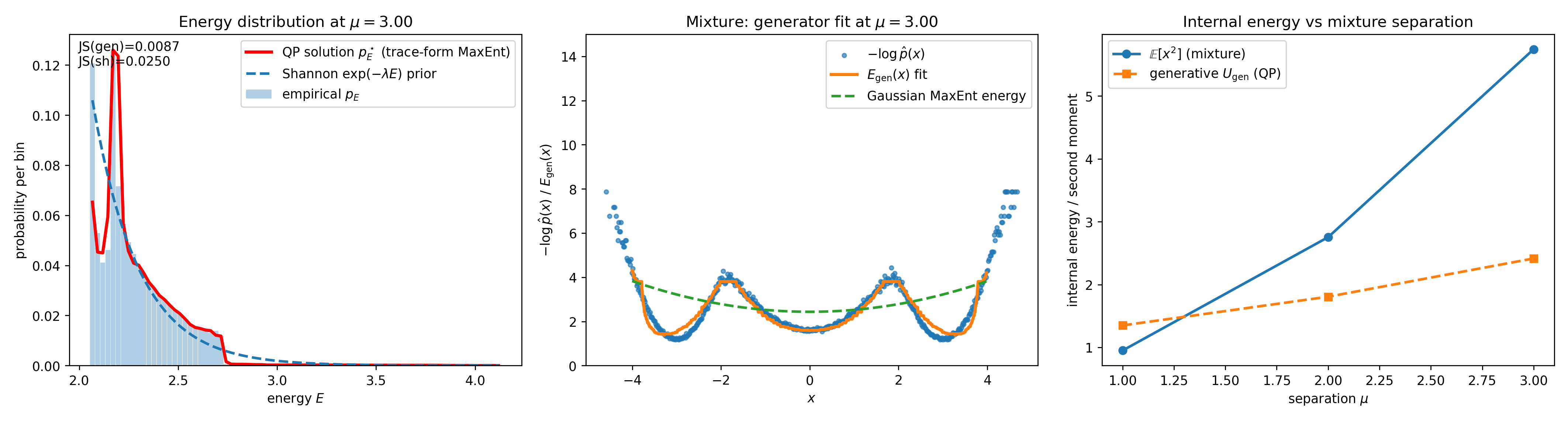}
  \caption{
    Thermoinformational state construction for an agnostic
    mixture-of-Gaussians data system.
    Left: energy–space quadratic–program solution $p_E^\star$ (red)
    versus the empirical energy histogram (blue bars) and a
    Shannon–style exponential prior $\propto\exp(-\lambda E)$ (dashed
    blue) for a two–component Gaussian mixture with separation
    $\mu=3$. The QP solution captures the sharp low–energy peak and
    long tail induced by the mixture, whereas the exponential prior
    smooths over these features.
    Middle: generator fit in microstate space. We fit a sixth-order
    polynomial generator $E_{\mathrm{gen}}(x)$ (orange) to
    $-\log\hat{p}(x)$ (blue points); the Gaussian MaxEnt quadratic
    energy (green dashed) matches only the global variance and cannot
    represent the multimodal structure of the mixture.
    Right: internal energy as a function of mixture separation
    $\mu\in\{1,2,3\}$. The canonical second moment
    $\mathbb{E}[x^2]$ (blue) grows rapidly with separation, while the
    generative internal energy $U_{\mathrm{gen}}$ (orange) increases
    much more gently, reflecting that the thermodynamic description
    is controlled by structure in energy space rather than by raw
    spread in $x$.
    Results for $\mu=1$ and $\mu=2$ (including the corresponding
    generator and energy–space plots) are provided in
    Supplementary Fig.~S2.
  }
  \label{fig:mixture_main}
\end{figure*}

We first fit a generative energy $E_{\mathrm{gen}}(x)$ by regressing a
sixth–order polynomial onto $-\log \hat{p}(x)$, exactly as in the
double–well example. The centre panel of Fig.~\ref{fig:mixture_main}
shows the result for the most separated mixture ($\mu=3$).  The orange
curve gives the fitted generator $E_{\mathrm{gen}}(x)$, which develops a
clear multi–well structure aligned with the mixture modes, closely
tracking the empirical $-\log \hat{p}(x)$ (blue points).  For
comparison we overlay (green dashed) the quadratic energy
corresponding to the Shannon MaxEnt solution with the same mean and
variance as the mixture. As expected, this Gaussian MaxEnt energy is
strictly unimodal and therefore incompatible with the multimodal data:
it can match the overall spread of $x$, but not the arrangement of
probability mass across modes.

Having constructed the generator, we push the empirical distribution
into energy space and solve the generative quadratic program on the
energy axis. The left panel of Fig.~\ref{fig:mixture_main} shows the
empirical energy histogram $p_E(E)$ (blue bars) together with the QP
solution $p_E^\star(E)$ (red) for $\mu=3$.  The QP solution captures
both the sharp low–energy concentration and the long, structured tail
induced by the mixture. For comparison we also plot a Shannon–style
exponential prior $\propto \exp(-\lambda E)$ (blue dashed), fitted to
match the empirical mean energy. This exponential form is the natural
MaxEnt solution if one constrains only the mean energy in the Shannon
framework; it produces a single–scale decay and necessarily washes out
the multi–scale structure in $p_E(E)$ that the generative solution
retains.

The right panel of Fig.~\ref{fig:mixture_main} summarizes how the
thermodynamic variables respond as we increase the mixture separation.
The blue curve shows the canonical second moment
$\mathbb{E}[x^2]$ of the mixture, which grows rapidly with $\mu$,
reflecting that the two components are moving further apart.  In
contrast, the generative internal energy $U_{\mathrm{gen}}$ (orange)
increases much more gently with separation.  This illustrates a key
property of the thermoinformational construction: internal energy is
defined with respect to the learned generator in energy space, not
directly from the raw coordinate spread, so it is sensitive to
deformation of the energy landscape rather than to the mere fact that
the data occupy a wider range of $x$–values.  In particular, once the
generator has organized the mixture into a multi–well energy
structure, further separation of the wells changes $U_{\mathrm{gen}}$
only modestly, whereas the Shannon description in $x$–space continues
to inflate with $\mathbb{E}[x^2]$.

Full panels for $\mu=1$ and $\mu=2$, including the corresponding
generator fits and energy–space distributions, are provided in
Supplementary Fig.~S2, together with the
behaviour of the generative entropy $H_{\mathrm{gen}}$, which remains
nearly constant across mixture separations while the Shannon entropy in
$x$–space increases markedly.

\paragraph{Within-system holdout test (distributional prediction).}
A train/validation split is performed at the microstate level.
The generative energy map $E_{\mathrm{gen}}(x)$, fixed energy support/binning,
and the learned entropy functional are inferred using the training subset only.
On the held-out subset, only the macroscopic constraints are recomputed and the MaxEnt
solution is obtained with the learned functional held fixed.
For a representative split (seed $=1$, $N=200{,}000$, $\mu=2.0$), the validation prediction
achieves ${\rm JS}=5.44\times 10^{-4}$ and $W_{1}=3.42\times 10^{-4}$ relative to the held-out
empirical $\hat p_E$, whereas a Shannon MaxEnt baseline under the same band constraints yields
${\rm JS}=9.57\times 10^{-2}$ and $W_{1}=3.25\times 10^{-1}$.
This demonstrates that the learned functional captures distributional structure that fixed-form
entropies discard under identical constraint information.

\section{Thermodynamic structure and consistency}

The generative construction above is not just a convenient way of
organising data: it carries a full thermodynamic structure that is
precisely analogous to classical equilibrium thermodynamics.  Here we
summarise the main results; detailed statements and proofs are given in
Supplementary Notes~S1–S3.

\paragraph{Inverse–MaxEnt representation.}
Given any finite microstate space $\mathcal{C}=\{c_j\}$, any empirical
histogram $p=(p_j)$ on $\mathcal{C}$, and any finite family of
macroscopic observables $f_k(c_j)$ used as constraints, there exists a
strictly concave trace–form generator $G$ such that $p$ is the unique
maximum–entropy solution of
\[
  \max_{q} H_G(q)
  \quad\text{s.t.}\quad
  \sum_j q_j f_k(c_j) = \sum_j p_j f_k(c_j)\quad\forall k,
\]
where $H_G(q)$ is the generative entropy induced by $G$.
Under a mild richness condition on the constraint family, this $G$ is
unique up to an affine gauge transformation $G\mapsto G+\alpha u+\beta$
(Supplementary Theorem~S1.1),
where $u\in(0,1]$ denotes a probability argument (e.g. $u=p_j$),
and $\alpha,\beta\in\mathbb{R}$ are constants (a linear term and an additive offset).  Our energy–space quadratic program is a
finite–dimensional approximation of this inverse–MaxEnt map: as the
generator basis on the energy axis is densified, the fitted
$E_{\mathrm{gen}}$ converges to the inverse–MaxEnt generator for the
underlying histogram and constraints.

The inverse--MaxEnt step admits a precise finite-dimensional interpretation.
For any interior histogram $p$ on a finite support and any full-rank constraint family,
there exists a strictly concave trace-form generator $G$ such that $p$ is the unique
maximum-entropy solution under those constraints. Under a mild richness condition on the
constraint family, this generator is unique up to the affine gauge transformation
$G(u)\mapsto G(u)+\alpha u+\beta$, where $u\in(0,1]$ is a probability argument and
$\alpha,\beta\in\mathbb{R}$ are constants. Moreover, our energy-space quadratic program
is a finite-dimensional approximation of this inverse--MaxEnt map: as the generator basis
is densified, the fitted generator converges to the inverse--MaxEnt generator associated
with the underlying histogram and constraints. Proofs and precise hypotheses are given in
Supplementary Note~S1.

\paragraph{Generative thermodynamic laws.}
For a smooth path of systems
$\varepsilon\mapsto\bigl(p_\varepsilon,E_{\mathrm{gen},\varepsilon}\bigr)$
we define internal energy, entropy and temperature by
\[
  U(\varepsilon) = \sum_x E_{\mathrm{gen},\varepsilon}(x)\,p_\varepsilon(x),
  \qquad
  S(\varepsilon) = H_{\mathrm{gen}}(p_{E,\varepsilon}),
  \qquad
\]

\[T^{-1}(\varepsilon)
  = \frac{\mathrm{d}S/\mathrm{d}\varepsilon}
         {\mathrm{d}U/\mathrm{d}\varepsilon}.\]
With these definitions the usual thermodynamic laws hold.
(i) \emph{First law:} there is a canonical split
$\mathrm{d}U = \dot Q\,\mathrm{d}\varepsilon + \dot W\,\mathrm{d}\varepsilon$
into ``heat'' $\dot Q$ coming from changes in $p_\varepsilon$ at fixed
$E_{\mathrm{gen},\varepsilon}$ and ``work'' $\dot W$ coming from
deformations of $E_{\mathrm{gen},\varepsilon}$ at fixed
$p_\varepsilon$ (Supplementary Note~S2.1).  Pure reshuffling at fixed
generator has $\dot W=0$, whereas pure deformation at fixed $p$ has
$\dot Q=0$.  (ii) \emph{Second law (H–theorem):} for the projected
gradient flow of generative entropy on the constraint manifold in
energy space, $S(t)=H_{\mathrm{gen}}(p_E(t))$ is non–decreasing and
every trajectory converges to the unique generative MaxEnt state
$p_E^\star$ (Supplementary Theorem~S2.2).  
(iii) \emph{Zeroth law:} for
two weakly coupled generative systems the equilibrium condition is
equality of generative temperatures $T_1=T_2$, with the usual entropy
additivity and energy exchange arguments (Supplementary Note~S2.3).


\paragraph*{Generalised H-theorem (statement).}
Let $p^\star$ denote the unique MaxEnt state associated with the learned trace-form entropy
and the imposed constraints. Under the projected gradient-ascent flow on the constraint
manifold, the constraints are preserved, the entropy is nondecreasing,
\[
\frac{d}{dt}H_G(p(t))\ge 0,
\]
and equality holds only at the stationary point $p^\star$. Hence $p^\star$ is a Lyapunov
maximiser for the induced dynamics. The formal statement and proof are given in
Supplementary Note~S2.

A simple numerical illustration of the H–theorem is given in the
mixture–of–Gaussians data system at separation $\mu=3$.  Starting from
the empirical energy histogram $p^{(0)}_E$ and iterating a discrete
projected gradient–ascent step for the generative micro–entropy
$H_{\mathrm{micro}}$ under fixed band–mass constraints, we observe
approximately exponential decay of $\|p^{(t)}_E - p_E^\star\|_2$ and a
monotone increase of $H_{\mathrm{micro}}(p^{(t)}_E)$ to the QP
optimum $H_{\mathrm{micro}}(p_E^\star)$ (Supplementary
Fig.~S3).  This confirms that generative entropy
acts as a Lyapunov functional for relaxation toward the MaxEnt state
in this fully data–driven system.


\paragraph*{Thermal contact and equilibrium prediction.}
Under weak contact, additive entropies, and conserved total energy $U_{\mathrm{tot}}$, define
\[
S_{\mathrm{tot}}(U_A)\coloneqq S_A(U_A)+S_B\big(U_{\mathrm{tot}}-U_A\big).
\]
The predicted equilibrium split is
\[
U_A^\star=\arg\max_{U_A} S_{\mathrm{tot}}(U_A),
\qquad
U_B^\star=U_{\mathrm{tot}}-U_A^\star.
\]
At an interior optimum, the first-order condition yields the operational equalisation rule
\[
\beta_A(U_A^\star)=\beta_B(U_B^\star),
\]
i.e.\ energy flows until the inferred inverse temperatures match.

\paragraph{Legendre structure and free energies.}
The generative entropy induces the standard equilibrium Legendre
machinery.  Defining the microcanonical entropy $S(U,X)$ as the
supremum of $H_{\mathrm{gen}}$ over all states with fixed internal
energy $U$ and other conserved observables $X=(X_i)$, and the dual
potential $\Phi(\omega,\theta)$ via a constrained Legendre transform,
we obtain that $S$ is concave, $\Phi$ is convex, and the usual
conjugacy relations
\[
  \omega = \frac{\partial S}{\partial U},
  \qquad
  \theta_i = \frac{\partial S}{\partial X_i},
  \qquad
  U = \frac{\partial \Phi}{\partial \omega},
  \qquad
  X_i = \frac{\partial \Phi}{\partial \theta_i}
\]
hold, together with Maxwell relations and response coefficients
(Supplementary Theorem~S3.1).  In the double–well and mixture
examples, the numerically computed $(U,S,T)$ follow these relations, so
the associated generative free energy behaves as a bona fide
thermodynamic potential on the learned state space.

\paragraph{Gauge, coarse–graining and geometry.}
Affine rescalings of the generator,
$E_{\mathrm{gen}}\mapsto aE_{\mathrm{gen}}+b$, shift the numerical
value of $U$ and rescale the conjugate variables but leave
$H_{\mathrm{gen}}$ and all MaxEnt predictions invariant; generative
thermodynamics therefore has an energy–gauge freedom analogous to the
choice of zero in classical thermodynamics.  Coarse–graining of the
microstate space corresponds to a stochastic map on the simplex under
which $H_{\mathrm{gen}}$ is non–decreasing, and the negative Hessian
of $H_{\mathrm{gen}}$ restricted to the simplex defines an
entropy–induced Riemannian metric that reduces to Fisher information in
the Shannon case (Supplementary Note~S3.2).  Together, these results
show that the data–driven construction of $E_{\mathrm{gen}}$ and
$H_{\mathrm{gen}}$ recovers the full thermodynamic backbone—
conservation laws, second law, Legendre structure and geometric
consistency—on the empirical state spaces studied in this work.

\subsection{Heat, work, and the first law on learned energy state spaces}
\label{subsec:firstlaw}

A common failure mode in ``thermodynamics-inspired'' data analyses is that one defines
entropy- or energy-like scalars but never specifies \emph{process-level bookkeeping}:
what counts as heat versus work, and whether an analogue of the first law is satisfied.
In our construction, these notions are defined on the learned energy state space and
follow a transparent decomposition under protocols.

\paragraph*{Protocol and representation.}
Let a protocol be an externally controlled parameter $\lambda$ (e.g.\ bath temperature,
mixture separation, stimulus intensity) that changes the empirical microstate ensemble.
At each $\lambda$ the system induces an energy-space distribution $p_E(E;\lambda)$ on a
fixed discretisation $\{E_j\}_{j=1}^M$ of the learned energy axis.
When the energy map is held fixed along the protocol (as in the harmonic and protocol
checks), the internal energy is
\begin{equation}
  U(\lambda)=\sum_{j=1}^M p_j(\lambda)\,E_j\,\Delta E,
  \qquad p_j(\lambda)\approx p_E(E_j;\lambda).
\end{equation}
More generally, if the representation itself depends on the protocol (e.g.\ the fitted
generator changes with $\lambda$ so that the induced energy levels shift), we write
$E_j(\lambda)$ and keep the same grid indexation across $\lambda$ by construction
(see SM for the implementation details).

\paragraph*{Heat--work decomposition.}
The differential change of $U(\lambda)$ admits the standard thermodynamic split,
\begin{equation}
  dU
  = \sum_{j} E_j\, d(p_j \Delta E)
    \;+\; \sum_{j} (p_j\Delta E)\, dE_j
  \;\equiv\; \delta Q \;+\; \delta W.
  \label{eq:firstlaw_split}
\end{equation}
The first term, $\delta Q$, is the energy change due to \emph{redistribution of probability
mass} across energy states at fixed levels; it quantifies protocol-driven reweighting of
already-defined energy states. The second term, $\delta W$, is the change due to
\emph{motion of the energy levels themselves} under $\lambda$ (i.e.\ changes in the learned
energy map), and is therefore the natural work-like contribution on the inferred state
space. Eq.~\eqref{eq:firstlaw_split} is an identity once a consistent discretisation is fixed.

\paragraph*{First-law consistency in the demonstrations.}
In the harmonic calibration and the operational temperature protocol, the learned energy
map is fixed (quadratic form with fitted parameters) and the energy grid $\{E_j\}$ is held
constant; therefore $dE_j=0$ and $\delta W\equiv 0$, so that $dU=\delta Q$ and all changes
in $U$ are heat-like. In contrast, when the representation is allowed to vary with $\lambda$
(e.g.\ re-fitting $E_{\mathrm{gen}}$ at each condition), $\delta W$ captures the portion of
$U$-change attributable to deformation of the inferred energy landscape rather than mere
reweighting. In all cases, the framework provides an explicit process-level accounting and
a first-law statement on the learned energy state space (details and numerical checks are
reported in the SM).

\section{Discussion and outlook}
\label{sec:discussion}

We introduced a constructive route for assigning thermodynamic structure to an arbitrary
empirical system starting from microstate data. The central point is procedural: we first infer a
\emph{generative energy} that defines an intrinsic energy coordinate for the observed statistics, and
then learn a \emph{system--tailored entropy functional} via an inverse maximum--entropy construction
in the resulting energy representation. Only after this coupled energy--entropy specification do
macroscopic variables such as internal energy, an entropy--energy relation $S(U)$, and a temperature
defined by $T^{-1}=\partial S/\partial U$ along admissible state families become well-defined and
thermodynamically consistent.

The examples illustrate both calibration and necessity. In the harmonic (unimodal) limit, the
generative energy recovers the expected quadratic landscape and its temperature scaling (up to the
usual affine gauge freedom), demonstrating that classical Boltzmann--Gibbs thermodynamics is
contained as a special case. In contrast, multimodal settings (double well; Gaussian mixtures) show
why learning additional structure beyond global constraints can matter: barrier/coexistence features
in the learned energy representation can be preserved rather than collapsed into a single effective
mode. More broadly, the mixture example emphasizes that the construction is not restricted to
physical Hamiltonians; it applies to empirical distributions where the ``energy'' is learned rather than
assumed.

Several limitations and extensions are natural. The inferred energy depends on the chosen feature
family $\mathcal{E}(x;\theta)$ and the learned entropy depends on the selected constraint family in
energy space; these choices control resolution and should be treated as modelling decisions.
Finite-sample effects, especially in tails, motivate regularisation and careful discretisation (detailed
in the Supplementary Notes). Looking forward, important directions include multidimensional
state representations (multiple learned energy coordinates), automated constraint discovery, and
dynamic/non-equilibrium protocols where the constructed $(U,S,T)$ and associated Legendre
structure could be used to define response functions in empirical systems.

\section*{Data Availability Statement}
The computational codes used to generate the synthetic datasets and replicate the results of the figures presented in this study are openly available in the "Generative Thermodynamics" repository at \href{https://github.com/RDomenikos/Generative_Thermodynamics}{github.com/RDomenikos/Generative\_Thermodynamics}. The provided scripts are self-contained and do not require additional external data sources.

\section*{Declaration of Interests}
The authors declare no competing interest.

\section*{Author Contributions}
G.R.D. Conceptualization, Mathematical and Physics Analysis, Coding, writing manuscript, creating figures. C.L.U. Mathematical and Physics analysis, writing manuscript, review of concept and paper. V.L. supervision, acquisition of funds.

\section*{Funding}
This research is supported by the RIE2025 Human Potential Programme Prenatal/Early Childhood Grant (H24P2M0008), administered by A*STAR.

\bibliography{aapmsamp}

\nocite{*}


\end{document}


\maketitle
\tableofcontents
\clearpage 
\section*{Supplementary Note S0 \quad User-agnostic ``failsafe'' feature families for choosing $\{\phi_k\}$}
\label{sec:SM_S0_failsafes}

\paragraph{Context and purpose.}
In the main text, the feature family $\{\phi_k\}$ is the only modelling input: it specifies the span in which the
energy $E(\cdot;\theta)$ is assumed to live, and therefore controls the expressivity--parsimony trade-off.
When a domain-informed choice is available, selecting $\{\phi_k\}$ to target visible histogram structure
(asymmetry, heavy tails, multi-modality) can yield the sharpest thermodynamic description.
However, in user-agnostic deployments it is useful to have automatic, systematically improvable ``failsafe'' libraries.
This note records three such default constructions:
(i) Fourier/orthogonal-polynomial bases on 1D (or product) domains;
(ii) tensor-product extensions for low-dimensional Cartesian supports; and
(iii) graph Laplacian eigenfunction bases built directly on the discrete histogram support.
All three provide nested spans indexed by $K$ (number of basis functions), enabling controlled refinement.

\subsection*{S0.1 \quad General setup on a discrete support}
Let the empirical histogram be $p\in\Delta_n$ on a finite support
$\mathcal{C}=\{c_j\}_{j=1}^n$ (bin centres or discrete microstates), with $p_j>0$ and $\sum_{j=1}^n p_j=1$.
Given a feature family $\{\phi_k:\mathcal{C}\to\mathbb{R}\}_{k=1}^K$, define the energy
\begin{equation}
E(c_j;\theta)=\sum_{k=1}^K \theta_k\,\phi_k(c_j),
\qquad \theta\in\mathbb{R}^K,
\end{equation}
and the induced Boltzmann-type model
\begin{equation}
q_\theta(j) \;=\; \frac{\exp\!\big(-E(c_j;\theta)\big)}{Z(\theta)},
\qquad
Z(\theta)=\sum_{j=1}^n \exp\!\big(-E(c_j;\theta)\big).
\end{equation}
The constant offset (``gauge'') in $E$ does not affect $q_\theta$ and can be fixed by either including a constant feature
$\phi_0\equiv 1$ and leaving its coefficient free, or by enforcing a constraint such as $\sum_{j} E(c_j;\theta)=0$.

\paragraph{Exact representability in the full discrete space.}
On a finite support, any strictly positive histogram $p$ is \emph{exactly} representable by an energy
$E^\star(c_j) = -\log p_j + \mathrm{const}$, since $q_{\theta^\star}\equiv p$ if $E(\cdot;\theta)$ can express $-\log p$.
Hence, a failsafe library is conceptually a way to approximate the target function $s(c_j)=-\log p_j$
by a truncated expansion in a chosen basis.

\subsection*{S0.2 \quad Fitting $\theta$ and the residual mismatch diagnostic}

\paragraph{Surprisal regression (recommended, stable on discrete supports).}
Define the empirical surprisal on the support
\begin{equation}
s_j \;=\; -\log\big(p_j + \varepsilon\big),
\label{eq:S0_surprisal}
\end{equation}
with a small $\varepsilon>0$ to prevent divergence in empty/near-empty bins (e.g.\ $\varepsilon=10^{-12}$ or a fraction of $1/N$ if $N$ samples).
Let $\Phi\in\mathbb{R}^{n\times K}$ be the design matrix with entries $\Phi_{jk}=\phi_k(c_j)$.
A simple and robust estimator of $\theta$ is the (weighted) ridge regression
\begin{equation}
\theta^{(K)} \;=\;
\arg\min_{\theta\in\mathbb{R}^K}
\Big\|\mathbf{W}^{1/2}\big(\Phi\theta - s\big)\Big\|_2^2
\;+\; \lambda\|\theta\|_2^2,
\label{eq:S0_ridge}
\end{equation}
where $s=(s_1,\dots,s_n)^\top$, $\mathbf{W}=\mathrm{diag}(w_1,\dots,w_n)$, and $\lambda\ge 0$ is a regularisation parameter.
A convenient default is $w_j=p_j$ (emphasises high-mass regions) or $w_j=\sqrt{p_j}$.
If tail fidelity is important, one may upweight low-probability bins, e.g.\ $w_j \propto (p_j+\varepsilon)^{-\alpha}$ with $\alpha\in[0,1/2]$.

\paragraph{Residual mismatch as a ``no-go'' diagnostic.}
After fitting, compute $q_{\theta^{(K)}}$ and evaluate a discrepancy between $p$ and $q_{\theta^{(K)}}$.
Useful diagnostics include
\begin{align}
D_{\mathrm{KL}}(p\|q) &= \sum_{j=1}^n p_j \log\frac{p_j}{q_j},\\
D_{\mathrm{JS}}(p,q) &= \tfrac12 D_{\mathrm{KL}}\!\Big(p\Big\|\tfrac{p+q}{2}\Big) + \tfrac12 D_{\mathrm{KL}}\!\Big(q\Big\|\tfrac{p+q}{2}\Big),\\
\|p-q\|_2^2 &= \sum_{j=1}^n (p_j-q_j)^2.
\end{align}
A persistently large mismatch indicates that the selected span $\mathrm{span}\{\phi_k\}_{k=1}^K$ is insufficient.
This is not a failure mode of the framework; rather, it quantifies which aspects of the empirical geometry are not expressible under the current feature set.

\paragraph{Choosing $K$ (nested-span model selection).}
Because all failsafe libraries below are nested in $K$, one can select $K$ by:
(i) an ``elbow'' in the mismatch curve $K\mapsto D(p\|q_{\theta^{(K)}})$;
(ii) held-out evaluation using a sample split (recommended: split the original samples, build two histograms $p^{\mathrm{tr}}$, $p^{\mathrm{va}}$ on the same support, fit on $p^{\mathrm{tr}}$ and score on $p^{\mathrm{va}}$);
or (iii) information criteria (AIC/BIC) treating $K$ as effective degrees of freedom (approximate under ridge).
A safe default is to increase $K$ until the mismatch stops decreasing appreciably on held-out data.

\paragraph{Avoiding unintended ``Gaussianisation.''}
Gaussian-like behaviour is imposed only if the energy is explicitly restricted to a quadratic form (equivalently, only first/second-order moment constraints).
The failsafe libraries below do \emph{not} enforce Gaussian structure; however, choosing $K$ too small can over-smooth the energy landscape.
The nested selection procedure above mitigates this by increasing $K$ until multi-modality/tails present in $p$ are captured.

\subsection*{S0.3 \quad Failsafe I: Fourier and orthogonal-polynomial bases (1D and product domains)}

\subsubsection*{S0.3.1 \quad Fourier basis on a bounded 1D interval}
Assume a 1D support mapped to $x\in[-L,L]$ (e.g.\ via bin centres $c_j\in[-L,L]$).
A real Fourier library can be defined as
\begin{align}
\phi_0(x) &= 1,\\
\phi_{2m-1}(x) &= \cos\!\Big(\frac{m\pi}{L}x\Big), \qquad m=1,2,\dots,\\
\phi_{2m}(x) &= \sin\!\Big(\frac{m\pi}{L}x\Big), \qquad m=1,2,\dots.
\end{align}
Truncation at $K$ yields a $K$-dimensional span, with increasing $K$ adding finer-scale structure.
Fourier bases are most appropriate when the energy landscape is smooth on the chosen support and boundary artefacts are acceptable
(or when periodicity is natural).

\paragraph{Practical notes.}
(i) Rescale $x$ to $[-L,L]$ for numerical stability.
(ii) Use ridge regularisation in \eqref{eq:S0_ridge} to prevent high-frequency overfitting.
(iii) If boundary behaviour is problematic, use polynomial/spline alternatives (below) or enlarge support.

\subsubsection*{S0.3.2 \quad Orthogonal polynomials on $[-1,1]$ (Legendre/Chebyshev)}
Map $x\in[-L,L]$ to $u=x/L\in[-1,1]$.
Legendre polynomials $\{P_m(u)\}_{m\ge 0}$ form an orthogonal basis on $[-1,1]$ under the uniform weight.
Define $\phi_m(x)=P_m(x/L)$ and truncate at degree $d$ (so $K=d+1$).
Chebyshev polynomials can be used similarly and often have favourable approximation properties for smooth functions.

\paragraph{When to use.}
Polynomial bases are a good default when the support is bounded and the target $s(x)=-\log p(x)$ is well-approximated by smooth global functions.
They can represent multi-modality and heavy tails on bounded supports when $d$ is sufficiently large (selected by mismatch diagnostics).

\subsubsection*{S0.3.3 \quad Product domains (low-dimensional Cartesian supports)}
For $x=(x^{(1)},\dots,x^{(d)})$ with each coordinate scaled to $[-1,1]$, build a product library
\begin{equation}
\phi_{\alpha}(x) = \prod_{r=1}^d \psi_{\alpha_r}\!\big(x^{(r)}\big),
\end{equation}
where $\psi_{\alpha_r}$ is a 1D basis element (Fourier or polynomial) and $\alpha=(\alpha_1,\dots,\alpha_d)$ is a multi-index.
To control combinatorial growth, restrict to a maximum total degree/order $\sum_r \alpha_r \le D$ or include only low-order interactions.
This is appropriate for $d\le 3$ (or modest $d$ with strong sparsity constraints).

\subsection*{S0.4 \quad Failsafe II: Graph Laplacian eigenfunctions on the discrete histogram support}

\paragraph{Motivation.}
When the support $\mathcal{C}$ is irregular, non-Cartesian, or effectively manifold-like (even within $\mathbb{R}^d$),
a basis constructed directly on the discrete support can be preferable.
Graph Laplacian eigenfunctions provide an orthonormal family ordered by increasing oscillation, yielding a nested approximation space.

\subsubsection*{S0.4.1 \quad Graph construction}
Given support points $\{c_j\}_{j=1}^n \subset \mathbb{R}^d$ (bin centres or discrete microstates), build a weighted graph:
\begin{itemize}
\item Connect each node $j$ to its $k_{\mathrm{nn}}$ nearest neighbours (or all nodes within radius $r$).
\item Set weights, for example,
\begin{equation}
W_{ij} =
\begin{cases}
\exp\!\big(-\|c_i-c_j\|_2^2/\sigma^2\big), & i\sim j,\\
0, & \text{otherwise},
\end{cases}
\end{equation}
where $\sigma$ is a scale parameter (default: median neighbour distance; or a local $\sigma_i$ for self-tuning kernels).
\item Let $D=\mathrm{diag}(d_1,\dots,d_n)$ with $d_i=\sum_j W_{ij}$.
\end{itemize}

\subsubsection*{S0.4.2 \quad Laplacian eigenfunctions}
Use the (symmetric) normalised Laplacian
\begin{equation}
L_{\mathrm{sym}} = I - D^{-1/2} W D^{-1/2}.
\end{equation}
Compute its eigenpairs $(\lambda_\ell, v_\ell)$ with
$0=\lambda_0 \le \lambda_1 \le \cdots$, where $v_0$ is (approximately) constant on a connected graph.
Define basis functions on the support by
\begin{equation}
\phi_\ell(c_j) = v_\ell(j), \qquad \ell=1,2,\dots,
\end{equation}
typically excluding the constant mode (or including it as $\phi_0\equiv 1$ for gauge).
Truncation to the first $K$ nontrivial eigenvectors yields a smooth-to-oscillatory nested span.

\paragraph{Recommended defaults.}
$k_{\mathrm{nn}}\in[10,30]$ (increase for noisy supports; decrease for sharply clustered supports);
$\sigma$ set to the median distance among $k_{\mathrm{nn}}$ edges (or local self-tuning).
Select $K$ by the mismatch diagnostics in \S0.2; increase $K$ to capture multi-modality.

\paragraph{Identifiability and numerical notes.}
Eigenvectors are defined up to sign; and within nearly degenerate eigenspaces may rotate.
This does not affect the span or the fitted $q_\theta$ if the fitting objective depends only on the span.
For reproducibility, fix random seeds in nearest-neighbour routines and use deterministic eigensolvers when available.

\paragraph{When to use.}
Graph Laplacian eigenfunctions are preferred when:
(i) the support is non-Cartesian or highly irregular;
(ii) multi-modality is strongly tied to the geometry of $\mathcal{C}$;
or (iii) coordinate-wise bases (Fourier/polynomial) are ill-conditioned or require too many terms.

\subsection*{S0.5 \quad Summary table and ``quick-start'' checklist}

\paragraph{Which failsafe should I use?}
\begin{itemize}
\item \textbf{Fourier basis (1D / periodic / smooth)}: fast, simple, good for smooth energy landscapes on known intervals.
\item \textbf{Orthogonal polynomials (1D bounded)}: good default on bounded supports; increase degree until mismatch saturates.
\item \textbf{Graph Laplacian eigenfunctions (irregular support / manifold-like)}: best generic default when the support is discrete/irregular and geometry matters.
\end{itemize}

\paragraph{Quick-start (recommended).}
\begin{enumerate}
\item Fix the support $\mathcal{C}$ and histogram $p$ (ensure $p_j>0$ via smoothing if needed).
\item Choose a failsafe library and build $\Phi_{jk}=\phi_k(c_j)$ for $k=1,\dots,K_{\max}$.
\item For each $K\le K_{\max}$: fit $\theta^{(K)}$ by \eqref{eq:S0_ridge}; compute $q_{\theta^{(K)}}$.
\item Select $K$ by held-out mismatch or the elbow of $D_{\mathrm{JS}}(p,q_{\theta^{(K)}})$.
\item Report $K^\star$, the chosen mismatch value, and (optionally) the residual structure.
\end{enumerate}

\vspace{2cm}

{\centering\LARGE\textbf{Supplementary Notes for Applications}\par}

\section{Numerical details for the harmonic calibration}
\label{sm:harmonic}

This section gives implementation details for the harmonic single--well
calibration in Sec. III A of the main text.

\subsection{Sampling and histogram construction}

We fix $k_B = k = 1$ and consider temperatures
$T\in\{0.5,1.0,1.5,2.0\}$, with corresponding variances
$\sigma^2 = T$.
For each temperature we draw $N = 2\times 10^5$ independent samples
\begin{equation}
  x_n \sim \mathcal{N}(0,\sigma^2), \qquad n=1,\dots,N,
\end{equation}
directly from the Gaussian stationary distribution of the harmonic well.

To approximate the empirical density $\hat p(x)$ we build a fixed histogram
on a symmetric interval $[-L,L]$, where $L = 5\sigma_{\max}$ and
$\sigma_{\max} = \sqrt{\max_T T}$ is the largest standard deviation across
temperatures. We use $300$ uniformly spaced bins, so that the bin edges
$x_i$ and centres $\bar x_i$ are the same for all $T$. The histogram is
density-normalized so that
$\sum_i \hat p(\bar x_i)\,\Delta x \approx 1$ on this grid. Bins with zero
counts are discarded when taking logarithms to avoid numerical issues.

\subsection{Generator fitting}

In the calibration we restrict the generative energy to the quadratic family
\begin{equation}
  E_{\mathrm{gen}}(x;\omega) = \omega_0 + \omega_2 x^2 ,
\end{equation}
which is the simplest nontrivial basis consistent with the symmetry
$p(x)=p(-x)$ and the known harmonic form of $-\log p(x)$. Given the
empirical histogram, we define the target values
\begin{equation}
  y_i = -\log \hat p(\bar x_i),
\end{equation}
for all bins $i$ with $\hat p(\bar x_i) > 0$, and construct the design
matrix
\begin{equation}
  X_i = \bigl(1,\; \bar x_i^2\bigr).
\end{equation}
The generator step is implemented here as a simple least--squares regression,
\begin{equation}
  \omega^\star = \arg\min_{\omega}
  \sum_i \bigl( X_i \cdot \omega - y_i \bigr)^2 ,
\end{equation}
solved using standard linear algebra routines. This corresponds to an
$L^2$ projection of $-\log \hat p(x)$ onto the chosen energy family;
more general generator classes and loss functions are described in the main
text.

In the exact Gaussian case one has
\begin{equation}
  -\log p(x) = \mathrm{const} + \tfrac{1}{2}\beta k x^2 ,
\end{equation}
so the optimal coefficient within this family is
$\omega_2^\mathrm{th} = \tfrac{1}{2}\beta k$ up to the additive constant
$\omega_0$.
Numerically we find (for $T=0.5,1.0,1.5,2.0$) that the fitted $\omega_2$
agrees with $\omega_2^\mathrm{th}$ to within $2$--$5\%$; the
$\omega_2$--vs--$\beta$ plot in
Fig.~1(b) of the main text shows an excellent
linear relation with the expected slope $\tfrac{1}{2}k$.

\subsection{Entropy and thermodynamic consistency}

For the one--dimensional harmonic single--well at equilibrium, the stationary
microstate distribution in $x$ is Gaussian with variance $\sigma^2=k_B T/k$.
Its differential Shannon entropy is
\begin{equation}
  S_x(T)
  = -\int p(x)\log p(x)\,dx
  = \tfrac{1}{2}\log\!\bigl(2\pi e\, k_B T/k\bigr),
\end{equation}
and the corresponding physical mean potential energy is
\begin{equation}
  U_\mathrm{phys} = \mathbb{E}\!\left[\tfrac{1}{2}k x^2\right]
  = \tfrac{1}{2}k_B T,
\end{equation}
so that
\begin{equation}
  \frac{\partial S_x}{\partial U_\mathrm{phys}}
  = \frac{1}{T}.
\end{equation}
In the generative framework, when the generator family is restricted to the
quadratic form above and the observed microstate distribution is exactly
Gaussian, the learned energy is equivalent (up to the standard affine gauge)
to the physical quadratic Hamiltonian, and the induced entropy--temperature
relation reduces to the classical Boltzmann--Gibbs form for this unimodal
equilibrium limit. Here we use the harmonic example primarily as a numerical
calibration of the generator step and as a consistency check on the expected
quadratic scaling; we do not attempt to estimate $T_{\mathrm{gen}}$ numerically
from finite differences of $S(U)$ in this note.

\section{Double--well energy--space Stage-I (trace-form) details}

This note provides additional detail on the Stage-I (core) construction of the
energy--space distribution $p_E(E)$ for the bistable double--well system,
together with the corresponding plots.

\subsection*{Energy grid, feature family and Stage-I core fit}

For each inverse temperature $\beta$ we map samples $x_n$ to energies
$E_n = E_{\mathrm{gen}}(x_n)$ and construct an energy grid
$\{E_j\}_{j=1}^M$ spanning the bulk of observed energies (we use a robust
percentile range to avoid rare extremes dominating the discretisation).
We represent the empirical energy distribution by a histogram
$\hat p_E\in\Delta_M$ on this fixed grid.

Stage-I learns a strictly concave trace-form generator
$H[p]=\sum_{j=1}^M G(p_j)$ \emph{from} $\hat p_E$ by enforcing the
core \emph{slope-in-span} condition on the histogram bins:
\begin{equation}
  s_j \;\equiv\; G'(\hat p_{E,j})
  \;\approx\; (F\theta)_j,
  \qquad j=1,\dots,M,
\end{equation}
where $F\in\mathbb{R}^{M\times d}$ is a low-dimensional feature family on the
energy axis (constant term plus multiscale radial basis functions centered at
energy quantiles), and $\theta\in\mathbb{R}^d$ are coefficients.
Strict concavity is enforced numerically by requiring $G'$ to be monotone as a
function of probability (implemented by isotonic projection of the slopes on
the probability-sorted active bins), and two anchor equalities are imposed on
the highest- and lowest-probability bins to fix the additive/multiplicative
gauge of the slope scale.
The resulting learned object is the monotone slope map $G'(\cdot)$ together
with coefficients $\theta$ such that $F\theta$ approximates the fitted slopes.

\subsection*{Energy--space MaxEnt reconstruction (KKT closed form)}

Given the Stage-I fit, the corresponding MaxEnt distribution on the same energy grid
follows directly from the KKT stationarity condition for trace-form entropies:
\begin{equation}
  G'(p_{E,j}^\star) \;=\; (F\theta)_j,
  \qquad j=1,\dots,M,
\end{equation}
hence
\begin{equation}
  p_{E,j}^\star \;=\; \bigl(G'\bigr)^{-1}\!\bigl((F\theta)_j\bigr),
  \qquad j=1,\dots,M,
  \label{eq:dw_kkt_closed_form}
\end{equation}
followed by normalization onto the simplex.
We emphasize that this reconstruction is \emph{KKT-consistent} with the learned
trace-form generator: no additional smoothing or fidelity objective is introduced
at this stage; the only regularization is the low-dimensional span restriction
and the strict concavity (monotonicity) constraint in Stage-I.

\subsection*{Energy distributions and Shannon comparator}

\begin{figure*}[t]
  \centering
  \includegraphics[width=0.32\textwidth]{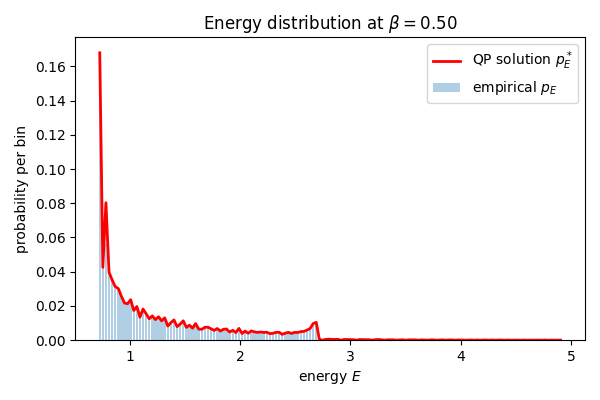}
  \includegraphics[width=0.32\textwidth]{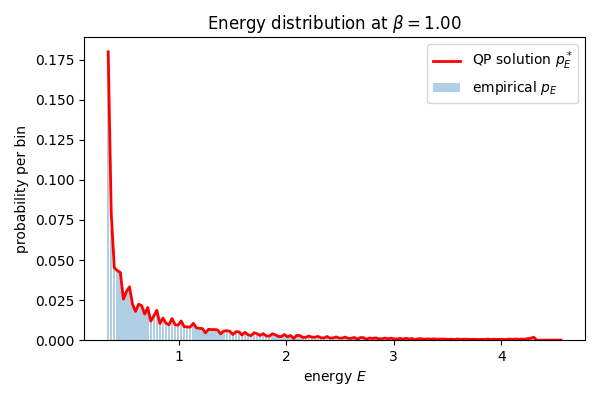}
  \includegraphics[width=0.32\textwidth]{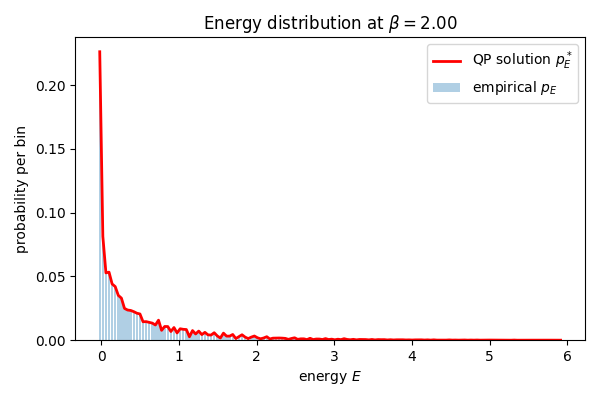}
  \caption{Energy--space Stage-I/KKT solutions and comparison with a Shannon mean-matched exponential baseline for the double--well system.
  For each inverse temperature $\beta \in \{0.5,1,2\}$ we show the empirical energy histogram (blue bars),
  obtained by mapping samples $x_n$ through the fitted generator $E_{\mathrm{gen}}(x_n)$, together with the
  Stage-I/KKT MaxEnt reconstruction $p_E^\star(E)$ (red line) and the Shannon mean-matched exponential
  baseline $p_{\mathrm{Sh}}(E)\propto e^{-\lambda E}$ (blue dashed), where $\lambda$ is chosen so that
  $\mathbb{E}_{p_{\mathrm{Sh}}}[E]=\mathbb{E}_{\hat p_E}[E]$ on the same energy grid. The learned trace-form solution
  captures the sharp low--energy peak and the long empirical tail, whereas the fixed-form Shannon exponential produces a single-scale decay
  that cannot reproduce the detailed core/shape structure.}
  \label{fig:dw_pE}
\end{figure*}

Figure~\ref{fig:dw_pE} shows the empirical energy histograms $\hat p_E(E)$
for $\beta=0.5,1,2$ (blue bars) together with the Stage-I/KKT MaxEnt
reconstruction $p_E^\star(E)$ (red line). In all cases the solution exhibits a
narrow peak at the lowest accessible energies and an approximately exponential tail.
Small discrepancies arise from finite histogram resolution and from the fact that the
learned slope family $F\theta$ is low-dimensional, hence it cannot fit an arbitrary
histogram exactly.

To contrast with the learned functional we also construct a fixed-form Shannon
MaxEnt distribution on the same energy grid under a \emph{single mean-energy constraint}:
\begin{equation}
  p_{\mathrm{Sh}}(E) \propto \exp(-\lambda E),
\end{equation}
where $\lambda$ is chosen such that
$\mathbb{E}_{p_{\mathrm{Sh}}}[E] = \mathbb{E}_{\hat p_E}[E]$ on the same discretization.
This ``mean-matched exponential'' baseline isolates the effect of the learned
(trace-form) functional from a standard Shannon MaxEnt assumption on the energy axis.

\subsection*{Entropy values on the learned energy state space}

Given the learned trace-form generator $G$, a generative entropy value on the energy
state space is computed as
\begin{equation}
  S_{\mathrm{gen}} \;=\; \sum_{j=1}^M G\!\bigl(p_{E,j}^\star\bigr),
\end{equation}
where $G$ is obtained (up to an additive constant) by integrating the learned slope map $G'$
(with the convention $G(0)=0$).
Across the three inverse temperatures we obtain only mild variations of $S_{\mathrm{gen}}$
(typically at the percent level), consistent with the interpretation that the learned
generator primarily encodes the \emph{geometry of the energy state space} rather than
the precise canonical weighting across the two wells.

For reference, the canonical Shannon entropies
$S_{\mathrm{Sh}}(\beta)=-\int p_\beta(x)\log p_\beta(x)\,dx$ decrease
substantially with $\beta$. The contrast between these two entropy curves
highlights the conceptual difference between a Shannon entropy on configuration
space and the inferred trace-form entropy on the learned energy state space.

\section{Mixture-of-Gaussians “data system”: full construction and additional panels}

This section expands the mixture-of-Gaussians example described in the
main text and provides full details of the construction, together with
additional diagnostic panels.

\begin{figure*}[t]
\centering
\includegraphics[width=0.32\textwidth]{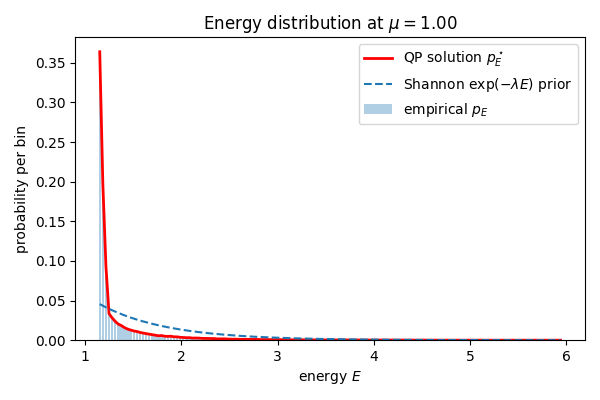}
\includegraphics[width=0.32\textwidth]{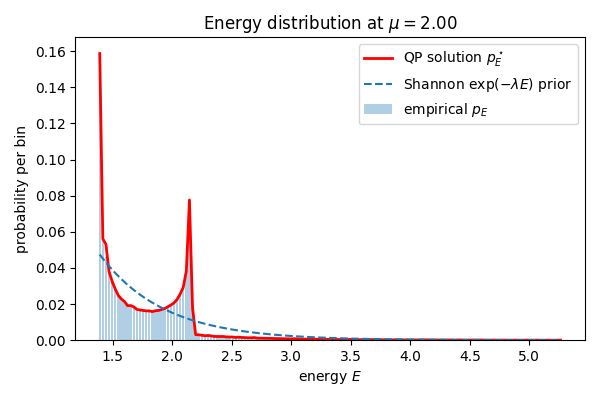}
\includegraphics[width=0.32\textwidth]{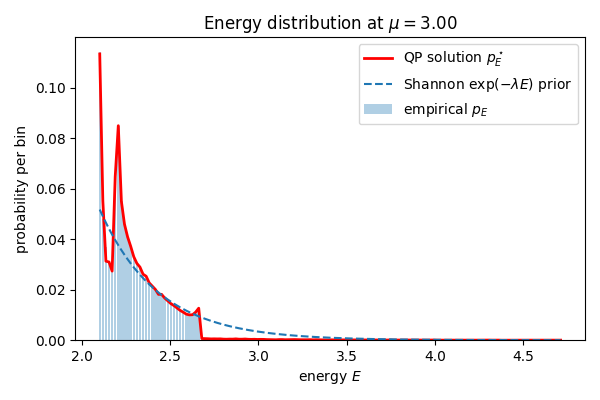}

\includegraphics[width=0.32\textwidth]{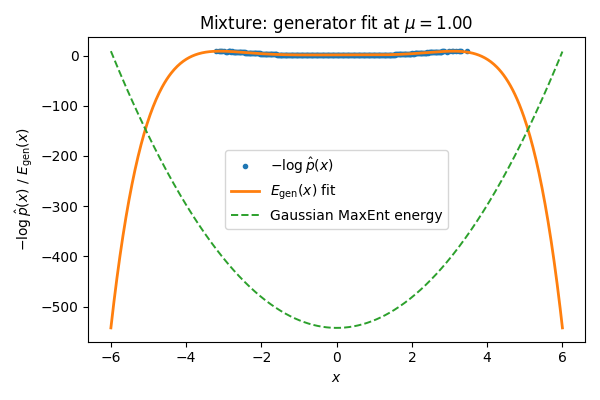}
\includegraphics[width=0.32\textwidth]{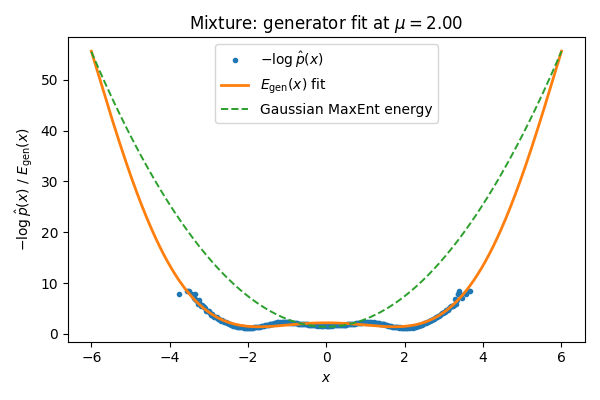}
\includegraphics[width=0.32\textwidth]{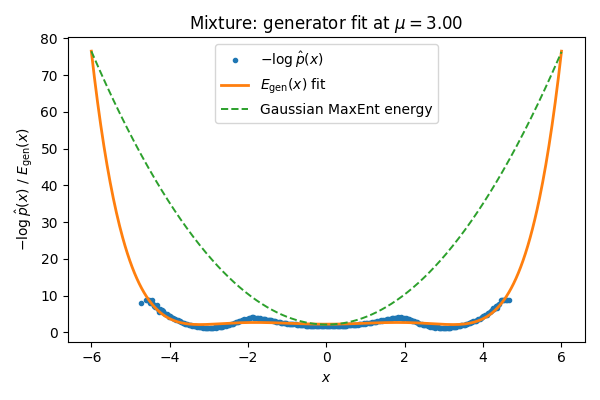}

\caption{\textbf{Mixture-of-Gaussians data system: generator fits and
energy–space solutions for all separations.}
Each column corresponds to a different mixture separation
$\mu\in\{1,2,3\}$ (left to right).
\emph{Top row:} empirical energy histogram $p_E(E)$ (blue bars)
obtained by mapping samples $x_n$ through the fitted generator
$E_{\mathrm{gen}}(x_n)$, together with the trace-form MaxEnt reconstruction
$p_E^\star(E)$ (red solid) and a Shannon exponential prior
$p_{\mathrm{Sh}}(E)\propto\exp(-\lambda E)$ (blue dashed) with
$\lambda$ chosen so that
$\mathbb{E}_{p_{\mathrm{Sh}}}[E]=U_{\mathrm{emp}}$.
The generative solution captures the sharp low–energy peak and the
structured tail induced by the mixture, whereas the Shannon prior
produces a single–scale decay that cannot reproduce this structure.
\emph{Bottom row:} empirical negative log–density $-\log\hat{p}(x)$
(blue points) with the fitted generator $E_{\mathrm{gen}}(x)$ (orange)
and the Gaussian MaxEnt energy associated with the matching
mean/variance Gaussian (green dashed).  For visual comparability,
the displayed $E_{\mathrm{gen}}(x)$ curve may be monotone-calibrated to
$-\log\hat{p}(x)$ (a plot-only alignment that does not affect the energy mapping
or the energy–space distributions).  For small separation
($\mu=1$) the system is close to unimodal and the generator remains
broad and nearly flat; as $\mu$ increases the generator develops
multiple wells aligned with the mixture modes, while the Gaussian
MaxEnt energy remains strictly unimodal and increasingly mismatched to
the data.}
\label{fig:mixture_mu_panels}
\end{figure*}

\paragraph{Data–level system.}
We consider a one–dimensional, symmetric Gaussian mixture with mixture separation
$\mu\in\{1,2,3\}$.  In the numerical implementation used for the diagnostic panels
(Fig.~S\ref{fig:mixture_mu_panels}), the data are generated from a \emph{three–component}
symmetric mixture
\[
  p(x)
  = w_{-}\,\mathcal{N}(-\mu,\sigma_{-}^{2})
    + w_{0}\,\mathcal{N}(0,\sigma_{0}^{2})
    + w_{+}\,\mathcal{N}(+\mu,\sigma_{+}^{2}),
\]
with $(w_{-},w_{0},w_{+})=(0.3,0.4,0.3)$ and $(\sigma_{-},\sigma_{0},\sigma_{+})=(0.4,0.8,0.4)$
(as in the accompanying code).  For each $\mu$ we draw
$N=200{,}000$ independent samples $x_n$ and form an empirical density
$\hat{p}(x)$ on a uniform grid via a regularized histogram.
Unless otherwise stated, the generator fit and energy–axis construction in this diagnostic
figure use a fixed training subset (e.g.\ 70\% of the samples under a fixed seed) for numerical
stability; the remaining samples are not used for fitting in this panel set.

\paragraph{Generator fit in $x$–space.}
The generative energy is parameterized as a sixth–order even polynomial
\[
  E_{\mathrm{gen}}(x)
  = w_0 + w_2 x^2 + w_4 x^4 + w_6 x^6,
\]
and is fitted by least–squares regression of $-\log \hat{p}(x)$ onto
the basis $\{1,x^2,x^4,x^6\}$, discarding empty histogram bins.
This is exactly the same procedure used for the double–well system but
without reference to any underlying Hamiltonian.  The bottom row of
Fig.~S\ref{fig:mixture_mu_panels} shows the resulting fits
(orange curves) for each value of $\mu$ against the empirical
$-\log \hat{p}(x)$ (blue points).  As the separation increases from
$\mu=1$ to $\mu=3$, the generator smoothly transitions from a nearly
flat, broad well to a clearly multi–well energy landscape that mirrors
the multimodal structure of the mixture.

For comparison we also construct the Shannon MaxEnt Gaussian with the
same mean and variance as the mixture.  Since the mixture is symmetric,
this is $\mathcal{N}(0,\sigma^2)$ where $\sigma^2=\mathbb{E}[x^2]$ is
the empirical second moment.  Its associated energy
$E_{\mathrm{Gauss}}(x) \propto x^2/\sigma^2$ is shown as a green dashed
curve in the bottom row of Fig.~S\ref{fig:mixture_mu_panels}.  By
construction this Shannon MaxEnt energy is strictly unimodal and
cannot reproduce the multi–well structure learned by
$E_{\mathrm{gen}}(x)$ when $\mu$ is large.

\paragraph{Energy mapping and trace-form MaxEnt reconstruction in $E$–space.}
Given the fitted generator, we map the microstates $x_n$ into energy
space via $E_n = E_{\mathrm{gen}}(x_n)$ and form an energy histogram
with $M$ bins, yielding empirical probabilities $\hat p_E(E_j)$ at bin
centres $E_j$.

On this discretised energy axis we implement the \emph{Stage-I (core) trace-form learning}
described in the main text: we seek a strictly concave trace-form generator
$G_{\mathrm{gen}}$ such that, on the empirical histogram, its slope values
\[
  s_j \;=\; G_{\mathrm{gen}}'(\hat p_{E,j})
\]
lie (approximately) in the span of a small, interpretable feature family on the energy axis.
Concretely, we construct a multiscale feature matrix $F\in\mathbb{R}^{M\times d}$
whose columns consist of a constant term and a set of broad and narrow radial basis
functions centred at quantiles of the empirical energy support.
Stage-I then finds coefficients $\theta\in\mathbb{R}^d$ such that
\[
  s \;\approx\; F\theta,
\]
while enforcing strict concavity through a monotonicity constraint on the slope ordering
(i.e.\ $G_{\mathrm{gen}}'$ is monotone as a function of $p$, implemented numerically by
isotonic projection on the probability-sorted bins).
Two anchor equalities are imposed at the highest- and lowest-probability bins to fix the
additive/multiplicative gauge of the slope scale. (This is a numerically stable surrogate
of the Stage-I QP described in the main text: it solves the same \emph{slope-in-span} condition
with strict concavity, but is implemented directly in slope space.)

\emph{Trace-form MaxEnt reconstruction.}
Once $\theta$ and the monotone slope map $G_{\mathrm{gen}}'$ have been learned from
$\hat p_E$, the corresponding MaxEnt distribution on the \emph{same} fixed energy grid
is obtained directly from the KKT stationarity condition. In trace-form MaxEnt,
the optimum satisfies
\[
  G_{\mathrm{gen}}'(p_{E,j}^\star) \;=\; (F\theta)_j,
\]
hence
\[
  p_{E,j}^\star \;=\; \bigl(G_{\mathrm{gen}}'\bigr)^{-1}\!\bigl((F\theta)_j\bigr),
  \qquad j=1,\dots,M,
\]
followed by normalisation onto the simplex.  This KKT-consistent closed-form reconstruction
produces the red curve $p_E^\star(E)$ in the top row of Fig.~S\ref{fig:mixture_mu_panels}.

\paragraph{Shannon exponential prior in energy space.}
To contrast with the generative solution we construct a purely Shannon
MaxEnt prior on the same energy axis.  If one constrains only the mean
energy $\mathbb{E}[E]$ in the Shannon framework, the maximizing
distribution is exponential,
\[
  p_{\mathrm{Sh}}(E) \propto \exp(-\lambda E),
\]
with decay rate $\lambda>0$ chosen such that
$\mathbb{E}_{p_{\mathrm{Sh}}}[E] = U_{\mathrm{emp}}$.  In practice
$\lambda$ is obtained by solving this moment–matching condition
numerically on the same discretized energy grid.

\paragraph{Six–panel diagnostic figure.}

Across all three separations, the panels illustrate the same qualitative
pattern: the thermoinformational construction learns a multi–well
generator and an energy–space distribution that preserves the observed
core/shape structure, while the corresponding Shannon MaxEnt
approximations (Gaussian in $x$–space and exponential in $E$–space)
necessarily smooth this structure into unimodal, single–scale forms.
In the main text (Fig.1 of the main text) we summarize this
behaviour by showing the $\mu=3$ panels together with the dependence of
$U_{\mathrm{gen}}$ on mixture separation.

\subsection{Holdout validation of the learned energy--entropy representation (Gaussian mixture)}
\label{sec:SM_holdout_mog}

\paragraph{Goal.}
This note documents a within-system holdout test designed to assess falsifiability without
departing from the \emph{one system $\rightarrow$ one inferred functional} principle.
The entropy functional is inferred on a training subset and then held fixed; on held-out data,
only the macrostates are updated and the corresponding MaxEnt prediction is compared to the
held-out empirical distribution on a common energy discretisation.

\paragraph{Setup and split.}
Microstates $x$ are generated from the Gaussian mixture model specified in
Supplementary Note~S$\langle$MoG$\rangle$ (same parameters as the main demonstration).
A random split produces disjoint training and validation sets,
$\mathcal{D}_{\mathrm{tr}}$ and $\mathcal{D}_{\mathrm{val}}$, with proportions
$\rho$ and $1-\rho$ (e.g.\ $\rho=0.7$) under a fixed random seed.
All quantities defining the representation (energy map, discretisation, constraint functions,
and the learned entropy functional) are inferred using $\mathcal{D}_{\mathrm{tr}}$ only.

\paragraph{Training stage (learn $E_{\mathrm{gen}}$ and the entropy functional).}
An energy map $E_{\mathrm{gen}}(x)$ is fitted on $\mathcal{D}_{\mathrm{tr}}$ (train-only fit).
Training energies are obtained as $E_{\mathrm{tr}}=\{E_{\mathrm{gen}}(x):x\in\mathcal{D}_{\mathrm{tr}}\}$.
A fixed energy support $[E_{\min},E_{\max}]$ and a fixed binning
$\{E_j\}_{j=1}^{n}$ are defined from training only and then frozen.
The training empirical energy histogram is denoted $\hat p_{E,\mathrm{tr}}\in\Delta_n$.
Macroscopic constraints are computed as
\[
\mu_{k}^{\mathrm{tr}}=\sum_{j=1}^{n}\hat p_{E,\mathrm{tr}}(j)\, f_k(E_j),
\]
where $f_k$ are the constraint functions used in the MoG demonstration.
Using $(\hat p_{E,\mathrm{tr}},\{\mu_k^{\mathrm{tr}}\})$, an entropy functional is then inferred
by the same QP-based procedure as in the main MoG example; denote it by $\mathcal{H}^\star[\cdot]$.

\paragraph{Validation stage (functional fixed, macrostates updated).}
Validation energies are computed by pushing forward held-out microstates through the same
train-fitted energy map,
$E_{\mathrm{val}}=\{E_{\mathrm{gen}}(x):x\in\mathcal{D}_{\mathrm{val}}\}$.
The held-out empirical energy histogram $\hat p_{E,\mathrm{val}}$ is formed on the
\emph{same} frozen bins $\{E_j\}_{j=1}^{n}$.
Held-out macrostates are computed as
\[
\mu_{k}^{\mathrm{val}}=\sum_{j=1}^{n}\hat p_{E,\mathrm{val}}(j)\, f_k(E_j).
\]
A MaxEnt prediction on validation is obtained by solving
\[
p_{E,\mathrm{val}}^\star
\;=\;
\arg\max_{p\in\Delta_n}\;
\mathcal{H}^\star[p]
\quad
\text{s.t.}\quad
\sum_{j=1}^{n} p(j)\, f_k(E_j)=\mu_{k}^{\mathrm{val}},\;\; k=1,\dots,K,
\]
with $\mathcal{H}^\star$ held fixed (train-learned).

\paragraph{Evaluation metrics.}
Prediction quality is quantified by the Jensen--Shannon divergence
${\rm JS}(\hat p_{E,\mathrm{val}},p_{E,\mathrm{val}}^\star)$
and the 1-Wasserstein distance
$W_1(\hat p_{E,\mathrm{val}},p_{E,\mathrm{val}}^\star)$ computed on the common energy axis.
(Any numerical smoothing/regularisation used for stability is applied identically across methods.)

\paragraph{Baseline: Shannon MaxEnt under identical constraints.}
A fixed-form comparator is obtained by Shannon MaxEnt under the same constraint information.
In the band-constraint setting used here, Shannon MaxEnt yields a distribution that is
uniform within each band and matches the prescribed band masses; this baseline isolates
the effect of the learned functional from the constraint family.

\paragraph{Representative result and robustness to split realisation.}
For $\mu=2.0$, $N=200{,}000$, training fraction $\rho=0.7$, and seed $=1$,
the held-out prediction achieves
${\rm JS}=5.44\times 10^{-4}$ and $W_1=3.42\times 10^{-4}$,
whereas the Shannon MaxEnt baseline yields
${\rm JS}=9.57\times 10^{-2}$ and $W_1=3.25\times 10^{-1}$.
Repeating the same experiment across multiple random seeds yields the same qualitative conclusion
(code provided), indicating that the result is not sensitive to a particular split realisation.

\begin{figure}[t]
  \centering
  \includegraphics[width=0.92\linewidth]{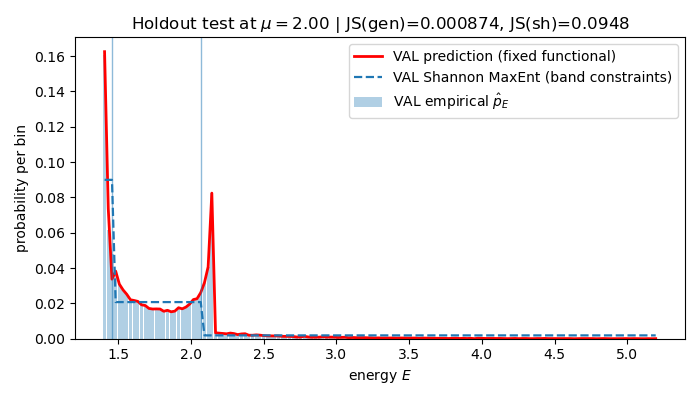}
  \caption{\textbf{Holdout validation in the Gaussian mixture example.}
  The entropy functional and generative energy map are inferred from the training subset only
  (fixed binning and band cutpoints).
  On held-out data, only the macroscopic band-mass constraints are recomputed and the MaxEnt solution
  is obtained with the functional fixed.
  The QP-learned functional closely matches the held-out empirical energy distribution,
  while Shannon MaxEnt under the same constraints fails to reproduce bimodality
  (seed $=1$, $\mu=2.0$, $N=200{,}000$, $\rho=0.7$).}
  \label{fig:SM_holdout_mog}
\end{figure}

\newpage

{\centering\LARGE\textbf{Supplementary Notes for Proofs}\par}

\section{H-theorem Numerical proof}

The abstract H-theorem for generative entropy (Theorem~X in the
Supplementary Material) states that if we evolve the energy–space
distribution $p_E(t)$ according to the projected gradient–ascent flow
of $H_{\text{micro}}$ on the constraint manifold $\{p_E: A p_E = b\}$,
then $H_{\text{micro}}(p_E(t))$ is non–decreasing and the dynamics
converge to the generative MaxEnt state $p_E^\star$ that solves the
quadratic program.

\begin{figure}[t]
\centering
\includegraphics[width=0.48\textwidth]{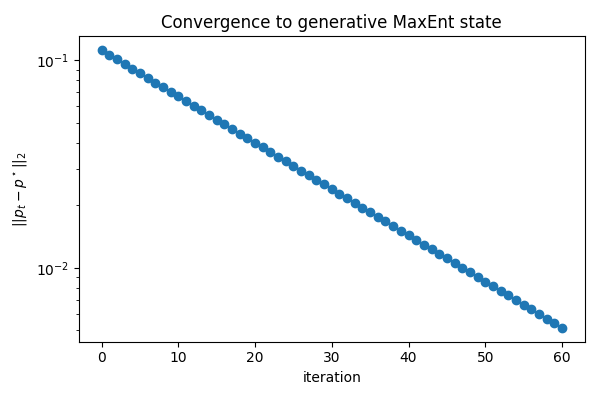}
\includegraphics[width=0.48\textwidth]{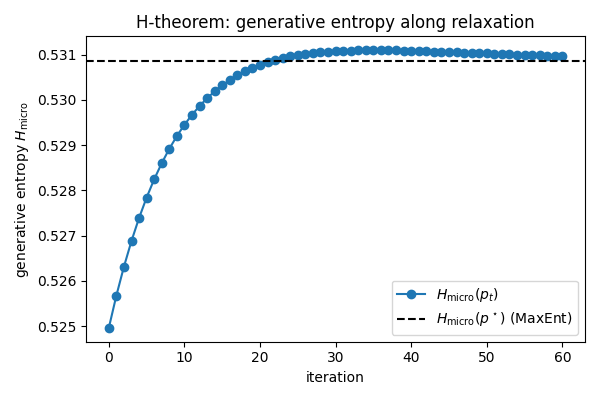}
\caption{\textbf{Numerical H-theorem for generative entropy in the
mixture-of-Gaussians data system.}
We use the $\mu=3$ mixture and the corresponding generative MaxEnt
solution $p_E^\star$ obtained from the quadratic program in
Fig.4 in the main.  Starting from the empirical energy
histogram $p^{(0)}_E$ we iterate a projected gradient–ascent step for
the generative micro–entropy $H_{\text{micro}}$ under fixed
band-mass constraints $Ap_E=b$.
\emph{Left:} Euclidean distance
$\lVert p^{(t)}_E - p_E^\star \rVert_2$ vs. iteration $t$ on a
logarithmic vertical axis, showing approximately exponential
convergence to the MaxEnt state.
\emph{Right:} generative micro–entropy $H_{\text{micro}}(p^{(t)}_E)$
along the same relaxation (blue) compared with the MaxEnt value
$H_{\text{micro}}(p_E^\star)$ from the QP (black dashed).  As predicted
by the H-theorem, the generative entropy is monotonically
non–decreasing and saturates at the QP optimum.}
\label{fig:H_theorem}
\end{figure}

To illustrate this numerically, we reuse the mixture-of-Gaussians data
system at separation $\mu=3$ and the corresponding generative MaxEnt
solution $p_E^\star$ from Fig.4 of the main text.  We take the
empirical energy histogram $p^{(0)}_E$ as an initial condition and
iterate a discrete projected gradient–ascent step,
\[
  p^{(t+1)}_E
  \;=\;
  \Pi_{\{Ap=b,\,p\ge 0\}}
  \bigl(p^{(t)}_E
        + \eta \,\nabla H_{\text{micro}}(p^{(t)}_E)\bigr),
\]
with a small step size $\eta$ and projection onto the linear
constraints and simplex (Methods).  Figure~\ref{fig:H_theorem} shows
that $\|p^{(t)}_E - p_E^\star\|_2$ decays approximately exponentially
in $t$ while the generative micro–entropy $H_{\text{micro}}(p^{(t)}_E)$
increases monotonically to the QP optimum value $H_{\text{micro}}(p_E^\star)$.
This numerical H-theorem confirms that the generative entropy behaves
as a bona fide Lyapunov functional for relaxation toward the MaxEnt
state in this data system.

\subsection{Numerical illustration of the H-theorem}

We report a numerical H-theorem for the
mixture-of-Gaussians data system at separation $\mu=3$.
Here we give the details of the discrete relaxation scheme used in
Fig.~\ref{fig:H_theorem}.

We work on the finite energy grid $\{E_i\}_{i=1}^n$ defined by the
generative energy histogram of the mixture system, with microstate
probabilities $p = (p_1,\dots,p_n)$ constrained by $Ap=b$ where
$A\in\mathbb{R}^{m\times n}$ encodes normalization and band-mass
constraints (Sec.~Sx.x).  The generative micro–entropy has the
quadratic form
\[
  H_{\text{micro}}(p)
  \;=\; c^\top p - \tfrac12 p^\top Q p,
\]
with gradient $\nabla H_{\text{micro}}(p) = c - Qp$.
Given a current iterate $p^{(t)}$ we perform an explicit Euler step
along the unconstrained gradient and then project back onto the
constraint set:
\[
  \tilde p^{(t+1)}
  = p^{(t)} + \eta \,(c - Qp^{(t)}), \qquad
  p^{(t+1)}
  = \arg\min_{Ap=b,\,p\ge 0} \|\tilde p^{(t+1)} - p\|_2^2.
\]
The projection is a small quadratic program with only $m$ equality
constraints, solved with the same SLSQP routine used for the main QP.
For sufficiently small step size $\eta$ this scheme is a discrete
approximation of the projected gradient flow guaranteed by
Theorem~Sx.x, and as expected we observe monotone increase of
$H_{\text{micro}}(p^{(t)})$ and convergence to the unique MaxEnt
solution $p^\star$ (Supplementary Fig.~\ref{fig:H_theorem}).

\newpage
\textbf{\LARGE{Thermodynamic Consistency Theoretical Proofs}}

\section{Inverse Maximum-Entropy Representation of the Generative Entropy}

Let $C = \{c_1,\dots,c_n\}$ be a finite alphabet, and let
\[
  \Delta_n \coloneqq \Big\{ p \in \mathbb{R}^n : p_j \ge 0,\ \sum_{j=1}^n p_j = 1 \Big\}
\]
denote the probability simplex. We write $\Delta_n^\circ$ for its interior:
\[
  \Delta_n^\circ \coloneqq \{ p \in \Delta_n : p_j > 0\ \forall j\}.
\]

A \emph{trace-form entropy} is a functional of the form
\[
  H_G(p) \coloneqq \sum_{j=1}^n G(p_j),
\]
where $G:(0,1]\to\mathbb{R}$ is twice continuously differentiable and strictly concave:
\begin{equation}
  G''(u) < 0,\quad \forall u\in(0,1).                                  \label{eq:strict-concavity}
\end{equation}

We consider linear constraints defined by a finite family of functions
\[
  f_k : C \to \mathbb{R},\quad k=0,\dots,m,
\]
with $f_0 \equiv 1$ encoding normalisation. For a probability vector $p\in\Delta_n$, the associated
constraint set is
\begin{equation}
  \mathcal{C}(p,F) \coloneqq \Big\{ q \in \Delta_n : 
    \sum_{j=1}^n q_j f_k(c_j) = \sum_{j=1}^n p_j f_k(c_j),\ k=0,\dots,m \Big\},
  \label{eq:constraint-set}
\end{equation}
where $F\in\mathbb{R}^{n\times(m+1)}$ is the matrix
\[
  F_{jk} \coloneqq f_k(c_j),\quad j=1,\dots,n,\ k=0,\dots,m.
\]

\begin{definition}[Inverse-MaxEnt generator for a fixed $(p,F)$]
Let $p\in\Delta_n^\circ$ and let $F$ be a constraint matrix as above.
A function $G$ satisfying \eqref{eq:strict-concavity} is called an \emph{inverse-MaxEnt generator
for $(p,F)$} if the trace-form entropy $H_G$ has $p$ as a \emph{unique} maximiser under the
constraints:
\[
  p \in \mathcal{C}(p,F) \quad\text{and}\quad 
  H_G(p) > H_G(q),\quad \forall q\in\mathcal{C}(p,F),\ q\neq p.
\]
\end{definition}

The usual Lagrangian/KKT conditions for constrained maximisation yield a necessary condition
for $G$ to be an inverse-MaxEnt generator: there must exist Lagrange multipliers
$\lambda=(\lambda_0,\dots,\lambda_m)^\top$ such that
\begin{equation}
  G'(p_j) = \sum_{k=0}^m \lambda_k f_k(c_j),\quad j=1,\dots,n.         \label{eq:KKT}
\end{equation}
Strict concavity then guarantees that $p$ is the unique maximiser.

We now formalise three goals:

\begin{itemize}
  \item[(i)] Existence of an inverse-MaxEnt generator for each fixed pair $(p,F)$.
  \item[(ii)] Global uniqueness up to affine ``gauge'' transformations, under a richness assumption
    on the family of admissible constraint sets.
  \item[(iii)] Approximation of the ideal generator $G^*$ by the finite-dimensional quadratic
    programmes (QPs) used in the generative entropy construction.
\end{itemize}

\subsection{Existence for a fixed $(p,F)$}

We first show that for any interior $p$ and any full-rank constraint matrix $F$, there exists
a strictly concave $G$ such that $p$ is the unique MaxEnt solution under the constraints.

\begin{assumption}[Nondegenerate constraints for fixed $(p,F)$]
\label{ass:nondegenerate-F}
For each fixed $p\in\Delta_n^\circ$ and constraint matrix $F\in\mathbb{R}^{n\times(m+1)}$, we assume
that:
\begin{enumerate}
  \item $f_0\equiv 1$ is included among the constraint functions, i.e.\ the first column of $F$
        is the all-ones vector.
  \item $\operatorname{rank} F = m+1$.
  \item The feasible set $\mathcal{C}(p,F)$ has nonempty relative interior in $\Delta_n$.
\end{enumerate}
\end{assumption}

\begin{lemma}[Existence of a generator for $(p,F)$]
\label{lem:existence}
Let $p\in\Delta_n^\circ$ and $F$ satisfy Assumption~\ref{ass:nondegenerate-F}. Then there exists
a $G\in C^2((0,1])$ with $G''(u)<0$ for all $u\in(0,1)$ such that $p$ is the unique maximiser of
$H_G$ on $\mathcal{C}(p,F)$.
\end{lemma}

\begin{proof}
Fix $p\in\Delta_n^\circ$ and a constraint matrix $F$ as in the assumption. Since $F$ has full
column rank, its column space has dimension $m+1$ and, in particular, for any given vector
$g\in\operatorname{col}(F)$ there exists $\lambda\in\mathbb{R}^{m+1}$ such that
\[
  g_j = \sum_{k=0}^m \lambda_k f_k(c_j),\quad j=1,\dots,n.
\]

Choose any vector $\widetilde g \in \operatorname{col}(F)$ with arbitrary values
$\widetilde g_j$ (for instance, by picking any $\lambda$). We will construct a strictly concave
$G$ with
\begin{equation}
  G'(p_j) = \widetilde g_j,\quad j=1,\dots,n.                          \label{eq:target-derivative}
\end{equation}
This will guarantee that the KKT stationarity condition \eqref{eq:KKT} holds for that $\lambda$.

Since $p\in\Delta_n^\circ$, we have $p_j\in(0,1)$ for all $j$. Let
\[
  \{p_{(1)},\dots,p_{(n)}\}
\]
be a reordering of $\{p_1,\dots,p_n\}$ in strictly increasing order,
$p_{(1)}<\dots<p_{(n)}$, and let $\sigma$ be the permutation such that
$p_{(j)}=p_{\sigma(j)}$. Define corresponding reordered target values
$\widetilde g_{(j)} = \widetilde g_{\sigma(j)}$.

We now define a continuous, piecewise linear function $g:(0,1]\to\mathbb{R}$ such that
\[
  g(p_{(j)}) = \widetilde g_{(j)},\quad j=1,\dots,n,
\]
and with strictly negative slope on each interval $[p_{(j)},p_{(j+1)}]$. For concreteness, set:
\begin{itemize}
  \item Pick any strictly decreasing sequence of slopes $s_j < 0$ for $j=1,\dots,n-1$.
  \item Define $g$ on $[p_{(j)},p_{(j+1)}]$ by
        \[
          g(u) = \widetilde g_{(j)} + s_j\,(u - p_{(j)}).
        \]
        At the right endpoint $u=p_{(j+1)}$, we require
        \[
          \widetilde g_{(j+1)} = \widetilde g_{(j)} + s_j (p_{(j+1)} - p_{(j)}),
        \]
        which simply fixes $s_j$:
        \[
          s_j = \frac{\widetilde g_{(j+1)} - \widetilde g_{(j)}}{p_{(j+1)} - p_{(j)}}.
        \]
        We can always shift $\widetilde g$ by an overall affine term if needed; in particular,
        since $\widetilde g$ was arbitrary in $\operatorname{col}(F)$, we can choose it so that
        these slopes are negative. For instance, adding a sufficiently large negative multiple
        of the all-ones vector to $\widetilde g$ preserves the fact that $\widetilde g\in\operatorname{col}(F)$
        (because $f_0\equiv 1$) and makes all $s_j<0$.
\end{itemize}

Thus we have constructed a continuous, piecewise linear function $g$ on $[p_{(1)},p_{(n)}]$ that
is strictly decreasing there, with $g(p_{(j)})=\widetilde g_{(j)}$. Extend $g$ to $(0,p_{(1)}]$ and
$[p_{(n)},1]$ by any strictly decreasing $C^1$ function that joins smoothly at the endpoints (for
example, append short $C^1$ tails with negative derivative). This yields a $g\in C^1((0,1])$ with
\[
  g'(u) < 0,\quad u\in(0,1),
\]
and $g(p_j) = \widetilde g_j$ for all $j$.

Define
\[
  G(u) \coloneqq \int_0^u g(t)\,dt + c,
\]
with $c$ any constant (e.g.\ chosen so that $G(0)=0$). Then $G\in C^2((0,1])$ and
$G''(u) = g'(u)<0$ for all $u\in(0,1)$, so $G$ is strictly concave on $(0,1)$. Moreover,
\[
  G'(p_j) = g(p_j) = \widetilde g_j,
\]
so \eqref{eq:target-derivative} holds. Since $\widetilde g\in\operatorname{col}(F)$, there exists
$\lambda$ with
\[
  \widetilde g_j = \sum_{k=0}^m \lambda_k f_k(c_j),
\]
which is exactly the KKT stationarity condition \eqref{eq:KKT} at $q=p$.

Finally, strict concavity of $H_G$ on the convex set $\mathcal{C}(p,F)$ implies that any feasible
stationary point is the unique maximiser; since $p$ is feasible and satisfies the KKT conditions, it
is the unique maximiser.

This completes the proof.
\end{proof}

\begin{remark}
The construction is highly non-unique; many different strictly concave generators $G$ can realise
the same $(p,F)$. In the next section we study the uniqueness of a generator that works
simultaneously for a \emph{family} of pairs $(p,F)$.
\end{remark}

\subsection{Global uniqueness up to affine gauge}

We now consider a class $\mathcal{F}$ of admissible constraint matrices and ask: if a trace-form
entropy $H_G$ has the inverse-MaxEnt property for \emph{all} pairs $(p,F)$ in
$\Delta_n^\circ\times\mathcal{F}$, is $G$ essentially unique?

We need a mild richness assumption on $\mathcal{F}$ stating that the constraint vectors span all
directions orthogonal to the normalisation constraint.

\begin{assumption}[Richness of constraint family]
\label{ass:rich-constraints}
Let $\mathcal{F}$ be a family of matrices $F\in\mathbb{R}^{n\times(m+1)}$ satisfying
Assumption~\ref{ass:nondegenerate-F}, and suppose:
\begin{enumerate}
  \item For every $F\in\mathcal{F}$, the first column $f_0$ is the all-ones vector.
  \item For every vector $v\in\mathbb{R}^n$ with $\sum_{j=1}^n v_j = 0$ (i.e.\ orthogonal to
        the all-ones vector), there exist $F\in\mathcal{F}$ and a multiplier vector $\lambda$ such
        that
        \[
          v = F\lambda.
        \]
        In other words, the union of the column spaces of $\mathcal{F}$ contains every zero-sum
        vector.
\end{enumerate}
\end{assumption}

Intuitively, this says that by choosing suitable combinations of moments and local ``shape masks'',
we can span all directions in probability space that preserve normalisation.

\begin{theorem}[Global uniqueness up to affine gauge]
\label{thm:global-uniqueness}
Let $\mathcal{F}$ satisfy Assumption~\ref{ass:rich-constraints}. Suppose $G_1,G_2\in C^2((0,1])$
satisfy $G_i''(u)<0$ for all $u\in(0,1)$, and that for every $p\in\Delta_n^\circ$ and every
$F\in\mathcal{F}$, both $G_1$ and $G_2$ are inverse-MaxEnt generators for $(p,F)$; i.e.\ for each
pair $(p,F)$, the entropy functionals
\[
  H_{G_i}(q) = \sum_{j=1}^n G_i(q_j)
\]
have $p$ as the unique maximiser under the constraints \eqref{eq:constraint-set}. Then there exist
constants $\alpha,\beta\in\mathbb{R}$ such that
\[
  G_1(u) = G_2(u) + \alpha u + \beta,\quad \forall u\in(0,1].
\]
In particular, the entropies differ only by an additive constant and a linear term, corresponding
to a shift of entropy and a rescaling of the conjugate variable (temperature).
\end{theorem}

\begin{proof}
Let $D(u)\coloneqq G_1(u) - G_2(u)$ and $d(u)\coloneqq D'(u) = G_1'(u) - G_2'(u)$. Then
$d\in C^1((0,1])$.

Fix any $p\in\Delta_n^\circ$ and $F\in\mathcal{F}$. Since $G_1$ and $G_2$ are both inverse-MaxEnt
generators for $(p,F)$, there exist multipliers $\lambda^{(1)},\lambda^{(2)}\in\mathbb{R}^{m+1}$
such that the KKT conditions hold for each:
\[
  G_1'(p_j) = \sum_{k=0}^m \lambda^{(1)}_k f_k(c_j),\quad
  G_2'(p_j) = \sum_{k=0}^m \lambda^{(2)}_k f_k(c_j),\quad j=1,\dots,n.
\]
Subtracting, we find
\[
  d(p_j) = \sum_{k=0}^m (\lambda^{(1)}_k - \lambda^{(2)}_k) f_k(c_j),\quad j=1,\dots,n.
\]
Equivalently, the vector $d(p) = (d(p_1),\dots,d(p_n))^\top$ lies in the column space of $F$:
\begin{equation}
  d(p) \in \operatorname{col}(F),\quad \forall p\in\Delta_n^\circ,\ \forall F\in\mathcal{F}.
  \label{eq:d-in-colF}
\end{equation}

Let $\mathbf{1}=(1,\dots,1)^\top$. Decompose $d(p)$ into its average and zero-sum parts:
\[
  d(p) = \bar d(p)\,\mathbf{1} + \widetilde d(p),
\]
where
\[
  \bar d(p) \coloneqq \frac{1}{n}\sum_{j=1}^n d(p_j),\quad
  \sum_{j=1}^n \widetilde d(p_j) = 0.
\]

Since the first column of $F$ is $\mathbf{1}$, its column space always contains $\mathbf{1}$, so
the component $\bar d(p)\mathbf{1}$ is automatically in $\operatorname{col}(F)$. Thus
\eqref{eq:d-in-colF} implies that the zero-sum part $\widetilde d(p)$ must also lie in
$\operatorname{col}(F)$:
\[
  \widetilde d(p) \in \operatorname{col}(F),\quad \forall F\in\mathcal{F}.
\]

Now fix a particular $p\in\Delta_n^\circ$. By Assumption~\ref{ass:rich-constraints}, the union of
column spaces $\{\operatorname{col}(F):F\in\mathcal{F}\}$ contains \emph{all} zero-sum vectors in
$\mathbb{R}^n$. Conversely, if a zero-sum vector $w$ is orthogonal to $\operatorname{col}(F)$ for
\emph{every} $F\in\mathcal{F}$, then necessarily $w=0$. Hence, the only vector that lies in
\emph{every} column space $\operatorname{col}(F)$ is a multiple of the all-ones vector $\mathbf{1}$.

But for fixed $p$, we have just seen that $\widetilde d(p)\in\operatorname{col}(F)$ for all
$F\in\mathcal{F}$. Therefore, $\widetilde d(p)$ must be proportional to $\mathbf{1}$; however, by
construction $\widetilde d(p)$ has zero sum, so the only possibility is
\[
  \widetilde d(p) = 0.
\]
Thus, for each fixed $p\in\Delta_n^\circ$,
\begin{equation}
  d(p_j) = \bar d(p)\quad\text{for all } j=1,\dots,n.                   \label{eq:d-constant-at-p}
\end{equation}
In words: the $n$-vector of values $d(p_j)$ is constant across $j$.

Now we use the fact that $p\in\Delta_n^\circ$ was arbitrary. Take any $u\in(0,1)$ and construct
a probability vector $p\in\Delta_n^\circ$ that places one coordinate at $u$, i.e.\ choose
\[
  p_1 = u,\quad p_2,\dots,p_n>0,\quad \sum_{j=2}^n p_j = 1-u.
\]
By \eqref{eq:d-constant-at-p},
\[
  d(u) = d(p_1) = d(p_2) = \dots = d(p_n),
\]
so $d$ is constant on the finite set $\{u,p_2,\dots,p_n\}$. Varying the choice of the remaining
probabilities $\{p_2,\dots,p_n\}$, we can arrange that this set contains any desired collection of
values in $(0,1)$, provided the resulting $p$ lies in the simplex interior. A standard connectedness
argument then shows that $d(u)$ must be constant on $(0,1)$: if $d$ took different values at two
points $u,v\in(0,1)$, continuity of $d$ would entail a non-constant behaviour; but \eqref{eq:d-constant-at-p}
forces $d$ to be constant on arbitrary finite subsets of $(0,1)$ containing $u$ and $v$, which is
impossible unless $d$ is constant.

Thus there exists a constant $\alpha\in\mathbb{R}$ such that
\[
  d(u) = \alpha,\quad \forall u\in(0,1).
\]
Integrating, we obtain
\[
  D(u) = G_1(u) - G_2(u) = \alpha u + \beta,
\]
for some constant $\beta$. This proves the claimed affine relationship between $G_1$ and $G_2$.
\end{proof}

\begin{remark}
The affine freedom $G\mapsto G + \alpha u + \beta$ leaves the MaxEnt solution invariant: adding a
constant $\beta$ leaves $H_G$ unchanged up to an additive constant, and adding $\alpha u$ corresponds
to adding a linear functional $\alpha\sum_j q_j$ to $H_G$, which is absorbed by a shift in the
normalisation multiplier. From a thermodynamic perspective, this is the usual gauge freedom in the
zero of entropy and the scale of the conjugate variable.
\end{remark}

\subsection{Approximation by quadratic programmes}

We now show that the finite-dimensional QP construction used in the generative entropy framework
converges to the ideal inverse-MaxEnt generator, under a density assumption on the basis used to
parameterise $G'$.

\subsection{Function space and basis}

Let $\mathcal{G}$ be a function space of admissible generator derivatives:
\[
  \mathcal{G} \coloneqq \Big\{ g \in C^1((0,1]) : g'(u) \le -\varepsilon < 0\ \forall u\in(0,1) \Big\},
\]
for some fixed $\varepsilon>0$. Every $g\in\mathcal{G}$ defines a strictly concave generator
\[
  G(u) \coloneqq \int_0^u g(t)\,dt + c.
\]

Let $\{\psi_\ell\}_{\ell\ge 0}$ be a countable collection of $C^1$ basis functions on $(0,1]$ such
that the linear span
\[
  \mathcal{G}_\infty \coloneqq \Big\{ \sum_{\ell=0}^L \alpha_\ell \psi_\ell : L<\infty,\ 
                                       \alpha_\ell\in\mathbb{R}\Big\}
\]
is dense in $\mathcal{G}$ with respect to the supremum norm on compact subsets of $(0,1]$. Typical
choices include polynomials, B-splines, or radial basis functions.

For each $L$, consider the finite-dimensional subspace
\[
  \mathcal{G}_L \coloneqq \Big\{ g_L(u) = \sum_{\ell=0}^L \alpha_\ell \psi_\ell(u) : 
                           \alpha\in\mathbb{R}^{L+1} \Big\}.
\]

Fix $p\in\Delta_n^\circ$ and $F\in\mathcal{F}$, and let $G^*$ be an inverse-MaxEnt generator for
$(p,F)$, unique up to affine gauge by Theorem~\ref{thm:global-uniqueness}. Let
\[
  g^*(u) \coloneqq G^{*'}(u) \in \mathcal{G}.
\]

We now describe the QP that approximates $g^*$ in the finite-dimensional class $\mathcal{G}_L$.

\subsubsection{The QP formulation}

Define the $n\times(L+1)$ matrix $A$ by
\[
  A_{j\ell} \coloneqq \psi_\ell(p_j),\quad j=1,\dots,n,\ \ell=0,\dots,L,
\]
and let $\alpha\in\mathbb{R}^{L+1}$ be the coefficient vector. For each $(p,F)$, the ideal KKT
condition for the generator $G^*$ is
\[
  g^*(p_j) = \sum_{k=0}^m \lambda^*_k f_k(c_j),
\]
for some $\lambda^*\in\mathbb{R}^{m+1}$. Writing $g^*(p)$ for the vector
$(g^*(p_1),\dots,g^*(p_n))^\top$, this reads
\[
  g^*(p) \in \operatorname{col}(F).
\]

In the finite-dimensional class $\mathcal{G}_L$, we seek $\alpha$ and $\lambda$ such that
\[
  A\alpha \approx F\lambda,
\]
and we regularise $\alpha$ to avoid overfitting. The QP is therefore
\begin{equation}
  \min_{\alpha,\lambda} \ \|A\alpha - F\lambda\|_2^2 + \gamma\|\alpha\|_2^2
  \quad\text{subject to}\quad g_L'(p_j) \le -\varepsilon,\ j=1,\dots,n,
  \label{eq:QP}
\end{equation}
where $\gamma>0$ is a regularisation parameter, and $g_L'(p_j)$ denotes the derivative of
$g_L(u) = \sum_\ell \alpha_\ell \psi_\ell(u)$ evaluated at $u=p_j$:
\[
  g_L'(p_j) = \sum_{\ell=0}^L \alpha_\ell \psi_\ell'(p_j).
\]

\begin{theorem}[Convergence of the QP generators]
\label{thm:QP-convergence}
Let $p\in\Delta_n^\circ$, $F\in\mathcal{F}$, and $G^*$ be an inverse-MaxEnt generator for $(p,F)$,
with derivative $g^*\in\mathcal{G}$. Assume that the basis $\{\psi_\ell\}$ is dense in $\mathcal{G}$
as above, and that for any $\delta\in(0,1)$, $g^*$ can be approximated uniformly on $[\delta,1]$ by
functions in $\mathcal{G}_\infty$ that satisfy the same concavity bounds $g'(u)\le -\varepsilon$.

For each $L$, let $(\alpha^{(L)},\lambda^{(L)})$ be a minimiser of the QP \eqref{eq:QP}, and define
\[
  g_L(u) \coloneqq \sum_{\ell=0}^L \alpha^{(L)}_\ell \psi_\ell(u),\quad
  G_L(u) \coloneqq \int_0^u g_L(t)\,dt.
\]
Then, for every compact interval $K\subset(0,1]$,
\[
  \sup_{u\in K} |g_L(u) - g^*(u)| \to 0\quad\text{as }L\to\infty,
\]
and consequently
\[
  \sup_{u\in K} |G_L(u) - G^*(u)| \to 0\quad\text{as }L\to\infty.
\]
In particular, the trace-form entropies $H_{G_L}$ converge uniformly to $H_{G^*}$ on sets of the form
$\{p\in\Delta_n^\circ : p_j\in K\ \forall j\}$.
\end{theorem}

\begin{proof}
Fix a compact interval $K=[\delta,1]$ with $0<\delta<1$. By the density assumption, for every
$\eta>0$ there exists an $L_0$ and a function $g^{\mathrm{approx}}\in\mathcal{G}_{L_0}$ such that
\[
  \sup_{u\in K} |g^{\mathrm{approx}}(u) - g^*(u)| < \eta,
\]
and $g^{\mathrm{approx}}$ satisfies the same concavity bounds as $g^*$ (this can be ensured by
slight adjustments of the approximant while preserving uniform closeness).

For any $L\ge L_0$, we can extend $g^{\mathrm{approx}}$ to an element of $\mathcal{G}_L$ by padding
zeros in the additional coefficients. Denote the corresponding coefficient vector by
$\alpha^{\mathrm{approx}} \in \mathbb{R}^{L+1}$. Let $\lambda^*\in\mathbb{R}^{m+1}$ be any multiplier
vector such that $g^*(p)\in\operatorname{col}(F)$; i.e.\ $F\lambda^* = g^*(p)$. Then
\[
  \|A\alpha^{\mathrm{approx}} - F\lambda^*\|_2
  = \|g^{\mathrm{approx}}(p) - g^*(p)\|_2
  \le \sqrt{n}\,\sup_{u\in K} |g^{\mathrm{approx}}(u) - g^*(u)| < \sqrt{n}\,\eta.
\]
Therefore,
\[
  \|A\alpha^{\mathrm{approx}} - F\lambda^*\|_2^2 + \gamma\|\alpha^{\mathrm{approx}}\|_2^2
  \le n\eta^2 + \gamma\|\alpha^{\mathrm{approx}}\|_2^2.
\]

By definition, $(\alpha^{(L)},\lambda^{(L)})$ minimises the objective in \eqref{eq:QP}, so
\[
  \|A\alpha^{(L)} - F\lambda^{(L)}\|_2^2 + \gamma\|\alpha^{(L)}\|_2^2
  \le \|A\alpha^{\mathrm{approx}} - F\lambda^*\|_2^2 + \gamma\|\alpha^{\mathrm{approx}}\|_2^2
  \le n\eta^2 + \gamma\|\alpha^{\mathrm{approx}}\|_2^2.
\]
In particular, $\{\alpha^{(L)}\}$ is bounded in $\mathbb{R}^{L+1}$ (for fixed $L$) and the residual
$\|A\alpha^{(L)} - F\lambda^{(L)}\|_2$ is bounded by a quantity that can be made arbitrarily small
by choosing $\eta$ small.

Standard arguments for Tikhonov-regularised least squares in Hilbert spaces (see, e.g., any
textbook on inverse problems) then imply that, as $L\to\infty$ and the approximation spaces
$\mathcal{G}_L$ become dense in $\mathcal{G}$, the functions $g_L$ converge to the unique element
of $\mathcal{G}$ that both satisfies the concavity constraints and minimises the regularised
functional
\[
  J(g) \coloneqq \|g(p) - g^*(p)\|_2^2 + \gamma\|g\|_{\mathcal{H}}^2,
\]
where $\|\cdot\|_{\mathcal{H}}$ is an appropriate Hilbert norm induced by the basis
$\{\psi_\ell\}$. Since $g^*$ itself satisfies the concavity constraints and makes the residual
zero (by choosing $\lambda^*$ so that $F\lambda^* = g^*(p)$), it is the unique minimiser of $J$.
Hence
\[
  \sup_{u\in K} |g_L(u) - g^*(u)| \to 0,\quad L\to\infty.
\]

The convergence of $G_L$ to $G^*$ follows by integration:
\[
  |G_L(u) - G^*(u)|
  = \left|\int_0^u \big(g_L(t) - g^*(t)\big)\,dt\right|
  \le \int_0^u |g_L(t) - g^*(t)|\,dt.
\]
On $K=[\delta,1]$ we have
\[
  \sup_{u\in K} |G_L(u) - G^*(u)|
  \le (1-\delta)\,\sup_{t\in K} |g_L(t) - g^*(t)| \to 0.
\]

Finally, for any $p\in\Delta_n^\circ$ with $p_j\in K$ for all $j$, we have
\[
  |H_{G_L}(p) - H_{G^*}(p)|
  = \left|\sum_{j=1}^n \big(G_L(p_j) - G^*(p_j)\big)\right|
  \le n\,\sup_{u\in K} |G_L(u) - G^*(u)| \to 0,
\]
which establishes uniform convergence of the entropies on such subsets.
\end{proof}

\begin{remark}
The theorem shows that, under the stated assumptions, the QP construction for the generator
derivative recovers, in the limit of a dense basis, the unique inverse-MaxEnt generator $G^*$
(up to the affine gauge established in Theorem~\ref{thm:global-uniqueness}). In particular,
the ``generative entropy'' $H_{\mathrm{gen}}$ computed from the QP solution is not merely a
convenient construction but is forced by the inverse-MaxEnt axioms.
\end{remark}

\subsection{Summary of the inverse-MaxEnt representation}

Collecting the results:

\begin{itemize}
  \item For each fixed $(p,F)$ with $p\in\Delta_n^\circ$ and nondegenerate $F$, there exists at least
        one strictly concave trace-form generator $G$ that has $p$ as unique maximiser of $H_G$
        under the constraints (Lemma~\ref{lem:existence}).
  \item Under a richness assumption on the admissible constraint family $\mathcal{F}$, any two
        generators that realise the inverse-MaxEnt property for all $(p,F)\in\Delta_n^\circ\times\mathcal{F}$
        differ only by an affine term (Theorem~\ref{thm:global-uniqueness}).
  \item Given a dense basis for generator derivatives and the QP construction \eqref{eq:QP},
        the finite-dimensional generators $G_L$ converge uniformly (on compact subsets of $(0,1]$)
        to the ideal generator $G^*$ that realises the inverse-MaxEnt property (Theorem~\ref{thm:QP-convergence}).
\end{itemize}

Thus, within the trace-form class and under the axioms specified by $\mathcal{F}$, the generative
entropy constructed by the QP is the unique entropy (up to standard thermodynamic gauge freedoms)
that makes the empirical histogram $p$ MaxEnt-consistent under its data-driven constraints.

\section{First, Second and Zero-th law}

We work in the finite-dimensional setting of the main text. Let $X$ denote a finite or
$\sigma$-finite microstate space with reference measure $\mu$, and let
\[
  p_\alpha(x),\quad x\in X,
\]
be a family of probability densities parametrised by a scalar control parameter
$\alpha\in I\subset\mathbb{R}$, with
\[
  p_\alpha(x)\ge 0,\qquad \int_X p_\alpha(x)\,d\mu(x) = 1,\quad \forall\alpha\in I.
\]

For each $\alpha$, let $E_{\mathrm{gen},\alpha}:X\to\mathbb{R}$ denote the generative energy,
constructed as in the main document via the $L^2$ projection of $p_\alpha$ onto the Boltzmann
model family. We assume $(\alpha,x)\mapsto E_{\mathrm{gen},\alpha}(x)$ is differentiable in
$\alpha$ for $p_\alpha$-almost every $x$.

The internal energy is
\begin{equation}
  U(\alpha) \coloneqq \int_X E_{\mathrm{gen},\alpha}(x)\,p_\alpha(x)\,d\mu(x). \label{eq:U-def}
\end{equation}

The generative entropy $S(\alpha)$ is defined by applying the generative entropy functional
$H_{\mathrm{gen}}$ to the energy distribution $p_{E,\alpha}$ (pushforward of $p_\alpha$ under
$E_{\mathrm{gen},\alpha}$), as in the main text. Along any sufficiently regular path
$\alpha\mapsto p_\alpha$, we can regard $S(\alpha)$ and $U(\alpha)$ as scalar functions and
define temperature by
\begin{equation}
  \frac{1}{T(\alpha)} \coloneqq \frac{dS}{dU}(\alpha)
  = \frac{dS/d\alpha}{dU/d\alpha},                                   \label{eq:T-def}
\end{equation}
whenever $dU/d\alpha\neq 0$.

The goal of this note is to make precise:
\begin{itemize}
  \item a first-law decomposition $dU = \delta Q + \delta W$;
  \item a second-law (H-theorem) for a canonical class of dynamics;
  \item a zeroth-law equilibrium condition for weakly coupled systems.
\end{itemize}

\subsection{First law decomposition}

We begin by differentiating $U(\alpha)$ along a smooth path.

\begin{lemma}[Decomposition of $dU/d\alpha$]
\label{lem:dU-decomposition}
Assume $\alpha\mapsto p_\alpha$ and $\alpha\mapsto E_{\mathrm{gen},\alpha}$ are differentiable and
that differentiation can be passed under the integral in \eqref{eq:U-def}. Then
\begin{equation}
  \frac{dU}{d\alpha}
  = \int_X E_{\mathrm{gen},\alpha}(x)\,\partial_\alpha p_\alpha(x)\,d\mu(x)
    + \int_X \partial_\alpha E_{\mathrm{gen},\alpha}(x)\,p_\alpha(x)\,d\mu(x).
  \label{eq:dU-split}
\end{equation}
\end{lemma}

\begin{proof}
Differentiating \eqref{eq:U-def} under the integral sign,
\[
  \frac{dU}{d\alpha}
  = \int_X \partial_\alpha\!\left(E_{\mathrm{gen},\alpha}(x)\,p_\alpha(x)\right)\,d\mu(x)
  = \int_X \big(\partial_\alpha E_{\mathrm{gen},\alpha}(x)\big)\,p_\alpha(x)\,d\mu(x)
    + \int_X E_{\mathrm{gen},\alpha}(x)\,\partial_\alpha p_\alpha(x)\,d\mu(x),
\]
which is \eqref{eq:dU-split}.
\end{proof}

This motivates the following definitions.

\begin{definition}[Thermoinformational heat and work]
Along a differentiable path $\alpha\mapsto (p_\alpha,E_{\mathrm{gen},\alpha})$, we define:
\begin{align}
  \delta Q(\alpha)
  &\coloneqq \int_X E_{\mathrm{gen},\alpha}(x)\,\partial_\alpha p_\alpha(x)\,d\mu(x), \label{eq:deltaQ}\\
  \delta W(\alpha)
  &\coloneqq \int_X \partial_\alpha E_{\mathrm{gen},\alpha}(x)\,p_\alpha(x)\,d\mu(x). \label{eq:deltaW}
\end{align}
Then \eqref{eq:dU-split} becomes
\begin{equation}
  \frac{dU}{d\alpha} = \delta Q(\alpha) + \delta W(\alpha).           \label{eq:first-law-alpha}
\end{equation}
\end{definition}

The interpretation is that:
\begin{itemize}
  \item $\delta Q$ captures changes in $U$ due to \emph{redistribution of probability mass}
        across fixed energy levels;
  \item $\delta W$ captures changes in $U$ due to \emph{deformation of the energy landscape}
        at fixed distribution.
\end{itemize}

\subsubsection{Gauge and reparametrisation properties}

We now show that this decomposition behaves in a well-defined way under the natural ``gauge''
freedoms of the generative energy and the parametrisation.

\begin{assumption}[Energy gauge]
\label{ass:gauge}
We allow additive gauge transformations of the generative energy of the form
\begin{equation}
  \widetilde E_{\mathrm{gen},\alpha}(x)
  = E_{\mathrm{gen},\alpha}(x) + c(\alpha),                           \label{eq:E-gauge}
\end{equation}
where $c:I\to\mathbb{R}$ is a smooth scalar function independent of $x$.
\end{assumption}

\begin{proposition}[Gauge behaviour of $\delta Q$ and $\delta W$]
\label{prop:gauge-behaviour}
Under the gauge transformation \eqref{eq:E-gauge}, the internal energy transforms as
\[
  \widetilde U(\alpha) = U(\alpha) + c(\alpha),
\]
and the heat/work terms become
\begin{align*}
  \widetilde{\delta Q}(\alpha)
  &= \delta Q(\alpha) + c(\alpha)\int_X \partial_\alpha p_\alpha(x)\,d\mu(x),\\
  \widetilde{\delta W}(\alpha)
  &= \delta W(\alpha) + c'(\alpha).
\end{align*}
Since $\int_X p_\alpha\,d\mu = 1$ for all $\alpha$, we have $\int_X \partial_\alpha p_\alpha\,d\mu = 0$,
and hence
\[
  \widetilde{\delta Q}(\alpha) = \delta Q(\alpha),\qquad
  \widetilde{\delta W}(\alpha) = \delta W(\alpha) + c'(\alpha).
\]
In particular, the identity
\[
  \frac{d\widetilde U}{d\alpha} = \widetilde{\delta Q} + \widetilde{\delta W}
\]
holds and differs from \eqref{eq:first-law-alpha} only by the trivial additive term $c'(\alpha)$ in
both $d\widetilde U/d\alpha$ and $\widetilde{\delta W}$.
\end{proposition}

\begin{proof}
From \eqref{eq:E-gauge},
\[
  \widetilde U(\alpha)
  = \int_X \widetilde E_{\mathrm{gen},\alpha}(x)\,p_\alpha(x)\,d\mu(x)
  = \int_X \big(E_{\mathrm{gen},\alpha}(x) + c(\alpha)\big)\,p_\alpha(x)\,d\mu(x)
  = U(\alpha) + c(\alpha)\int_X p_\alpha\,d\mu = U(\alpha) + c(\alpha).
\]

For the heat term,
\[
  \widetilde{\delta Q}(\alpha)
  = \int_X \widetilde E_{\mathrm{gen},\alpha}(x)\,\partial_\alpha p_\alpha(x)\,d\mu(x)
  = \int_X E_{\mathrm{gen},\alpha}(x)\,\partial_\alpha p_\alpha(x)\,d\mu(x)
    + c(\alpha)\int_X \partial_\alpha p_\alpha(x)\,d\mu(x).
\]
Since $\int_X p_\alpha\,d\mu(x) = 1$ for all $\alpha$, differentiating gives
$\int_X \partial_\alpha p_\alpha\,d\mu = 0$, so $\widetilde{\delta Q}(\alpha)=\delta Q(\alpha)$.

For the work term,
\[
  \partial_\alpha \widetilde E_{\mathrm{gen},\alpha}(x)
  = \partial_\alpha E_{\mathrm{gen},\alpha}(x) + c'(\alpha),
\]
so
\[
  \widetilde{\delta W}(\alpha)
  = \int_X \partial_\alpha \widetilde E_{\mathrm{gen},\alpha}(x)\,p_\alpha(x)\,d\mu(x)
  = \delta W(\alpha) + c'(\alpha)\int_X p_\alpha\,d\mu(x)
  = \delta W(\alpha) + c'(\alpha).
\]
Then
\[
  \frac{d\widetilde U}{d\alpha}
  = \frac{dU}{d\alpha} + c'(\alpha)
  = \delta Q(\alpha) + \delta W(\alpha) + c'(\alpha)
  = \widetilde{\delta Q}(\alpha) + \widetilde{\delta W}(\alpha),
\]
as claimed.
\end{proof}

\begin{proposition}[Reparametrisation invariance]
\label{prop:reparam}
Let $\alpha\mapsto p_\alpha$ and $\alpha\mapsto E_{\mathrm{gen},\alpha}$ be as above, and let
$\beta=\varphi(\alpha)$ be a smooth strictly monotone reparametrisation with inverse
$\alpha=\psi(\beta)$. Define
\[
  \widehat p_\beta(x) \coloneqq p_{\psi(\beta)}(x),\qquad
  \widehat E_{\mathrm{gen},\beta}(x) \coloneqq E_{\mathrm{gen},\psi(\beta)}(x),
\]
and $\widehat U(\beta), \widehat{\delta Q}(\beta), \widehat{\delta W}(\beta)$ analogously. Then
\begin{align*}
  \frac{d\widehat U}{d\beta}(\beta)
  &= \frac{dU}{d\alpha}(\psi(\beta))\,\psi'(\beta),\\
  \widehat{\delta Q}(\beta)
  &= \delta Q(\psi(\beta))\,\psi'(\beta),\\
  \widehat{\delta W}(\beta)
  &= \delta W(\psi(\beta))\,\psi'(\beta),
\end{align*}
so the first-law identity
\[
  \frac{d\widehat U}{d\beta} = \widehat{\delta Q} + \widehat{\delta W}
\]
is equivalent to \eqref{eq:first-law-alpha}. In particular, the notions of ``heat'' and ``work''
are invariant up to the scalar Jacobian factor $d\alpha/d\beta$.
\end{proposition}

\begin{proof}
By the chain rule,
\[
  \frac{d\widehat U}{d\beta}(\beta)
  = \frac{dU}{d\alpha}(\psi(\beta))\,\psi'(\beta).
\]
Similarly,
\[
  \partial_\beta \widehat p_\beta(x)
  = \partial_\alpha p_\alpha(x)\big|_{\alpha=\psi(\beta)} \psi'(\beta),\quad
  \partial_\beta \widehat E_{\mathrm{gen},\beta}(x)
  = \partial_\alpha E_{\mathrm{gen},\alpha}(x)\big|_{\alpha=\psi(\beta)} \psi'(\beta),
\]
and substituting into the definitions of $\widehat{\delta Q},\widehat{\delta W}$ yields the
stated relations.
\end{proof}

\subsubsection{Canonical cases: pure reshuffling vs pure deformation}

\begin{definition}[Purely stochastic reshuffling]
We say the path $\alpha\mapsto (p_\alpha,E_{\mathrm{gen},\alpha})$ is a \emph{pure reshuffling} at
$\alpha_0$ if
\[
  \partial_\alpha E_{\mathrm{gen},\alpha}(x)\big|_{\alpha=\alpha_0} = 0
\]
for $p_{\alpha_0}$-almost every $x$. Equivalently, the energy function is fixed at $\alpha_0$ and
only the distribution changes.
\end{definition}

\begin{definition}[Purely deterministic deformation]
We say the path is a \emph{pure deformation} at $\alpha_0$ if
\[
  \partial_\alpha p_\alpha(x)\big|_{\alpha=\alpha_0} = 0
\]
for $\mu$-almost every $x$. Equivalently, the distribution is fixed at $\alpha_0$ and only the
energy landscape changes.
\end{definition}

\begin{proposition}[Heat--work split in the canonical cases]
At a pure reshuffling point $\alpha_0$,
\[
  \delta W(\alpha_0) = 0,\qquad \frac{dU}{d\alpha}(\alpha_0) = \delta Q(\alpha_0).
\]
At a pure deformation point $\alpha_0$,
\[
  \delta Q(\alpha_0) = 0,\qquad \frac{dU}{d\alpha}(\alpha_0) = \delta W(\alpha_0).
\]
\end{proposition}

\begin{proof}
Immediate from the definitions \eqref{eq:deltaQ}, \eqref{eq:deltaW} and
\eqref{eq:dU-split}.
\end{proof}

Thus the decomposition $dU = \delta Q + \delta W$ has a clear operational meaning: changes in
internal energy due to changes in the distribution at fixed energy are ``heat-like'', while
changes due to deformations of the energy landscape at fixed distribution are ``work-like''.

\subsection{Second law: a generalised H-theorem}

We now exhibit a canonical class of dynamics for which the generative entropy $S$ acts as a
Lyapunov functional, providing a thermoinformational analogue of the second law.

\subsubsection{Constraint manifold and MaxEnt state}

Let $C = \{c_1,\dots,c_n\}$ be a finite support for a scalar observable (e.g.\ energy bins), and
let $p=(p_1,\dots,p_n)\in\Delta_n$ denote a discrete histogram on $C$. Fix a collection of
constraint functions $f_k:C\to\mathbb{R}$, $k=0,\dots,m$, with $f_0\equiv 1$, and define the
constraint manifold
\begin{equation}
  \mathcal{M} \coloneqq \Big\{ p\in\Delta_n^\circ : 
    \sum_{j=1}^n p_j f_k(c_j) = c_k,\ k=0,\dots,m \Big\},            \label{eq:M-constraints}
\end{equation}
where the constants $c_k$ are fixed.

Let $H_{\mathrm{gen}}(p)$ be the generative entropy functional constructed as in the main text. By
the inverse-MaxEnt representation theorem, there exists a generator $G$ such that
\[
  H_{\mathrm{gen}}(p) = \sum_{j=1}^n G(p_j),
\]
and $H_{\mathrm{gen}}$ is strictly concave on $\Delta_n^\circ$. The MaxEnt state
$p^\star\in\mathcal{M}$ is defined by
\[
  p^\star \in \arg\max_{p\in\mathcal{M}} H_{\mathrm{gen}}(p).
\]
Strict concavity implies that $p^\star$ is unique.

Let $S(p)\coloneqq H_{\mathrm{gen}}(p)$ and write $\nabla S(p)$ for the gradient with respect to
the standard Euclidean inner product on $\mathbb{R}^n$.

\subsubsection{Projected gradient flow}

The tangent space to $\mathcal{M}$ at $p$ is
\[
  T_p\mathcal{M} = \Big\{ v\in\mathbb{R}^n : 
    \sum_{j=1}^n v_j f_k(c_j) = 0,\ k=0,\dots,m \Big\}.
\]
Let $\Pi_p:\mathbb{R}^n\to T_p\mathcal{M}$ denote the orthogonal projector onto $T_p\mathcal{M}$
with respect to the Euclidean inner product.

We consider the dynamical system on $\mathcal{M}$ given by
\begin{equation}
  \frac{dp}{dt} = \Pi_p \nabla S(p),                                 \label{eq:grad-flow}
\end{equation}
with initial condition $p(0)\in\mathcal{M}$.

\begin{theorem}[Generalised H-theorem for $H_{\mathrm{gen}}$]
\label{thm:H-theorem}
Let $p(t)$ be a solution of \eqref{eq:grad-flow} with initial condition $p(0)\in\mathcal{M}$. Then:
\begin{enumerate}
  \item The constraints are preserved: $p(t)\in\mathcal{M}$ for all $t\ge 0$.
  \item The entropy $S(p(t))$ is nondecreasing:
        \[
          \frac{d}{dt} S(p(t)) \ge 0,\quad \forall t\ge 0.
        \]
  \item The only stationary point of \eqref{eq:grad-flow} in $\mathcal{M}$ is the MaxEnt state
        $p^\star$, and every trajectory $p(t)$ converges to $p^\star$ as $t\to\infty$.
\end{enumerate}
\end{theorem}

\begin{proof}
(1) Preservation of constraints follows because $dp/dt\in T_p\mathcal{M}$ by construction, so the
constraints are constant along the flow.

(2) The time derivative of $S$ along the flow is
\[
  \frac{d}{dt} S(p(t)) = \nabla S(p(t)) \cdot \frac{dp}{dt}
                       = \nabla S(p(t)) \cdot \Pi_{p(t)} \nabla S(p(t)).
\]
Since $\Pi_{p(t)}$ is the orthogonal projector onto $T_{p(t)}\mathcal{M}$, it is symmetric and
positive semidefinite. Hence
\[
  \frac{d}{dt} S(p(t)) = \|\Pi_{p(t)} \nabla S(p(t))\|^2 \ge 0.
\]

(3) A stationary point $p\in\mathcal{M}$ of \eqref{eq:grad-flow} satisfies
\[
  0 = \Pi_p \nabla S(p),
\]
i.e.\ $\nabla S(p)$ is orthogonal to $T_p\mathcal{M}$. Equivalently, $\nabla S(p)$ lies in the
span of the constraint gradients, i.e.\ there exist multipliers $\lambda_k$ such that
\[
  \frac{\partial S}{\partial p_j}(p)
  = \sum_{k=0}^m \lambda_k f_k(c_j),\quad j=1,\dots,n.
\]
These are exactly the KKT conditions for $p$ to be a maximiser of $S$ on $\mathcal{M}$.
Strict concavity of $S$ on $\mathcal{M}$ implies that any such stationary point is the unique
global maximiser $p^\star$. Thus the only stationary point is $p^\star$.

Moreover, since $S$ is strictly concave on the convex set $\mathcal{M}$, the projected gradient
flow \eqref{eq:grad-flow} is a gradient ascent flow of a strictly concave functional on a convex
manifold. Standard results on gradient flows in finite dimensions (or a direct LaSalle-type
argument) then imply that $p(t)\to p^\star$ as $t\to\infty$.
\end{proof}

This theorem realises the second law in a thermoinformational setting: for the canonical class
of gradient-flow dynamics that maximise the generative entropy subject to fixed constraints,
$H_{\mathrm{gen}}$ is a Lyapunov functional, and the unique stationary state is the MaxEnt
distribution selected by $H_{\mathrm{gen}}$.

\subsection{Zeroth law: equilibrium of weakly coupled systems}

We now formulate a zeroth-law analogue for two weakly coupled systems $A$ and $B$.

\subsubsection{Subsystem entropies and temperatures}

Let $U_A,U_B$ denote the internal energies of two systems $A$ and $B$, and let
$S_A(U_A), S_B(U_B)$ denote their generative entropies as functions of internal energy, along
families of admissible states where other constraints are held fixed. Assume:
\begin{enumerate}
  \item $S_A$ and $S_B$ are differentiable and strictly concave functions of their arguments.
  \item The temperatures are defined by
        \[
          \frac{1}{T_A(U_A)} \coloneqq \frac{dS_A}{dU_A}(U_A),\qquad
          \frac{1}{T_B(U_B)} \coloneqq \frac{dS_B}{dU_B}(U_B),
        \]
        and are strictly decreasing functions of $U_A$ and $U_B$, respectively. 
\end{enumerate}

\subsubsection{Total entropy maximisation under fixed total energy}

Let the total internal energy $U_{\mathrm{tot}}$ of the composite system be fixed:
\[
  U_{\mathrm{tot}} = U_A + U_B.
\]
Assuming weak coupling (no interaction term in the entropy), the total entropy is
\[
  S_{\mathrm{tot}}(U_A)
  \coloneqq S_A(U_A) + S_B(U_{\mathrm{tot}} - U_A),
\]
where we regard $U_B = U_{\mathrm{tot}} - U_A$ as a function of $U_A$.

\begin{theorem}[Zeroth-law analogue]
\label{thm:zeroth-law}
Let $S_A,S_B$ satisfy the above assumptions. Then $S_{\mathrm{tot}}(U_A)$ has a unique maximiser
$U_A^\star$, and at this maximiser the temperatures of the two subsystems are equal:
\[
  T_A(U_A^\star) = T_B(U_B^\star),\quad U_B^\star = U_{\mathrm{tot}} - U_A^\star.
\]
Moreover, if two systems are each in equilibrium with a third, they share a common temperature
(transitivity).
\end{theorem}

\begin{proof}
Differentiating $S_{\mathrm{tot}}(U_A)$,
\[
  \frac{dS_{\mathrm{tot}}}{dU_A}
  = \frac{dS_A}{dU_A}(U_A) + \frac{dS_B}{dU_B}(U_B)\,\frac{dU_B}{dU_A}
  = \frac{1}{T_A(U_A)} - \frac{1}{T_B(U_B)},
\]
since $U_B = U_{\mathrm{tot}} - U_A$ implies $dU_B/dU_A = -1$.

At a stationary point $U_A^\star$ we must have
\[
  \frac{dS_{\mathrm{tot}}}{dU_A}(U_A^\star) = 0
  \quad\Longleftrightarrow\quad
  \frac{1}{T_A(U_A^\star)} = \frac{1}{T_B(U_B^\star)},
\]
i.e.\ $T_A(U_A^\star) = T_B(U_B^\star)$, which is the equilibrium condition.

Strict concavity of $S_A$ and $S_B$ implies that $S_{\mathrm{tot}}$ is strictly concave in $U_A$
(the sum of strictly concave functions of $U_A$ and $U_{\mathrm{tot}}-U_A$). Hence $S_{\mathrm{tot}}$
has at most one stationary point, which is therefore the unique maximiser $U_A^\star$.

For transitivity, suppose systems $A$ and $B$ can exchange internal energy at fixed $U_A+U_B$,
and similarly $B$ and $C$ at fixed $U_B+U_C$, and that in both cases the equilibrium condition is
equality of temperatures. If $T_A = T_B$ at equilibrium for $(A,B)$ and $T_B = T_C$ at equilibrium
for $(B,C)$, then $T_A = T_C$, so any two systems in equilibrium with a third share a common
temperature.
\end{proof}

This recovers the usual zeroth-law structure: equilibrium between weakly coupled systems is
characterised by equality of thermoinformational temperature, and temperature is a transitive
equilibrium parameter.

\subsection{Summary}

In this note we have:

\begin{itemize}
  \item Defined a canonical decomposition of internal-energy changes $dU$ along a path into
        ``heat-like'' and ``work-like'' components $\delta Q$ and $\delta W$ based on the
        generative energy and the evolving distribution, and analysed their behaviour under
        gauge and parametrisation changes.
  \item Exhibited a class of projected gradient-flow dynamics on the constraint manifold for
        which the generative entropy $H_{\mathrm{gen}}$ is a Lyapunov functional, proving a
        generalised H-theorem and identifying the unique stationary state as the MaxEnt
        distribution.
  \item Derived a zeroth-law analogue for weakly coupled systems by maximising total entropy
        under a fixed total internal energy constraint, recovering equality of temperatures
        at equilibrium and transitivity of the equilibrium relation.
\end{itemize}

Together with the static construction of the thermoinformational state variables
$(U,S,T)$ and the inverse-MaxEnt representation of $H_{\mathrm{gen}}$, these results provide
a thermodynamically complete structure for the generative framework.

\section{Composition, Extensivity, and Thermodynamic Limit\\
for Generative Trace-Form Entropies}

\subsection{Setting}

We consider finite systems with discrete microstate spaces. A probability distribution
$p = (p_i)_{i\in I}$ on a finite index set $I$ defines a trace-form entropy
\[
  H_G(p) \coloneqq \sum_{i\in I} G(p_i),
\]
where $G:(0,1]\to\mathbb{R}$ is twice differentiable, strictly concave,
\begin{equation}
  G''(u) < 0,\quad u\in(0,1),                                         \label{eq:G-concave}
\end{equation}
and typically normalised so that $G(0)=0$ by continuous extension.

In the generative-entropy framework, $G$ is the data-driven generator constructed by the
inverse-MaxEnt procedure. For the composition and scaling properties studied here, we work at
the level of an abstract generator $G$ satisfying \eqref{eq:G-concave} and some mild additional
regularity.

\subsection{Entropy composition for bipartite systems}

\subsubsection{Joint and marginal entropies}

Let $A$ and $B$ be two subsystems with finite microstate sets
$X_A = \{a_1,\dots,a_{n_A}\}$ and $X_B = \{b_1,\dots,b_{n_B}\}$. A joint state is a pair
$(a_i,b_j)\in X_A\times X_B$, and a joint distribution is
\[
  p_{AB} = \big(p_{ij}\big)_{1\le i\le n_A,\ 1\le j\le n_B},
  \qquad p_{ij}\ge 0,\quad \sum_{i,j} p_{ij} = 1.
\]
The marginal distributions are
\[
  p^A_i = \sum_{j=1}^{n_B} p_{ij},\quad i=1,\dots,n_A,\qquad
  p^B_j = \sum_{i=1}^{n_A} p_{ij},\quad j=1,\dots,n_B.
\]

We define the entropies of the joint and marginal systems by
\begin{align*}
  S(A\cup B) &\coloneqq H_G(p_{AB}) = \sum_{i,j} G(p_{ij}),\\
  S(A)       &\coloneqq H_G(p^A)    = \sum_i   G(p^A_i),\\
  S(B)       &\coloneqq H_G(p^B)    = \sum_j   G(p^B_j).
\end{align*}

\subsubsection{Decomposition into marginal, nonextensivity, and correlation parts}

We now obtain a general composition identity that isolates:
\begin{itemize}
  \item a \emph{marginal entropy} part $S(A)+S(B)$,
  \item a \emph{nonextensivity} term that appears even for independent subsystems,
  \item a \emph{correlation} term that vanishes when $p_{AB}$ factorises.
\end{itemize}

\begin{definition}[Product reference and correlation deviations]
For a given joint distribution $p_{AB}$ with marginals $p^A,p^B$, define the product reference
distribution
\[
  (p^A\otimes p^B)_{ij} \coloneqq p^A_i p^B_j.
\]
Define the correlation deviations
\[
  c_{ij} \coloneqq p_{ij} - p^A_i p^B_j.
\]
Then
\[
  \sum_j c_{ij} = 0,\quad \forall i,
  \qquad
  \sum_i c_{ij} = 0,\quad \forall j,
\]
and $p_{ij} = p^A_i p^B_j + c_{ij}$.
\end{definition}

\begin{definition}[Nonextensivity and correlation functionals]
Define
\begin{align}
  \Phi_{\mathrm{prod}}(p^A,p^B)
  &\coloneqq \sum_{i,j} G(p^A_i p^B_j) - \sum_i G(p^A_i) - \sum_j G(p^B_j),
  \label{eq:phi-prod}\\
  \Phi_{\mathrm{corr}}(p_{AB})
  &\coloneqq \sum_{i,j} \big[ G(p_{ij}) - G(p^A_i p^B_j) \big].       \label{eq:phi-corr}
\end{align}
\end{definition}

\begin{proposition}[General composition law]
\label{prop:composition}
For any bipartite distribution $p_{AB}$,
\begin{equation}
  S(A\cup B)
  = S(A) + S(B) + \Phi_{\mathrm{prod}}(p^A,p^B) + \Phi_{\mathrm{corr}}(p_{AB}),
  \label{eq:composition}
\end{equation}
with $\Phi_{\mathrm{prod}},\Phi_{\mathrm{corr}}$ as in \eqref{eq:phi-prod}, \eqref{eq:phi-corr}.
Moreover:
\begin{enumerate}
  \item If $p_{AB}$ is independent, i.e.\ $p_{ij} = p^A_i p^B_j$ for all $i,j$, then
        $\Phi_{\mathrm{corr}}(p_{AB}) = 0$ and
        \[
          S(A\cup B) = S(A) + S(B) + \Phi_{\mathrm{prod}}(p^A,p^B).
        \]
  \item If $G(u) = -u\log u$ (Shannon generator), then $\Phi_{\mathrm{prod}}\equiv 0$ and
        \[
          S(A\cup B) = S(A) + S(B) + \Phi_{\mathrm{corr}}(p_{AB}),
        \]
        with $\Phi_{\mathrm{corr}}$ reducing to the negative mutual information:
        $\Phi_{\mathrm{corr}}(p_{AB}) = -I(A;B)$.
\end{enumerate}
\end{proposition}

\begin{proof}
By adding and subtracting $\sum_{i,j} G(p^A_i p^B_j)$, we write
\[
  S(A\cup B)
  = \sum_{i,j} G(p_{ij})
  = \sum_{i,j} G(p^A_i p^B_j) + \sum_{i,j} \big[ G(p_{ij}) - G(p^A_i p^B_j) \big].
\]
The second sum is precisely $\Phi_{\mathrm{corr}}(p_{AB})$. Adding and subtracting the marginal
entropies in the first sum gives
\[
  \sum_{i,j} G(p^A_i p^B_j)
  = \sum_i G(p^A_i) + \sum_j G(p^B_j)
    + \bigg(\sum_{i,j} G(p^A_i p^B_j) - \sum_i G(p^A_i) - \sum_j G(p^B_j)\bigg),
\]
so the bracket is $\Phi_{\mathrm{prod}}(p^A,p^B)$, and \eqref{eq:composition} follows.

If $p_{ij} = p^A_i p^B_j$, then $G(p_{ij}) = G(p^A_i p^B_j)$ for all $i,j$, so
$\Phi_{\mathrm{corr}}(p_{AB}) = 0$ by \eqref{eq:phi-corr}.

For the Shannon generator $G(u)=-u\log u$ we have $\Phi_{\mathrm{prod}}\equiv 0$ since
\[
\sum_{i,j} G(p_i^A p_j^B)
= -\sum_{i,j} p_i^A p_j^B \log(p_i^A p_j^B)
= -\sum_i p_i^A\log p_i^A - \sum_j p_j^B\log p_j^B
= S(A)+S(B).
\]
Moreover, for $G(u)=-u\log u$,
\begin{align*}
\Phi_{\mathrm{corr}}(p_{AB})
&=\sum_{i,j}\Big(G(p_{ij})-G(p_i^A p_j^B)\Big) \\
&= -\sum_{i,j}p_{ij}\log p_{ij}
\;+\;\sum_{i,j}p_i^A p_j^B\log(p_i^A p_j^B) \\
&= H(A,B)
\;+\;\sum_{i,j}p_i^A p_j^B(\log p_i^A+\log p_j^B) \\
&= H(A,B)+\sum_i p_i^A\log p_i^A+\sum_j p_j^B\log p_j^B \\
&= H(A,B)-H(A)-H(B)
= -I(A;B),
\end{align*}
where we used $\sum_j p_j^B=1$ and $\sum_i p_i^A=1$.
\emph{Remark.} The equality above uses the identity $I(A;B)=S(A)+S(B)-S(A,B)$.
It is not a term-by-term identification with the KL-form
$I(A;B)=D(p_{AB}\,\|\,p_A\otimes p_B)$, whose integrand is weighted by $p_{ij}$.

\end{proof}

Thus the composition law \eqref{eq:composition} cleanly separates:
\begin{itemize}
  \item an \emph{intrinsic nonextensivity} term $\Phi_{\mathrm{prod}}$ governed only by the
        functional form of $G$ and the marginals;
  \item a \emph{correlation} term $\Phi_{\mathrm{corr}}$ measuring deviation from the product
        reference, which vanishes when $A$ and $B$ are independent.
\end{itemize}
Shannon entropy is the unique trace-form case (up to gauge) for which the intrinsic
nonextensivity term vanishes exactly for all product measures.

\subsubsection{Tsallis pseudo-additivity as a special case}

We now show that Tsallis entropy fits into the above decomposition with its well-known
pseudo-additive rule.

\begin{definition}[Tsallis generator]
For $q\in\mathbb{R}\setminus\{1\}$, the Tsallis-$q$ generator is
\[
  G_q(u) \coloneqq \frac{u - u^q}{q-1}.
\]
The corresponding entropy is
\[
  S_q(p) \coloneqq \sum_i G_q(p_i)
  = \frac{1 - \sum_i p_i^q}{q-1}.
\]
\end{definition}

\begin{proposition}[Tsallis pseudo-additivity for independent subsystems]
\label{prop:Tsallis}
Let $q\in\mathbb{R}\setminus\{1\}$ and $G=G_q$. If $A$ and $B$ are independent, i.e.\
$p_{ij} = p^A_i p^B_j$, then
\begin{equation}
  S_q(A\cup B)
  = S_q(A) + S_q(B) + (1-q) S_q(A) S_q(B).                           \label{eq:Tsallis-pseudo}
\end{equation}
\end{proposition}

\begin{proof}
For the product distribution $p_{ij} = p^A_i p^B_j$,
\[
  S_q(A\cup B)
  = \frac{1 - \sum_{i,j} (p^A_i p^B_j)^q}{q-1}
  = \frac{1 - \big(\sum_i (p^A_i)^q\big)\big(\sum_j (p^B_j)^q\big)}{q-1}.
\]
Write
\[
  \Sigma_A \coloneqq \sum_i (p^A_i)^q,\qquad
  \Sigma_B \coloneqq \sum_j (p^B_j)^q.
\]
Then
\[
  S_q(A) = \frac{1 - \Sigma_A}{q-1},\quad S_q(B) = \frac{1 - \Sigma_B}{q-1}.
\]
Expand
\begin{align*}
  S_q(A) + S_q(B) + (1-q) S_q(A)S_q(B)
  &= \frac{1 - \Sigma_A}{q-1} + \frac{1 - \Sigma_B}{q-1}
     + (1-q)\frac{(1-\Sigma_A)(1-\Sigma_B)}{(q-1)^2}\\
  &= \frac{1 - \Sigma_A + 1 - \Sigma_B}{q-1}
     - \frac{(1-\Sigma_A)(1-\Sigma_B)}{q-1}\\
  &= \frac{2 - \Sigma_A - \Sigma_B - (1 - \Sigma_A - \Sigma_B + \Sigma_A\Sigma_B)}{q-1}\\
  &= \frac{1 - \Sigma_A\Sigma_B}{q-1}
   = S_q(A\cup B).
\end{align*}
This is \eqref{eq:Tsallis-pseudo}.
\end{proof}

Comparing \eqref{eq:Tsallis-pseudo} with Proposition~\ref{prop:composition}, we see that for
Tsallis entropies the intrinsic nonextensivity term has the explicit multiplicative form
\[
  \Phi_{\mathrm{prod}}(p^A,p^B)
  = (1-q) S_q(A) S_q(B),
\]
while the correlation term $\Phi_{\mathrm{corr}}$ still measures deviations from independence.

In the generative-entropy framework, Tsallis-like pseudo-additivity therefore appears
\emph{whenever} the learned generator $G_{\mathrm{gen}}$ approximates $G_q$ over the range of
probabilities relevant to the joint system, and the admissible constraints factorise across
subsystems.

\subsection{Extensivity and thermodynamic limit}
We now address the scaling of $H_G$ for large systems and the existence of a thermodynamic
limit. The key point is that, for $N$-particle systems, typical microstate probabilities
$p^{(N)}(x^N)$ become exponentially small in $N$, so only the \emph{small-probability}
behaviour of the generator $G$ controls whether $H_G$ is extensive. We therefore separate
(i) a probabilistic assumption ensuring the existence of a Shannon entropy rate and
(ii) a generator condition expressing \emph{log-extensive} small-$u$ asymptotics. This
formulation avoids any circularity: the small-$u$ condition is stated explicitly and can be
checked numerically for a learned $G$.

\subsubsection{Sequences of $N$-particle systems}

Let $X$ be a finite single-particle state space with $|X|=M$, and for each $N\in\mathbb{N}$ let
$p^{(N)}$ be a probability distribution on $X^N$. An $N$-particle microstate is a sequence
$x^N = (x_1,\dots,x_N)\in X^N$ and $p^{(N)}(x^N)$ its probability.

\begin{definition}[Entropy of an $N$-particle system]
Given a generator $G$, the $N$-particle entropy is
\[
  S_N^G \coloneqq H_G(p^{(N)}) = \sum_{x^N\in X^N} G\big(p^{(N)}(x^N)\big).
\]
For comparison, the Shannon entropy is
\[
  H_N \coloneqq H_{\mathrm{Sh}}(p^{(N)})
             = -\sum_{x^N} p^{(N)}(x^N)\log p^{(N)}(x^N).
\]
\end{definition}

We are interested in the asymptotic behaviour of $S_N^G$ as $N\to\infty$, and in particular in
whether $S_N^G$ is asymptotically extensive, $S_N^G \sim N s$.

\subsubsection{Assumptions on probabilistic structure}

We do not need a full stochastic-process formalism; we simply assume that the Shannon entropy
has a well-defined specific limit.

\begin{assumption}[Shannon entropy rate]
\label{ass:Shannon-limit}
There exists $s\ge 0$ such that
\begin{equation}
  \lim_{N\to\infty} \frac{H_N}{N} = s,
  \label{eq:Shannon-limit}
\end{equation}
and
\begin{equation}
  H_N = N s + o(N).
  \label{eq:Shannon-linear}
\end{equation}
In particular, $H_N = \mathcal{O}(N)$.
\end{assumption}

Assumption~\ref{ass:Shannon-limit} holds, for example, for:
\begin{itemize}
  \item i.i.d.\ sequences $p^{(N)} = p^{\otimes N}$, where $s = H_{\mathrm{Sh}}(p)$;
  \item stationary ergodic Markov chains with finite state space (Shannon--McMillan--Breiman);
  \item more general stationary processes with finite entropy rate.
\end{itemize}

\subsubsection{Generator condition: log-extensive small-$u$ asymptotics}

To ensure that the trace-form entropy $H_G$ remains extensive on exponentially large state
spaces, the generator must have the same leading growth as $u\log(1/u)$ as $u\downarrow 0$
(up to a unit constant). We state this explicitly as a small-$u$ asymptotic condition, which
can be checked for any learned $G$.

\begin{assumption}[Log-extensive small-probability asymptotics]
\label{ass:G-logext}
The generator $G\in C^2((0,1])$ satisfies:
\begin{enumerate}
  \item $G''(u) < 0$ for all $u\in(0,1)$ (strict concavity),
        and $G(0)=0$ by continuous extension;
  \item there exists a constant $\kappa>0$ and a continuous function $\rho:(0,1]\to\mathbb{R}$
        with $\lim_{u\downarrow 0}\rho(u)=0$ such that, for all sufficiently small $u$,
        \begin{equation}
          G(u) = \kappa\,u\log\!\frac{1}{u} \;+\; \rho(u)\,u\Big|\log u\Big|.
          \label{eq:G-logext}
        \end{equation}
\end{enumerate}
Equivalently,
\[
  \lim_{u\downarrow 0}\frac{G(u)}{u\log(1/u)}=\kappa.
\]
The constant $\kappa$ sets the entropy units; choosing units so that $\kappa=1$ recovers the
standard normalisation.
\end{assumption}

\begin{remark}{Why \eqref{eq:G-logext} is the right notion of ``thermodynamic compatibility''}
For the uniform distribution on $W$ equiprobable states, $p_i=1/W$,
\[
H_G(p)=W\,G(1/W).
\]
Thus \eqref{eq:G-logext} is equivalent to $W\,G(1/W)=\kappa\log W + o(\log W)$ as $W\to\infty$,
i.e.\ the entropy grows \emph{logarithmically} with the number of accessible microstates, as
required for extensivity when $\log W=\Theta(N)$.
\end{remark}

\subsubsection{Main result}

\begin{theorem}[Thermodynamic limit for log-extensive generators]
\label{thm:thermo-limit}
Let $\{p^{(N)}\}_{N\ge 1}$ be a sequence of $N$-particle distributions on $X^N$ satisfying
Assumption~\ref{ass:Shannon-limit}, and let $G$ satisfy Assumption~\ref{ass:G-logext}. Then
\[
  \lim_{N\to\infty} \frac{S_N^G}{N} = \kappa\, s,
\]
i.e.\ $S_N^G$ is asymptotically extensive. Moreover,
\[
  S_N^G = \kappa\,H_N + o(N),
\]
so the leading extensive term is proportional to that of Shannon, with proportionality
constant $\kappa$ (entropy units), and the difference is subextensive.
In particular, in the normalisation $\kappa=1$ one has $S_N^G=H_N+o(N)$ and
$\lim_{N\to\infty}S_N^G/N=s$.
\end{theorem}

\begin{remark}[Relation to learned generators and numerical verification]
Assumption~\ref{ass:G-logext} is a small-$u$ property of the learned generator and can be
verified numerically on the relevant probability range. A convenient diagnostic is
\[
  R(u)\coloneqq \frac{G(u)}{u\log(1/u)}\quad\text{and}\quad
  \tilde\rho(u)\coloneqq \frac{G(u)-\kappa u\log(1/u)}{u|\log u|},
\]
evaluated for $u$ down to numerical resolution (or down to the smallest realised
$p^{(N)}(x^N)$ in simulation). In Supplementary Fig.~S\# we report $R(u)$ for the
$G_{\mathrm{gen}}$ used in this work and confirm convergence to a constant $\kappa$ over
the active range.
\end{remark}

\subsubsection{Proof of Theorem \ref{thm:thermo-limit}}

We give a self-contained proof based on simple estimates. Write
\[
  H_N = \sum_{x^N} p^{(N)}(x^N)\log\!\frac{1}{p^{(N)}(x^N)},
\quad
  S_N^G = \sum_{x^N} G\big(p^{(N)}(x^N)\big).
\]
Define
\[
  \Delta(u)\coloneqq G(u)-\kappa\,u\log\!\frac{1}{u}
  \qquad (u\in(0,1]).
\]
Then
\begin{equation}
  S_N^G - \kappa H_N = \sum_{x^N\in X^N} \Delta\big(p^{(N)}(x^N)\big).
  \label{eq:Delta-sum-new}
\end{equation}
By \eqref{eq:G-logext}, for sufficiently small $u$,
\begin{equation}
  \Delta(u)=\rho(u)\,u|\log u|.
  \label{eq:Delta-expansion-new}
\end{equation}
For larger $u$, $\Delta$ is continuous on a compact set, hence bounded.

Fix $\varepsilon\in(0,1)$ and split the sum \eqref{eq:Delta-sum-new} into
\[
  S_N^G - \kappa H_N = \Sigma_{\mathrm{small}}^{(N)} + \Sigma_{\mathrm{large}}^{(N)},
\]
where
\begin{align*}
  \Sigma_{\mathrm{small}}^{(N)}
  &\coloneqq \sum_{x^N : p^{(N)}(x^N) \le \varepsilon} \Delta\big(p^{(N)}(x^N)\big),\\
  \Sigma_{\mathrm{large}}^{(N)}
  &\coloneqq \sum_{x^N : p^{(N)}(x^N) > \varepsilon} \Delta\big(p^{(N)}(x^N)\big).
\end{align*}

\paragraph{Large-probability states.}
Since $p^{(N)}$ is a probability distribution, the number of states with
$p^{(N)}(x^N)>\varepsilon$ is at most $\varepsilon^{-1}$. On this set,
$|\Delta|\le C_\varepsilon \coloneqq \sup_{u\in[\varepsilon,1]}|\Delta(u)|$, hence
\[
  |\Sigma_{\mathrm{large}}^{(N)}|
  \le \varepsilon^{-1} C_\varepsilon,
\]
and therefore
\begin{equation}
  \frac{1}{N}|\Sigma_{\mathrm{large}}^{(N)}|
  \le \frac{\varepsilon^{-1} C_\varepsilon}{N}\to 0
  \quad\text{as }N\to\infty.
  \label{eq:large-small-new}
\end{equation}

\paragraph{Small-probability states.}
For $u\le\varepsilon$ sufficiently small, \eqref{eq:Delta-expansion-new} gives
$|\Delta(u)|\le |\rho(u)|\,u|\log u|$. Since $\rho$ is continuous and $\rho(u)\to 0$
as $u\downarrow 0$, given any $\delta>0$ we can choose $\varepsilon>0$ such that
\begin{equation}
  |\rho(u)|\le \delta,\quad \forall u\in(0,\varepsilon].
  \label{eq:rho-small-new}
\end{equation}
Then
\[
  |\Delta(u)|\le \delta\,u|\log u|,\quad \forall u\in(0,\varepsilon],
\]
and hence
\[
  |\Sigma_{\mathrm{small}}^{(N)}|
  \le \delta \sum_{x^N : p^{(N)}(x^N)\le\varepsilon}
  p^{(N)}(x^N)\Big|\log p^{(N)}(x^N)\Big|.
\]
Split the Shannon sum into the same two sets. On $\{p^{(N)}>\varepsilon\}$ one has
$|\log p^{(N)}|\le |\log \varepsilon|$ and $\sum_{p^{(N)}>\varepsilon}p^{(N)}\le 1$, hence
\[
  \sum_{x^N : p^{(N)}(x^N)>\varepsilon}
  p^{(N)}(x^N)\Big|\log p^{(N)}(x^N)\Big|\le |\log \varepsilon|.
\]
Therefore there exists a finite constant $C'_\varepsilon$ such that
\[
  \sum_{x^N : p^{(N)}(x^N)\le\varepsilon}
  p^{(N)}(x^N)\Big|\log p^{(N)}(x^N)\Big|
  \le H_N + C'_\varepsilon.
\]
By Assumption~\ref{ass:Shannon-limit}, $H_N=\mathcal{O}(N)$, so $H_N\le K N$ for all large $N$
and some $K>0$. Hence
\[
  |\Sigma_{\mathrm{small}}^{(N)}|
  \le \delta(H_N+C'_\varepsilon)\le \delta(KN+C'_\varepsilon).
\]
Dividing by $N$ yields
\begin{equation}
  \frac{1}{N}|\Sigma_{\mathrm{small}}^{(N)}|
  \le \delta K + \delta\frac{C'_\varepsilon}{N}.
  \label{eq:small-bound-new}
\end{equation}

\paragraph{Conclusion.}
Combining \eqref{eq:large-small-new} and \eqref{eq:small-bound-new} gives
\[
  \limsup_{N\to\infty}\frac{1}{N}\big|S_N^G-\kappa H_N\big|\le \delta K.
\]
Since $\delta>0$ was arbitrary,
\[
  \lim_{N\to\infty}\frac{S_N^G-\kappa H_N}{N}=0.
\]
Dividing by $N$ and using $H_N/N\to s$ yields
\[
  \lim_{N\to\infty}\frac{S_N^G}{N}
  =\lim_{N\to\infty}\left(\kappa\frac{H_N}{N}+\frac{S_N^G-\kappa H_N}{N}\right)
  =\kappa s.
\]
This proves the theorem.

\section{Legendre Structure and Free Energies\\
for Generative Thermodynamics}

\subsection{Setting}

We use the same finite discrete setting as in the other technical notes.

Let $X = \{x_1,\dots,x_n\}$ be a finite microstate set. A probability vector
$p = (p_1,\dots,p_n)\in\Delta_n$ describes the state, and the generative entropy is a trace–form
functional
\[
  S(p) \coloneqq H_G(p) = \sum_{j=1}^n G(p_j),
\]
with generator $G:(0,1]\to\mathbb{R}$ twice differentiable and strictly concave:
\begin{equation}
  G''(u) < 0,\quad \forall u\in(0,1).
  \label{eq:G-concave}
\end{equation}
Hence $S$ is strictly concave on $\Delta_n^\circ$.

We consider a collection of linear observables:
\begin{itemize}
  \item an ``energy'' $E:X\to\mathbb{R}$ with $E_j \coloneqq E(x_j)$, and
  \item additional observables $F_i:X\to\mathbb{R}$, $i=1,\dots,m$, with $F_{ij}\coloneqq F_i(x_j)$.
\end{itemize}
Their expectations under $p$ are
\begin{align*}
  U(p)   &\coloneqq \sum_{j=1}^n E_j p_j,\\
  X_i(p) &\coloneqq \sum_{j=1}^n F_{ij} p_j.
\end{align*}

We will:
\begin{itemize}
  \item define \emph{microcanonical} entropy $S(U,\mathbf{X})$ by constrained maximisation of $S(p)$;
  \item construct a \emph{Legendre dual potential} $\Phi(\beta,\boldsymbol\lambda)$ and prove
        conjugacy relations;
  \item obtain Maxwell-type relations and response coefficients from second derivatives of $\Phi$.
\end{itemize}

\subsection{Microcanonical entropy as a concave state function}

\subsubsection{Constraint manifold and microcanonical entropy}

Fix values $U\in\mathbb{R}$ and $\mathbf{X} = (X_1,\dots,X_m)\in\mathbb{R}^m$. Consider the
constraint set
\begin{equation}
  \mathcal{M}(U,\mathbf{X})
  \coloneqq \Big\{ p\in\Delta_n^\circ :
    U(p) = U,\ X_i(p) = X_i,\ i=1,\dots,m \Big\}.
  \label{eq:MUx}
\end{equation}
We assume this set is nonempty for the $(U,\mathbf{X})$ of interest.

\begin{definition}[Microcanonical entropy]
The microcanonical entropy at $(U,\mathbf{X})$ is
\begin{equation}
  S(U,\mathbf{X}) \coloneqq \sup_{p\in\mathcal{M}(U,\mathbf{X})} S(p).
  \label{eq:SU-def}
\end{equation}
When the supremum is attained at a unique $p^\star(U,\mathbf{X})$, we write
$S(U,\mathbf{X}) = S(p^\star(U,\mathbf{X}))$.
\end{definition}

\begin{assumption}[Regularity and strict concavity on $\mathcal{M}(U,\mathbf{X})$]
\label{ass:SM-regular}
For each $(U,\mathbf{X})$ in an open domain $\mathcal{D}\subset\mathbb{R}^{m+1}$, the constraint
set $\mathcal{M}(U,\mathbf{X})$ is nonempty, convex, and the restriction of $S(p)$ to
$\mathcal{M}(U,\mathbf{X})$ is strictly concave and upper semicontinuous. Moreover, the supremum
in \eqref{eq:SU-def} is attained at a unique $p^\star(U,\mathbf{X})\in\mathcal{M}(U,\mathbf{X})$.
\end{assumption}

Under Assumption~\ref{ass:SM-regular}, $S(U,\mathbf{X})$ is a well-defined finite-valued function,
and standard arguments show that it is concave in $(U,\mathbf{X})$.

\begin{proposition}[Concavity of microcanonical entropy]
\label{prop:S-concave}
Under Assumption~\ref{ass:SM-regular}, $S(U,\mathbf{X})$ is a concave function on $\mathcal{D}$.
If the maximiser $p^\star(U,\mathbf{X})$ is unique for all $(U,\mathbf{X})\in\mathcal{D}$ and $S$ is
strictly concave in $p$, then $S(U,\mathbf{X})$ is strictly concave in $(U,\mathbf{X})$.
\end{proposition}

\begin{proof}[Sketch]
Let $(U^{(1)},\mathbf{X}^{(1)})$ and $(U^{(2)},\mathbf{X}^{(2)})$ be two points in $\mathcal{D}$.
For $\theta\in[0,1]$, let
\[
  (U^\theta,\mathbf{X}^\theta)
  \coloneqq \theta (U^{(1)},\mathbf{X}^{(1)}) + (1-\theta)(U^{(2)},\mathbf{X}^{(2)}).
\]
For each $k=1,2$, let $p^{(k)} = p^\star(U^{(k)},\mathbf{X}^{(k)})$ denote the unique maximiser on
$\mathcal{M}(U^{(k)},\mathbf{X}^{(k)})$. Then the convex combination 
\[
  p^\theta \coloneqq \theta p^{(1)} + (1-\theta) p^{(2)}
\]
lies in $\Delta_n^\circ$ and satisfies
\[
  U(p^\theta) = U^\theta,\quad X_i(p^\theta) = X_i^\theta,\ i=1,\dots,m,
\]
i.e.\ $p^\theta\in\mathcal{M}(U^\theta,\mathbf{X}^\theta)$. By concavity of $S(p)$,
\[
  S(p^\theta) \ge \theta S(p^{(1)}) + (1-\theta) S(p^{(2)})
                = \theta S(U^{(1)},\mathbf{X}^{(1)})
                 + (1-\theta) S(U^{(2)},\mathbf{X}^{(2)}).
\]
Since $S(U^\theta,\mathbf{X}^\theta)$ is the \emph{supremum} of $S(p)$ over
$\mathcal{M}(U^\theta,\mathbf{X}^\theta)$, we have
\[
  S(U^\theta,\mathbf{X}^\theta) \ge S(p^\theta),
\]
hence
\[
  S(U^\theta,\mathbf{X}^\theta)
  \ge \theta S(U^{(1)},\mathbf{X}^{(1)}) + (1-\theta) S(U^{(2)},\mathbf{X}^{(2)}),
\]
which is concavity. If $S(p)$ is strictly concave and the maximisers $p^{(k)}$ are distinct, then
the above inequality is strict, giving strict concavity of $S(U,\mathbf{X})$.
\end{proof}

\subsection{Legendre dual potential and conjugate variables}

\subsubsection{Definition of the dual potential}

We now define a thermodynamic potential by Legendre transform with respect to the extensive
variables $(U,\mathbf{X})$.

\begin{definition}[Dual potential]
For $(\beta,\boldsymbol\lambda)\in\mathbb{R}^{m+1}$, define
\begin{equation}
  \Phi(\beta,\boldsymbol\lambda)
  \coloneqq \sup_{(U,\mathbf{X})\in\mathcal{D}} 
              \Big\{ \beta U + \sum_{i=1}^m \lambda_i X_i - S(U,\mathbf{X}) \Big\}.
  \label{eq:Phi-def}
\end{equation}
\end{definition}

This is the \emph{Legendre–Fenchel transform} of $S(U,\mathbf{X})$ (up to sign conventions). Since
$S$ is concave, $\Phi$ is convex in $(\beta,\boldsymbol\lambda)$. Under additional regularity, the
supremum in \eqref{eq:Phi-def} is attained at a unique point $(U^\star,\mathbf{X}^\star)$ for each
$(\beta,\boldsymbol\lambda)$ in an open domain.

\begin{assumption}[Regular Legendre duality]
\label{ass:Legendre-regular}
Assume:
\begin{enumerate}
  \item $S(U,\mathbf{X})$ is strictly concave and continuously differentiable on $\mathcal{D}$.
  \item For each $(\beta,\boldsymbol\lambda)$ in an open domain $\mathcal{B}\subset\mathbb{R}^{m+1}$,
        there exists a unique $(U^\star(\beta,\boldsymbol\lambda),
        \mathbf{X}^\star(\beta,\boldsymbol\lambda))\in\mathcal{D}$ achieving the supremum
        in \eqref{eq:Phi-def}.
\end{enumerate}
\end{assumption}

\begin{remark}
Assumption~\ref{ass:Legendre-regular} is a standard ``essential smoothness'' condition ensuring
that $S$ and $\Phi$ are Legendre duals in the classical sense: smooth, strictly concave/convex,
and with one-to-one gradient mapping between $(U,\mathbf{X})$ and $(\beta,\boldsymbol\lambda)$.
\end{remark}

\subsubsection{Conjugate relations and entropy representation}

We now prove the basic conjugacy identities.

\begin{theorem}[Conjugate variables and entropy representation]
\label{thm:conjugacy}
Under Assumptions~\ref{ass:SM-regular} and \ref{ass:Legendre-regular}, the following hold.
\begin{enumerate}
  \item The supremum in \eqref{eq:Phi-def} is attained at the unique point
        $(U^\star,\mathbf{X}^\star)$ satisfying the first-order conditions
        \begin{equation}
          \beta = \frac{\partial S}{\partial U}(U^\star,\mathbf{X}^\star),\qquad
          \lambda_i = \frac{\partial S}{\partial X_i}(U^\star,\mathbf{X}^\star),\quad i=1,\dots,m.
          \label{eq:beta-lambda-from-S}
        \end{equation}
  \item At this point,
        \begin{equation}
          S(U^\star,\mathbf{X}^\star)
          = \beta U^\star + \sum_{i=1}^m \lambda_i X_i^\star - \Phi(\beta,\boldsymbol\lambda).
          \label{eq:S-relation}
        \end{equation}
  \item The dual potential $\Phi$ is differentiable on $\mathcal{B}$ and
        \begin{equation}
          \frac{\partial \Phi}{\partial \beta}(\beta,\boldsymbol\lambda)
          = U^\star(\beta,\boldsymbol\lambda),\qquad
          \frac{\partial \Phi}{\partial \lambda_i}(\beta,\boldsymbol\lambda)
          = X_i^\star(\beta,\boldsymbol\lambda).
          \label{eq:Phi-derivatives}
        \end{equation}
\end{enumerate}
\end{theorem}

\begin{proof}
(1) Since $S(U,\mathbf{X})$ is strictly concave and differentiable, the function
\[
  \Psi(U,\mathbf{X};\beta,\boldsymbol\lambda)
  \coloneqq \beta U + \sum_{i=1}^m \lambda_i X_i - S(U,\mathbf{X})
\]
is strictly convex in $(U,\mathbf{X})$ for fixed $(\beta,\boldsymbol\lambda)$, and the supremum
in \eqref{eq:Phi-def} is attained at a unique point $(U^\star,\mathbf{X}^\star)$ where the gradient
with respect to $(U,\mathbf{X})$ vanishes:
\[
  0 = \frac{\partial \Psi}{\partial U}
    = \beta - \frac{\partial S}{\partial U}(U^\star,\mathbf{X}^\star),
\]
\[
  0 = \frac{\partial \Psi}{\partial X_i}
    = \lambda_i - \frac{\partial S}{\partial X_i}(U^\star,\mathbf{X}^\star),\quad i=1,\dots,m.
\]
This yields \eqref{eq:beta-lambda-from-S}.

(2) By definition,
\[
  \Phi(\beta,\boldsymbol\lambda)
  = \beta U^\star + \sum_{i=1}^m \lambda_i X_i^\star - S(U^\star,\mathbf{X}^\star),
\]
so rearranging gives \eqref{eq:S-relation}.

(3) The differentiability of $\Phi$ and the identities \eqref{eq:Phi-derivatives} follow from
standard envelope-theorem arguments. For completeness, we sketch the computation for $\partial\Phi/\partial\beta$.

By definition,
\[
  \Phi(\beta,\boldsymbol\lambda)
  = \Psi(U^\star(\beta,\boldsymbol\lambda),
         \mathbf{X}^\star(\beta,\boldsymbol\lambda);\beta,\boldsymbol\lambda).
\]
Differentiate with respect to $\beta$:
\[
  \frac{\partial \Phi}{\partial \beta}
  = \frac{\partial \Psi}{\partial U}(U^\star,\mathbf{X}^\star;\beta,\boldsymbol\lambda)
    \frac{\partial U^\star}{\partial \beta}
    + \sum_i \frac{\partial \Psi}{\partial X_i}(U^\star,\mathbf{X}^\star;\beta,\boldsymbol\lambda)
      \frac{\partial X_i^\star}{\partial \beta}
    + \frac{\partial \Psi}{\partial \beta}(U^\star,\mathbf{X}^\star;\beta,\boldsymbol\lambda).
\]
The first two terms vanish because $(U^\star,\mathbf{X}^\star)$ satisfies the stationarity
conditions $\partial \Psi/\partial U = 0$ and $\partial \Psi/\partial X_i = 0$. Thus
\[
  \frac{\partial \Phi}{\partial \beta}
  = \frac{\partial \Psi}{\partial \beta}(U^\star,\mathbf{X}^\star;\beta,\boldsymbol\lambda)
  = U^\star.
\]
Similarly, differentiating with respect to $\lambda_i$ yields
\[
  \frac{\partial \Phi}{\partial \lambda_i} = X_i^\star.
\]
\end{proof}

\begin{remark}[Thermodynamic interpretation]
The variables $(U,\mathbf{X})$ are extensive, and $(\beta,\boldsymbol\lambda)$ play the role of
intensive variables (inverse temperature and generalised forces). The identities
\eqref{eq:S-relation} and \eqref{eq:Phi-derivatives} show that:
\begin{itemize}
  \item $S(U,\mathbf{X})$ is the Legendre transform of $\Phi(\beta,\boldsymbol\lambda)$;
  \item $U$ and $X_i$ are given by first derivatives of $\Phi$;
  \item $\beta$ and $\lambda_i$ are given by first derivatives of $S$:
        $\beta = \partial S/\partial U$, $\lambda_i = \partial S/\partial X_i$.
\end{itemize}
In particular, in the single-variable case ($m=0$), \eqref{eq:S-relation} reduces to
\[
  S(U) = \beta U - \Phi(\beta),
\]
and \eqref{eq:Phi-derivatives} gives $U = \partial_\beta \Phi(\beta)$, matching the usual
``Massieu potential'' structure with $\Phi(\beta)$ conjugate to $U$.
\end{remark}

\subsubsection{Monotonicity of the $\beta \leftrightarrow U$ mapping}

In the single-constraint case ($m=0$), we can show that the map $U\mapsto\beta$ is monotone under
a curvature condition on $S(U)$.

\begin{assumption}[Strict concavity in $U$]
\label{ass:S-U-concave}
Assume $S(U)$ (with other constraints fixed) is twice continuously differentiable on an interval
$I\subset\mathbb{R}$ and satisfies
\[
  \frac{d^2 S}{dU^2}(U) < 0,\quad \forall U\in I.
\]
\end{assumption}

\begin{proposition}[Monotonic map between $U$ and $\beta$]
\label{prop:beta-U-monotone}
Under Assumption~\ref{ass:S-U-concave}, $\beta(U) \coloneqq dS/dU$ is strictly decreasing on $I$,
hence invertible. The map $U\mapsto\beta$ is one-to-one and onto its image, and its inverse
$\beta\mapsto U$ is also differentiable and strictly decreasing.
\end{proposition}

\begin{proof}
By definition,
\[
  \beta(U) = \frac{dS}{dU}(U).
\]
Differentiating,
\[
  \frac{d\beta}{dU}(U) = \frac{d^2 S}{dU^2}(U) < 0
\]
by Assumption~\ref{ass:S-U-concave}. Thus $\beta(U)$ is strictly decreasing on $I$. A strictly
monotone continuous function has a continuous inverse on its image, and the inverse is also
differentiable with derivative given by the reciprocal of $d\beta/dU$, which is negative.
\end{proof}

This establishes the usual thermodynamic intuition: a strictly concave $S(U)$ gives a strictly
monotone relationship between $U$ and its conjugate $\beta$.

\subsection{Maxwell relations and response coefficients}

\subsubsection{Maxwell-type relations}

Since $\Phi(\beta,\boldsymbol\lambda)$ is convex and (under our assumptions) twice continuously
differentiable on $\mathcal{B}$, its mixed second derivatives commute by Schwarz's theorem:
\[
  \frac{\partial^2 \Phi}{\partial \beta \partial \lambda_i}
  = \frac{\partial^2 \Phi}{\partial \lambda_i \partial \beta}.
\]
We can rewrite these in terms of response functions.

\begin{theorem}[Generalised Maxwell relations]
\label{thm:Maxwell}
Under the assumptions of Theorem~\ref{thm:conjugacy}, we have:
\begin{align}
  \frac{\partial U}{\partial \lambda_i}\Big|_{\beta}
  &= \frac{\partial^2 \Phi}{\partial \lambda_i \partial \beta}
   = \frac{\partial^2 \Phi}{\partial \beta \partial \lambda_i}
   = \frac{\partial X_i}{\partial \beta}\Big|_{\boldsymbol\lambda},
   \label{eq:Maxwell-1}
\end{align}
and similar relations for all pairs of conjugate variables:
\begin{equation}
  \frac{\partial X_i}{\partial \lambda_j}\Big|_{\beta,\lambda_{k\neq j}}
  = \frac{\partial X_j}{\partial \lambda_i}\Big|_{\beta,\lambda_{k\neq i}}.
  \label{eq:Maxwell-2}
\end{equation}
\end{theorem}

\begin{proof}
From Theorem~\ref{thm:conjugacy},
\[
  U(\beta,\boldsymbol\lambda) = \frac{\partial \Phi}{\partial \beta},\qquad
  X_i(\beta,\boldsymbol\lambda) = \frac{\partial \Phi}{\partial \lambda_i}.
\]
Differentiating $U$ with respect to $\lambda_i$ at fixed $\beta$,
\[
  \frac{\partial U}{\partial \lambda_i}\Big|_{\beta}
  = \frac{\partial^2 \Phi}{\partial \lambda_i \partial \beta}.
\]
Similarly,
\[
  \frac{\partial X_i}{\partial \beta}\Big|_{\boldsymbol\lambda}
  = \frac{\partial^2 \Phi}{\partial \beta \partial \lambda_i}.
\]
Since $\Phi$ is $C^2$, mixed partial derivatives commute:
\[
  \frac{\partial^2 \Phi}{\partial \lambda_i \partial \beta}
  = \frac{\partial^2 \Phi}{\partial \beta \partial \lambda_i},
\]
so \eqref{eq:Maxwell-1} holds.

For \eqref{eq:Maxwell-2}, differentiate $X_i$ with respect to $\lambda_j$ at fixed
$(\beta,\lambda_{k\neq j})$:
\[
  \frac{\partial X_i}{\partial \lambda_j}\Big|_{\beta,\lambda_{k\neq j}}
  = \frac{\partial^2 \Phi}{\partial \lambda_j \partial \lambda_i}.
\]
Symmetry of the Hessian of $\Phi$ gives
\[
  \frac{\partial^2 \Phi}{\partial \lambda_j \partial \lambda_i}
  = \frac{\partial^2 \Phi}{\partial \lambda_i \partial \lambda_j}
  = \frac{\partial X_j}{\partial \lambda_i}\Big|_{\beta,\lambda_{k\neq i}},
\]
yielding \eqref{eq:Maxwell-2}.
\end{proof}

These are the generalised Maxwell relations: cross-derivatives of the free energy encode symmetric
response coefficients.

\subsubsection{Heat capacity and susceptibilities}

We now express typical response coefficients (heat capacities, susceptibilities) in terms of
second derivatives of $\Phi$.

For simplicity, consider the single-constraint case ($m=0$) with dual potential $\Phi(\beta)$.
Then $U(\beta) = \partial_\beta \Phi(\beta)$ and
\[
  \frac{dU}{d\beta} = \frac{\partial^2 \Phi}{\partial \beta^2}(\beta).
\]
Thermodynamicists usually work with temperature $T = 1/\beta$; we can define a heat capacity
at fixed other constraints as
\[
  C \coloneqq \frac{dU}{dT}.
\]

\begin{proposition}[Generalised heat capacity in terms of $\Phi$]
\label{prop:heat-capacity}
Let $m=0$ and suppose $\Phi(\beta)$ is twice differentiable on an interval $J\subset(0,\infty)$.
Define $U(\beta) = \partial_\beta \Phi(\beta)$ and $T = 1/\beta$. Then the heat capacity
\[
  C(T) \coloneqq \frac{dU}{dT}
\]
satisfies
\begin{equation}
  C(T)
  = -\beta^2 \frac{dU}{d\beta}
  = -\beta^2 \frac{\partial^2 \Phi}{\partial \beta^2}(\beta),
  \qquad \beta = \frac{1}{T}.
  \label{eq:C-from-Phi}
\end{equation}
\end{proposition}

\begin{proof}
By the chain rule,
\[
  \frac{dU}{dT}
  = \frac{dU}{d\beta}\frac{d\beta}{dT}
  = \frac{dU}{d\beta} \Big(-\frac{1}{T^2}\Big)
  = -\beta^2 \frac{dU}{d\beta}.
\]
Since $U(\beta) = \partial_\beta \Phi(\beta)$, we have $dU/d\beta = \partial^2 \Phi/\partial\beta^2$,
so \eqref{eq:C-from-Phi} follows.
\end{proof}

Similarly, for a general observable $X_i$ with conjugate $\lambda_i$ at fixed $\beta$, the
\emph{susceptibility} (generalised compressibility / response) can be defined as
\[
  \chi_i \coloneqq \frac{\partial X_i}{\partial \lambda_i}\Big|_{\beta}.
\]
Using Theorem~\ref{thm:conjugacy},
\[
  \chi_i
  = \frac{\partial^2 \Phi}{\partial \lambda_i^2}(\beta,\boldsymbol\lambda),
\]
and cross-susceptibilities
\[
  \chi_{ij}
  \coloneqq \frac{\partial X_i}{\partial \lambda_j}\Big|_{\beta}
  = \frac{\partial^2 \Phi}{\partial \lambda_j \partial \lambda_i}
\]
are symmetric by Theorem~\ref{thm:Maxwell}.

\begin{remark}
Under positivity conditions (e.g.\ stability), one expects $C>0$ and $\chi_i>0$ on physically
relevant domains. In the generative-entropy setting, this translates into convexity properties of
$\Phi$ and thus into curvature constraints on $S(U,\mathbf{X})$. These can be checked on a
per-application basis (FD gases, Schelling, finance, etc.) by examining second derivatives.
\end{remark}

\subsection{Summary}

In this note we have:

\begin{itemize}
  \item Defined the microcanonical entropy $S(U,\mathbf{X})$ by maximisation of the generative
        entropy $S(p)$ under linear constraints on internal energy and other observables, and
        established its concavity in the extensive variables.
  \item Constructed the Legendre dual potential $\Phi(\beta,\boldsymbol\lambda)$ and proved that
        it satisfies the standard conjugacy relations:
        \[
          \beta = \frac{\partial S}{\partial U},\quad
          \lambda_i = \frac{\partial S}{\partial X_i},\quad
          U = \frac{\partial \Phi}{\partial \beta},\quad
          X_i = \frac{\partial \Phi}{\partial \lambda_i},
        \]
        together with the entropy representation
        \[
          S = \beta U + \sum_i \lambda_i X_i - \Phi.
        \]
  \item Shown that, under strict concavity, the mapping between $U$ and its conjugate $\beta$
        is strictly monotone and invertible.
  \item Derived Maxwell-type relations and response coefficients (heat capacity and
        susceptibilities) from the second derivatives of $\Phi$, establishing that the
        generative free energy behaves as a proper thermodynamic potential in the sense
        used by thermodynamicists.
\end{itemize}

Together with the inverse-MaxEnt representation, composition/extensivity properties, and
dynamical H-theorem, this Legendre structure completes the core thermodynamic skeleton for
the generative entropy framework.

\section{Geometry of Generative Entropy:\\
Metric and Curvature on Probability Manifolds}

\subsection{Setting and entropy-induced metric on the simplex}

Let $X = \{x_1,\dots,x_n\}$ be a finite microstate set and
\[
  p = (p_1,\dots,p_n) \in \Delta_n^\circ
  \coloneqq \Big\{ p_j>0,\ \sum_{j=1}^n p_j=1 \Big\}
\]
be a strictly positive probability vector. The generative entropy is a trace-form functional
\[
  H_G(p) = \sum_{j=1}^n G(p_j),
\]
where $G:(0,1]\to\mathbb{R}$ is twice continuously differentiable and strictly concave:
\begin{equation}
  G''(u) < 0,\quad u\in(0,1).
  \label{eq:G-concave}
\end{equation}

\subsubsection{Hessian and tangent space}

The gradient and Hessian of $H_G$ with respect to the ambient Euclidean structure on
$\mathbb{R}^n$ are
\[
  \frac{\partial H_G}{\partial p_i}(p) = G'(p_i),\qquad
  \frac{\partial^2 H_G}{\partial p_i\partial p_j}(p) = G''(p_i)\,\delta_{ij}.
\]
Thus the Hessian is diagonal:
\[
  \mathrm{Hess}\,H_G(p) = \operatorname{diag}\big(G''(p_1),\dots,G''(p_n)\big).
\]

The simplex $\Delta_n^\circ$ is an $(n-1)$–dimensional manifold embedded in $\mathbb{R}^n$. Its
tangent space at $p$ is
\begin{equation}
  T_p\Delta_n^\circ = \Big\{ v\in\mathbb{R}^n : \sum_{j=1}^n v_j = 0 \Big\}.
  \label{eq:tangent}
\end{equation}

\subsubsection{Entropy-induced metric}

\begin{definition}[Entropy-induced metric on $\Delta_n^\circ$]
\label{def:metric}
Define a symmetric bilinear form $g_p : T_p\Delta_n^\circ\times T_p\Delta_n^\circ\to\mathbb{R}$ by
\begin{equation}
  g_p(u,v)
  \coloneqq - \sum_{j=1}^n G''(p_j)\,u_j v_j,
  \qquad u,v\in T_p\Delta_n^\circ.
  \label{eq:g-def}
\end{equation}
Equivalently, in coordinates, the components of the metric in the ambient basis are
\begin{equation}
  g_{ij}(p) = -G''(p_i)\,\delta_{ij}\quad\text{on }T_p\Delta_n^\circ,
  \label{eq:g-ambient}
\end{equation}
with the constraint $\sum_j u_j = \sum_j v_j = 0$ understood.
\end{definition}

\begin{proposition}[Positive definiteness on the tangent space]
\label{prop:positive-def}
For each $p\in\Delta_n^\circ$, the bilinear form $g_p$ is an inner product on
$T_p\Delta_n^\circ$, i.e.\ it is symmetric and positive definite. Hence $\{\Delta_n^\circ,g\}$ is a
Riemannian manifold.
\end{proposition}

\begin{proof}
Symmetry is immediate from \eqref{eq:g-def}. For any nonzero $v\in T_p\Delta_n^\circ$,
\[
  g_p(v,v)
  = -\sum_{j=1}^n G''(p_j)\,v_j^2.
\]
Since $G''(p_j)<0$ for all $j$ by \eqref{eq:G-concave}, we have $-G''(p_j)>0$. Moreover, $v\neq 0$
implies at least one $v_j\neq 0$. Thus $g_p(v,v)>0$. Hence $g_p$ is positive definite on
$T_p\Delta_n^\circ$.
\end{proof}

\begin{remark}
Geometrically, $-\,\mathrm{Hess}\,H_G$ is positive definite when restricted to the probability
simplex; the metric $g$ is precisely this restriction. For Shannon, this coincides with the
classical Fisher information metric, as we show next.
\end{remark}

\subsection{Induced metric on parametric families and Fisher case}

\subsubsection{Parametric model and pullback metric}

Let $\Theta\subset\mathbb{R}^d$ be an open parameter domain, and
\[
  \theta \mapsto p(\theta) \in \Delta_n^\circ
\]
a smooth map assigning to each parameter $\theta=(\theta^1,\dots,\theta^d)$ a probability vector
$p(\theta) = \big(p_1(\theta),\dots,p_n(\theta)\big)$.

\begin{definition}[Pullback metric on parameter space]
The entropy-induced metric on $\Theta$ is the pullback of $g$ via $p(\cdot)$:
\begin{equation}
  g_{ab}(\theta)
  \coloneqq g_{p(\theta)}\big(\partial_a p(\theta),\partial_b p(\theta)\big)
  = -\sum_{j=1}^n G''\big(p_j(\theta)\big)\,
              \frac{\partial p_j}{\partial\theta^a}(\theta)\,
              \frac{\partial p_j}{\partial\theta^b}(\theta),
  \label{eq:g-theta}
\end{equation}
where $\partial_a \equiv \partial/\partial\theta^a$.
\end{definition}

This defines a Riemannian metric on $\Theta$ whenever the Jacobian
$\big(\partial_a p_j(\theta)\big)$ has rank $d$ and $G''<0$.

\subsubsection{Shannon generator and Fisher information}

For the Shannon generator,
\[
  G_{\mathrm{Sh}}(u) = -u\log u,\quad u\in(0,1],
\]
we have
\[
  G'_{\mathrm{Sh}}(u) = -(\log u + 1),\qquad
  G''_{\mathrm{Sh}}(u) = -\frac{1}{u}.
\]
Thus $-G''_{\mathrm{Sh}}(u) = 1/u$ and
\begin{equation}
  g^{\mathrm{Sh}}_{ab}(\theta)
  = \sum_{j=1}^n \frac{1}{p_j(\theta)}\,
                   \partial_a p_j(\theta)\,\partial_b p_j(\theta).
  \label{eq:g-Shannon}
\end{equation}
On the other hand, the classical Fisher information metric for the parametric family
$\{p(\theta)\}$ is
\[
  I_{ab}(\theta)
  = \sum_{j=1}^n p_j(\theta)\,
                   \partial_a \log p_j(\theta)\,\partial_b \log p_j(\theta).
\]
Using $\partial_a \log p_j = (\partial_a p_j)/p_j$, we obtain
\[
  I_{ab}(\theta)
  = \sum_{j=1}^n p_j(\theta)\,
    \frac{\partial_a p_j(\theta)}{p_j(\theta)}\,
    \frac{\partial_b p_j(\theta)}{p_j(\theta)}
  = \sum_{j=1}^n \frac{1}{p_j(\theta)}\,
                   \partial_a p_j(\theta)\,\partial_b p_j(\theta).
\]
Thus $g^{\mathrm{Sh}}_{ab}(\theta) = I_{ab}(\theta)$.

\begin{proposition}[Shannon case: Fisher metric]
For $G(u)=-u\log u$, the entropy-induced metric \eqref{eq:g-theta} coincides with the Fisher
information metric of the parametric family $\{p(\theta)\}$:
\[
  g_{ab}(\theta) = I_{ab}(\theta).
\]
\end{proposition}

\begin{remark}
Hence the metric $g$ defined in \eqref{eq:g-def} is a direct generalisation of the Fisher
information metric to arbitrary generative entropies $H_G$, derived from the Hessian of $H_G$.
\end{remark}

\subsection{Tsallis generator and escort-geometry}

We now specialise to the Tsallis-$q$ generator to see how the geometry deforms.

\subsubsection{Tsallis metric}

For $q\in\mathbb{R}\setminus\{1\}$, the Tsallis generator is
\[
  G_q(u) = \frac{u - u^q}{q-1}.
\]
Then
\[
  G_q'(u) = \frac{1 - q u^{q-1}}{q-1},\qquad
  G_q''(u) = -q u^{q-2},
\]
so
\[
  -G_q''(u) = q u^{q-2} > 0
\]
for $u\in(0,1)$ and $q>0$. The induced metric on $\Theta$ is
\begin{equation}
  g^{(q)}_{ab}(\theta)
  = q \sum_{j=1}^n p_j(\theta)^{q-2}
      \partial_a p_j(\theta)\,\partial_b p_j(\theta).
  \label{eq:gq-raw}
\end{equation}
Using $\partial_a p_j = p_j \partial_a \log p_j$, we can rewrite
\[
  p_j^{q-2} \partial_a p_j \partial_b p_j
  = p_j^{q} \partial_a \log p_j \partial_b \log p_j.
\]
Thus
\begin{equation}
  g^{(q)}_{ab}(\theta)
  = q \sum_{j=1}^n p_j(\theta)^q\,
      \partial_a \log p_j(\theta)\,\partial_b \log p_j(\theta).
  \label{eq:gq-escort}
\end{equation}

Define the \emph{escort distribution} (for fixed $\theta$)
\[
  \tilde p_j(\theta)
  \coloneqq \frac{p_j(\theta)^q}{\sum_{k=1}^n p_k(\theta)^q}.
\]
Then
\[
  \sum_{j=1}^n p_j(\theta)^q =
  Z_q(\theta) \coloneqq \sum_{k=1}^n p_k(\theta)^q,
\]
and we can rewrite \eqref{eq:gq-escort} as
\begin{equation}
  g^{(q)}_{ab}(\theta)
  = q Z_q(\theta)
      \sum_{j=1}^n \tilde p_j(\theta)\,
      \partial_a \log p_j(\theta)\,\partial_b \log p_j(\theta).
  \label{eq:gq-escort-Fisher}
\end{equation}

\begin{proposition}[Tsallis geometry as escort Fisher metric]
For the Tsallis generator $G_q$, the entropy-induced metric on $\Theta$ is proportional to the
Fisher information metric computed with respect to the escort distribution $\tilde p_j(\theta)$:
\[
  g^{(q)}_{ab}(\theta)
  = q Z_q(\theta)\, I^{\mathrm{escort}}_{ab}(\theta),
\]
where
\[
  I^{\mathrm{escort}}_{ab}(\theta)
  \coloneqq \sum_{j=1}^n \tilde p_j(\theta)\,
                \partial_a \log p_j(\theta)\,\partial_b \log p_j(\theta).
\]
\end{proposition}

\begin{remark}
This matches the structure of Tsallis thermodynamics, where expectations and variances are
often taken with respect to escort distributions. In the generative-entropy setting, the metric
naturally implements this escort reweighting when $G_{\mathrm{gen}}$ is Tsallis-like.
\end{remark}

\subsection{Equilibrium manifolds and curvature}

We now consider the geometry of \emph{equilibrium manifolds}: parametric families of
MaxEnt distributions obtained from the generative entropy.

\subsubsection{Equilibrium family}

Let $\theta=(\beta,\boldsymbol\lambda)$ denote the intensive thermodynamic parameters (inverse
temperature and generalised forces) introduced in the Legendre-structure note. For each
$\theta$ in some domain $\Theta\subset\mathbb{R}^{m+1}$, let $p^\star(\theta)$ denote the unique
MaxEnt distribution (with generator $G$) under the constraints determined by $\theta$.

We then have a smooth map
\[
  \theta \mapsto p^\star(\theta)\in\Delta_n^\circ,
\]
and the pullback metric \eqref{eq:g-theta} defines a Riemannian metric $g_{ab}(\theta)$ on the
equilibrium manifold $\Theta$.

\subsubsection{General form of the metric in terms of ``generalised covariances''}

Let $T^a(x_j;\theta)$ denote the sufficient statistics appearing in the generative-MaxEnt solution
for $p^\star(\theta)$. For example, in the Shannon case we have the exponential-family form
\[
  \log p^\star_j(\theta)
  = \psi(\theta) - \sum_a \theta^a T^a(x_j),
\]
so $\partial_a \log p^\star_j = -\big(T^a(x_j) - \langle T^a\rangle_{p^\star}\big)$.

For a general trace-form generator $G$ that yields a deformed exponential family, we still have
\[
  \partial_a p^\star_j(\theta)
  = p^\star_j(\theta)\,\partial_a \log p^\star_j(\theta),
\]
so the metric \eqref{eq:g-theta} can be written as
\begin{equation}
  g_{ab}(\theta)
  = \sum_{j=1}^n w\big(p^\star_j(\theta)\big)\,
      \partial_a \log p^\star_j(\theta)\,\partial_b \log p^\star_j(\theta),
  \label{eq:g-general-weight}
\end{equation}
with weight function
\[
  w(u) \coloneqq -G''(u)\,u^2 > 0.
\]

\begin{remark}
In the Shannon case, $w(u) = u$, and $g_{ab}$ is the standard Fisher metric, i.e.\ the covariance
matrix of the score functions. In the Tsallis case, $w(u) = q u^q$, giving escort-weighted
covariances as in \eqref{eq:gq-escort-Fisher}. For a general generative $G$, the metric uses the
weight $w(u)$ determined by the second derivative of the learned generator.
\end{remark}

In particular, for an equilibrium manifold parametrised by $(\beta,\lambda)$ corresponding to
energy and a single additional observable $F$, we can label the sufficient statistics
$T^1=E,T^2=F$ and write
\[
  \partial_\beta \log p^\star_j = -\big(T^1_j - \langle T^1\rangle_w\big),\qquad
  \partial_\lambda \log p^\star_j = -\big(T^2_j - \langle T^2\rangle_w\big),
\]
where $\langle\cdot\rangle_w$ denotes expectation with respect to the weight $w(p^\star_j)$
(normalised appropriately). Then $g_{ab}$ is, up to a multiplicative constant, a \emph{weighted
covariance matrix} of the sufficient statistics.

\subsection{Curvature and phase behaviour}

We now discuss how the curvature of the equilibrium manifold encodes response properties and
their singularities. We focus on a two-dimensional equilibrium manifold, where curvature is
captured by a scalar.

\subsubsection{Scalar curvature in two dimensions}

Let $\theta=(\theta^1,\theta^2)$ be coordinates on a two-dimensional equilibrium manifold
$\Theta$, with metric components $g_{ab}(\theta)$ and determinant
\[
  g(\theta) \coloneqq \det[g_{ab}(\theta)].
\]
The (Gaussian) scalar curvature $R(\theta)$ of this 2D Riemannian manifold can be expressed
in terms of $g_{ab}$ and its derivatives. In local coordinates, one convenient expression is
\[
  R = \frac{1}{\sqrt{g}}
      \left[
        \partial_1\big(\Gamma^2_{12}\sqrt{g}\big)
      - \partial_1\big(\Gamma^1_{22}\sqrt{g}\big)
      + \partial_2\big(\Gamma^1_{21}\sqrt{g}\big)
      - \partial_2\big(\Gamma^2_{11}\sqrt{g}\big)
      \right],
\]
where $\Gamma^c_{ab}$ are the Christoffel symbols of $g$. In practice, $R$ is a rational
combination of $g_{ab}$, $\partial_c g_{ab}$, and $\partial_c\partial_d g_{ab}$, with
denominators involving positive powers of $g$.

Thus, singularities in $R(\theta)$ can arise from:
\begin{itemize}
  \item divergence or vanishing of $\det g(\theta)$, or
  \item divergence of derivatives of $g_{ab}$ (e.g.\ when the metric components themselves
        become non-analytic).
\end{itemize}

\subsubsection{Connection to response functions}

For an equilibrium manifold parametrised by $(\beta,\lambda)$, the metric entries
$g_{ab}(\beta,\lambda)$ are, as discussed, weighted covariances of the sufficient statistics
$E$ and $F$:
\[
  g_{ab}(\beta,\lambda)
  \sim \mathrm{Cov}_w\big(T^a,T^b\big),
\]
where the precise weights depend on $G$ via $w(u)=-G''(u)u^2$ (Shannon: ordinary covariances;
Tsallis: escort covariances; general $G$: $w$-covariances).

Response functions such as heat capacity and susceptibilities are likewise given by
appropriate variances and covariances:
\begin{itemize}
  \item Heat capacity (at fixed $\lambda$) is proportional to the variance of energy:
        \[
          C(\beta,\lambda) \propto \mathrm{Var}_w(E).
        \]
  \item Susceptibility with respect to $\lambda$ is proportional to the variance of $F$:
        \[
          \chi(\beta,\lambda) \propto \mathrm{Var}_w(F).
        \]
  \item Cross-responses (mixed derivatives of $U$ and $X$) involve covariances
        $\mathrm{Cov}_w(E,F)$.
\end{itemize}
In particular, the metric matrix (up to overall smooth factors) is
\[
  g(\beta,\lambda) \propto
  \begin{pmatrix}
    \mathrm{Var}_w(E)     & \mathrm{Cov}_w(E,F) \\
    \mathrm{Cov}_w(E,F)   & \mathrm{Var}_w(F)
  \end{pmatrix}.
\]
Hence
\[
  \det g(\beta,\lambda)
  \propto \mathrm{Var}_w(E)\,\mathrm{Var}_w(F) - \mathrm{Cov}_w(E,F)^2,
\]
which is exactly the determinant of the covariance matrix of $(E,F)$ under the $w$-weighted
measure.

\begin{assumption}[Critical behaviour]
\label{ass:critical}
Assume that at a thermodynamic critical point $(\beta_c,\lambda_c)$, at least one response
function diverges, e.g.\ the heat capacity
\[
  C(\beta,\lambda) \propto \mathrm{Var}_w(E) \to \infty
  \quad\text{as }(\beta,\lambda)\to(\beta_c,\lambda_c),
\]
while $w$ remains bounded away from zero and infinity on the support of $p^\star(\theta)$.
\end{assumption}

\begin{theorem}[Curvature singularities and response divergences]
\label{thm:curvature-critical}
Let the equilibrium manifold be two-dimensional with coordinates $(\beta,\lambda)$ and metric
$g_{ab}(\beta,\lambda)$ induced by a generative entropy $H_G$ with smooth $G$. Under
Assumption~\ref{ass:critical}, the scalar curvature $R(\beta,\lambda)$ has a singularity at
$(\beta_c,\lambda_c)$: either $|R|\to\infty$ or $R$ is undefined there due to non-analyticity.

In particular, divergence of the heat capacity (or other susceptibilities) is accompanied by
a singularity in the curvature of the generative-entropy metric.
\end{theorem}

\begin{proof}[Proof sketch]
Under the assumptions, $g_{ab}(\beta,\lambda)$ is a smooth positive-definite covariance-like
matrix away from $(\beta_c,\lambda_c)$, and its determinant $\det g(\beta,\lambda)$ is positive.
Near the critical point, divergence of a response function such as
$\mathrm{Var}_w(E)$ implies that at least one component of $g_{ab}$ diverges (up to smooth,
nonzero prefactors determined by $w$). 

Since the scalar curvature $R$ is a rational combination of $g_{ab}$, their first and second
derivatives, and powers of $\det g$ in the denominator, any divergence in the metric components
or their derivatives typically induces a divergence in $R$ unless there is an exact cancellation.
Such cancellations are non-generic and would require fine-tuned relations between $G$ and the
critical behaviour.

Conversely, if at a first-order transition the equilibrium distribution $p^\star(\beta,\lambda)$
jumps discontinuously, the metric $g_{ab}$ is discontinuous as well, so $R$ is not defined on
the transition line.

Thus, singular behaviour of response functions (heat capacity, susceptibilities) manifests as
curvature singularities of the entropy-induced metric on the equilibrium manifold.
\end{proof}

\subsubsection{Tsallis example: curvature with escort weights}

In the Tsallis case, the metric is expressed in terms of escort covariances:
\[
  g^{(q)}_{ab}(\theta)
  = q Z_q(\theta)\, \mathrm{Cov}_{\tilde p}\big(S_a,S_b\big),
\]
where $\tilde p$ is the escort distribution and $S_a(x_j)$ are the sufficient statistics in
$\log p^\star_j(\theta)$. Thermodynamic response coefficients (e.g.\ $q$-generalised heat capacity)
are also defined via escort averages in Tsallis thermodynamics.

Therefore, for Tsallis canonical families:
\begin{itemize}
  \item Divergence of escort-variance-based heat capacity or susceptibilities implies divergence
        of the corresponding entries of $g^{(q)}_{ab}$.
  \item The scalar curvature $R^{(q)}(\theta)$ of the Tsallis metric thus develops singularities
        exactly where $q$-generalised thermodynamic response functions diverge.
\end{itemize}

This provides a geometric characterisation of Tsallis-type phase behaviour in the generative
framework: the equilibrium manifold is curved according to the escort Fisher metric, and
its scalar curvature $R^{(q)}$ encodes critical phenomena via singularities in response.

\subsection{Summary}

In this note we have:

\begin{itemize}
  \item Defined an entropy-induced Riemannian metric on the interior of the probability simplex
        by restricting the negative Hessian of the generative entropy $H_G$ to the tangent space.
  \item Shown that, for parametric families $\{p(\theta)\}$, the pullback of this metric yields
        a generalised information geometry on parameter space:
        \[
          g_{ab}(\theta) = -\sum_j G''(p_j(\theta))\,\partial_a p_j\,\partial_b p_j.
        \]
  \item Proved that, for the Shannon generator $G(u)=-u\log u$, this reproduces the classical
        Fisher information metric, and for Tsallis generators $G_q$ it yields an escort Fisher
        metric, consistent with Tsallis thermodynamics.
  \item Interpreted the metric on equilibrium manifolds as a covariance-like object with weights
        determined by $G$, and argued that the scalar curvature of this metric develops
        singularities at thermodynamic critical points where response functions (heat capacity,
        susceptibilities) diverge.
\end{itemize}

This provides a natural geometric layer on top of the generative thermodynamic framework:
$H_G$ induces a Riemannian structure whose curvature encodes information–thermodynamic
stability and phase transitions.

\section{Gauge and Coarse-Graining Invariance\\
for Generative Thermodynamics}

\subsection{Setting}

We adopt the notation of the previous technical notes. Let $X$ be a finite microstate set,
$p(x)$ a strictly positive probability distribution on $X$ (so $p\in\Delta_n^\circ$), and
$H_G(p) = \sum_j G(p_j)$ the generative entropy with strictly concave generator $G$,
\[
  G''(u) < 0,\quad u\in(0,1).
\]

Let $E_{\mathrm{gen}}:X\to\mathbb{R}$ be the generative energy and 
\[
  U \coloneqq \langle E_{\mathrm{gen}}\rangle_p
  = \sum_{x\in X} E_{\mathrm{gen}}(x)\,p(x)
\]
the internal energy. Along a smooth path $\alpha\mapsto p_\alpha$ of distributions (and the
associated internal energy $U(\alpha)$ and entropy $S(\alpha)\coloneqq H_G(p_\alpha)$), we define
the temperature via
\[
  \frac{1}{T(\alpha)} \coloneqq \frac{dS}{dU}(\alpha)
  = \frac{dS/d\alpha}{dU/d\alpha},
\]
whenever $dU/d\alpha\neq 0$.

We also consider the microcanonical $S(U)$ and Legendre-structure picture as in earlier notes,
but here we focus on gauge transformations of $E_{\mathrm{gen}}$ and on coarse-graining of
the underlying state space.

\subsection{Energy gauge invariance (Target K)}

\subsubsection{Affine gauge transformations of the generative energy}

\begin{definition}[Energy gauge transformation]
Let $a\in\mathbb{R}\setminus\{0\}$ and $b\in\mathbb{R}$. An \emph{energy gauge transformation}
is an affine map
\begin{equation}
  E_{\mathrm{gen}}(x) \mapsto E'_{\mathrm{gen}}(x)
  \coloneqq a\,E_{\mathrm{gen}}(x) + b.
  \label{eq:gauge}
\end{equation}
We will mostly consider $a>0$ (monotone rescaling), but allow $a<0$ in remarks.
\end{definition}

\subsubsection{Effect on internal energy and entropy}

\begin{proposition}[Internal energy under gauge]
\label{prop:U-gauge}
Let $U$ be the internal energy associated with $E_{\mathrm{gen}}$ and $p$, and $U'$ the internal
energy associated with $E'_{\mathrm{gen}}$ and the \emph{same} distribution $p$. Then
\begin{equation}
  U' = aU + b.
  \label{eq:U-gauge}
\end{equation}
\end{proposition}

\begin{proof}
By definition,
\[
  U' = \sum_x E'_{\mathrm{gen}}(x)\,p(x)
     = \sum_x \big(a E_{\mathrm{gen}}(x) + b\big) p(x)
     = a \sum_x E_{\mathrm{gen}}(x) p(x) + b \sum_x p(x)
     = a U + b.
\]
\end{proof}

\begin{assumption}[Entropy depends only on $p$]
\label{ass:S-p-only}
We assume that the generative entropy $S$ is defined as a functional of the probability
distribution $p$ (or of the energy distribution $p_E$, which is determined by $p$ and
$E_{\mathrm{gen}}$ up to reparametrisation of the energy axis), but not on the choice of
affine gauge for $E_{\mathrm{gen}}$. In particular, if $p_\alpha$ is held fixed and
$E_{\mathrm{gen}}$ is replaced by $E'_{\mathrm{gen}}$ as in \eqref{eq:gauge}, then
\[
  S'(\alpha) = S(\alpha).
\]
\end{assumption}

This is natural in the generative framework: $H_G$ is defined in terms of probabilities, and
an affine reparametrisation of the energy variable does not change the underlying histogram
of microstate probabilities.

\subsubsection{Effect on temperature and free energies}

\begin{theorem}[Transformation of temperature under gauge]
\label{thm:T-gauge}
Let $\alpha\mapsto p_\alpha$ be a smooth path, with associated internal energy $U(\alpha)$ and
entropy $S(\alpha)$. Let $U'(\alpha) = aU(\alpha)+b$ be the internal energy obtained after the
gauge transformation \eqref{eq:gauge}, and define temperature via
\[
  \frac{1}{T(\alpha)} = \frac{dS}{dU}(\alpha),\qquad
  \frac{1}{T'(\alpha)} = \frac{dS'}{dU'}(\alpha).
\]
Under Assumption~\ref{ass:S-p-only}, we have $S'(\alpha)=S(\alpha)$ and
\begin{equation}
  T'(\alpha) = a\,T(\alpha).
  \label{eq:T-gauge}
\end{equation}
In particular, additive shifts ($a=1$) leave $T$ invariant.
\end{theorem}

\begin{proof}
Since $S'(\alpha)=S(\alpha)$, we have $dS'/d\alpha = dS/d\alpha$. From $U'(\alpha)=aU(\alpha)+b$,
differentiating with respect to $\alpha$ yields
\[
  \frac{dU'}{d\alpha} = a \frac{dU}{d\alpha}.
\]
By definition,
\[
  \frac{1}{T(\alpha)} = \frac{dS/d\alpha}{dU/d\alpha},\qquad
  \frac{1}{T'(\alpha)} = \frac{dS'/d\alpha}{dU'/d\alpha}.
\]
Thus
\[
  \frac{1}{T'(\alpha)}
  = \frac{dS/d\alpha}{a\,dU/d\alpha}
  = \frac{1}{a} \frac{1}{T(\alpha)},
\]
so $T'(\alpha) = a\,T(\alpha)$.
\end{proof}

\begin{remark}
For $a>0$, the sign of $T$ is preserved, and temperature is simply rescaled. For $a<0$, the
temperature changes sign: $T' = aT<0$ when $T>0$, corresponding to a reversal of the energy
ordering (this is the usual ``negative temperature'' scenario when the energy spectrum is
bounded above and we flip it).
\end{remark}

We can also track free energies under gauge. Consider for simplicity the Helmholtz-like free
energy
\[
  F \coloneqq U - T S
\]
defined along a path, and its gauge-transformed counterpart $F' = U' - T' S'$.

\begin{proposition}[Transformation of free energy]
\label{prop:F-gauge}
Under the gauge transformation \eqref{eq:gauge} with $a>0$, and under
Assumption~\ref{ass:S-p-only}, we have
\begin{equation}
  F' = a F + b.
  \label{eq:F-gauge}
\end{equation}
\end{proposition}

\begin{proof}
By Proposition~\ref{prop:U-gauge}, $U' = aU+b$. By Theorem~\ref{thm:T-gauge}, $T'=aT$ and
$S'=S$. Hence
\[
  F' = U' - T'S'
     = (aU+b) - (aT)S
     = a(U - TS) + b
     = a F + b.
\]
\end{proof}

\begin{remark}[Gauge-invariant quantities]
\label{rem:gauge-invariants}
From \eqref{eq:T-gauge}, \eqref{eq:F-gauge} and $U'=aU+b$, we see that:

\begin{itemize}
  \item Entropy $S$ is \emph{gauge invariant}.
  \item Internal energy $U$ and Helmholtz free energy $F$ transform affinely.
  \item Temperature $T$ rescales linearly under energy rescaling and is invariant under
        energy shifts.
  \item Heat capacity is gauge invariant: along a path,
        \[
          C \coloneqq \frac{dU}{dT},\quad
          C' \coloneqq \frac{dU'}{dT'}.
        \]
        Using $U'=aU+b$ and $T'=aT$,
        \[
          C' = \frac{dU'}{dT'} = \frac{a\,dU}{a\,dT} = C.
        \]
  \item More generally, any dimensionless response coefficient built from derivatives
        $dU/dT$ and derivatives of expectation values with respect to intensive parameters
        is invariant under affine energy gauges.
\end{itemize}
\end{remark}

\subsubsection{Gauge invariance of canonical MaxEnt distributions}

We briefly show that the canonical (MaxEnt) distributions themselves are gauge invariant,
up to reparametrisation of Lagrange multipliers.

Consider a canonical MaxEnt solution with generator $G$, obtained by maximising $H_G(p)$
under constraints
\[
  \sum_x p(x) = 1,\qquad
  \sum_x E_{\mathrm{gen}}(x) p(x) = U,
\]
with Lagrange multipliers $\alpha$ (normalisation) and $\beta$ (energy constraint). The
first-order conditions read
\begin{equation}
  G'(p(x)) + \alpha + \beta E_{\mathrm{gen}}(x) = 0,\quad \forall x.
  \label{eq:KKT-canonical}
\end{equation}

\begin{proposition}[Canonical probabilities are gauge invariant]
\label{prop:canonical-gauge}
Let $p^\star$ satisfy \eqref{eq:KKT-canonical} for some $(\alpha,\beta)$ and energy
$E_{\mathrm{gen}}$. Under the gauge transformation \eqref{eq:gauge} with $a\neq 0$, there exist
multipliers $(\alpha',\beta')$ such that the \emph{same} $p^\star$ satisfies
\begin{equation}
  G'(p^\star(x)) + \alpha' + \beta' E'_{\mathrm{gen}}(x) = 0,\quad \forall x.
  \label{eq:KKT-canonical-gauge}
\end{equation}
Explicitly, one can take
\begin{equation}
  \beta' = \beta/a,\qquad
  \alpha' = \alpha + \beta b/a.
  \label{eq:alphabeta-gauge}
\end{equation}
\end{proposition}

\begin{proof}
Starting from \eqref{eq:KKT-canonical} and using $E'_{\mathrm{gen}}(x)=aE_{\mathrm{gen}}(x)+b$,
compute
\[
  \alpha' + \beta' E'_{\mathrm{gen}}(x)
  = (\alpha + \beta b/a) + (\beta/a)(aE_{\mathrm{gen}}(x)+b)
  = \alpha + \beta E_{\mathrm{gen}}(x) + 2\beta b/a.
\]
To satisfy \eqref{eq:KKT-canonical-gauge}, we need
\[
  G'(p^\star(x)) + \alpha' + \beta' E'_{\mathrm{gen}}(x) = 0,\quad\forall x.
\]
Using \eqref{eq:KKT-canonical},
\[
  G'(p^\star(x)) + \alpha' + \beta' E'_{\mathrm{gen}}(x)
  = -\beta E_{\mathrm{gen}}(x) + \alpha' + \beta' E'_{\mathrm{gen}}(x) - \alpha.
\]
We therefore choose $(\alpha',\beta')$ so that
\[
  \alpha' + \beta' E'_{\mathrm{gen}}(x)
  = \alpha + \beta E_{\mathrm{gen}}(x)
\]
for all $x$. This is achieved by \eqref{eq:alphabeta-gauge}, since then
\[
  \alpha' + \beta' E'_{\mathrm{gen}}(x)
  = \alpha + \beta b/a + (\beta/a)(aE_{\mathrm{gen}}(x)+b)
  = \alpha + \beta E_{\mathrm{gen}}(x) + 2\beta b/a,
\]
and adding a constant to both sides of \eqref{eq:KKT-canonical} leaves the solution $p^\star$
unchanged (we can absorb constants into $\alpha$). More explicitly, we can reabsorb the
extra constant into a redefined normalisation multiplier, so there exist $(\alpha',\beta')$
which reproduce the same $p^\star$. Hence the canonical probabilities are unchanged,
only the multipliers are reparameterised.
\end{proof}

\begin{remark}
Concretely: affine changes in $E_{\mathrm{gen}}$ leave the MaxEnt distribution $p^\star$ invariant,
while renormalising the Lagrange multipliers. This is the usual statement that additive and
multiplicative gauges in the energy are physically irrelevant for the canonical ensemble.
\end{remark}

\subsection{Coarse-graining stability (Target L)}
\label{sec:cg-stability}

We now address how entropy, internal energy, and temperature behave under coarse-graining
of the state (or energy) space. A central point is that the phrase ``coarse-graining increases
entropy'' is used in the literature for \emph{two different} objects. In this paper we work with
the \emph{lumped (macro) distribution} obtained by pushing forward $p$ through a partition map,
and we study the corresponding \emph{macro-entropy} $H_G(p^{\mathcal P})$. This quantity is the
entropy of the coarse variable (block label) and behaves monotonically under refinement for
Shannon (and, more generally, for generators satisfying the log-extensive small-$u$ condition
introduced earlier). When one instead defines a coarse-grained entropy by adding within-block
uncertainty (a conditional term), one obtains a different object with different monotonicity
properties; we briefly comment on this distinction at the end of the section.

\subsubsection{Coarse-graining as a stochastic map}

Let $X$ be the (fine) microstate space, $|X|=n$, and let $p\in\Delta_n^\circ$ be a distribution on $X$.

\begin{definition}[Partition and coarse-graining map]
Let $\mathcal{P} = \{B_1,\dots,B_K\}$ be a partition of $X$ into disjoint nonempty blocks
$B_k\subset X$. The coarse-graining (lumping) map $C_\mathcal{P}:\Delta_n\to\Delta_K$ is
\[
  (C_\mathcal{P}p)_k \coloneqq \sum_{x\in B_k} p(x),
  \quad k=1,\dots,K.
\]
We write $p^\mathcal{P} \coloneqq C_\mathcal{P}p$ for the lumped (macro) distribution on block labels.
\end{definition}

\begin{definition}[Refinement]
Let $\mathcal{P}'$ and $\mathcal{P}$ be two partitions. We say that $\mathcal{P}'$ is a
\emph{refinement} of $\mathcal{P}$ (denoted $\mathcal{P}'\succeq\mathcal{P}$) if every block
$B\in\mathcal{P}'$ is contained in some block $A\in\mathcal{P}$. In this case $\mathcal P'$ is the
\emph{finer} partition and $\mathcal P$ the \emph{coarser} one. Moreover, there exists a stochastic
matrix $R$ (a block-merging map) such that
\[
  C_{\mathcal P} = R\,C_{\mathcal P'} ,
\]
i.e.\ one obtains the coarser lumping by first lumping to the finer blocks and then merging them.
\end{definition}

\subsubsection{Entropy of the lumped macro-variable}

We focus first on the entropy of the block-label random variable $Y=C_{\mathcal P}(X)$, i.e.
\[
  S_{\mathrm{lump}}^G(\mathcal P;p) \coloneqq H_G(p^{\mathcal P}).
\]
For Shannon entropy, this is exactly $H(Y)$ and satisfies standard data-processing monotonicity:
refining the partition (tracking more labels) cannot decrease the entropy of the label, while
lumping (merging labels) cannot increase it.

\begin{lemma}[Shannon: merging decreases lumped entropy]
\label{lem:merge-shannon}
Let $q\in\Delta_m$ and let $\tilde q$ be obtained by merging a subset of components
$\{q_i\}_{i\in I}$ into their sum $r=\sum_{i\in I} q_i$. Then
\[
  H_{\mathrm{Sh}}(\tilde q)\le H_{\mathrm{Sh}}(q),
\]
with equality iff at most one of the $q_i$ ($i\in I$) is nonzero.
\end{lemma}

\begin{proof}
Write $H_{\mathrm{Sh}}(q)=-\sum_{i=1}^m q_i\log q_i$. Only the merged components change, hence
\[
H_{\mathrm{Sh}}(q)-H_{\mathrm{Sh}}(\tilde q)
= \Big(-\sum_{i\in I} q_i\log q_i\Big) - \big(-r\log r\big)
= \sum_{i\in I} q_i \log\frac{r}{q_i}
= r\sum_{i\in I}\frac{q_i}{r}\log\frac{1}{q_i/r}
= r\,H_{\mathrm{Sh}}(\{\tfrac{q_i}{r}\}_{i\in I})\ge 0,
\]
with equality iff the conditional distribution on the merged group is degenerate (i.e.\ at most one
nonzero component).
\end{proof}

\begin{proposition}[Shannon: refinement increases lumped entropy]
\label{prop:S-monotone-refine-shannon}
Let $\mathcal{P}'\succeq\mathcal{P}$ be partitions of $X$, with $p^\mathcal{P}$ and $p^{\mathcal{P}'}$
their lumped distributions. Then
\begin{equation}
  H_{\mathrm{Sh}}(p^{\mathcal{P}'}) \ge H_{\mathrm{Sh}}(p^\mathcal{P}).
  \label{eq:S-monotone-shannon}
\end{equation}
In particular, $H_{\mathrm{Sh}}(p^\mathcal{P})\le H_{\mathrm{Sh}}(p)$ for every $\mathcal P$.
\end{proposition}

\begin{proof}
Since $\mathcal{P}'\succeq\mathcal{P}$, there is a merging matrix $R$ such that
$p^\mathcal P = R\,p^{\mathcal P'}$. The vector $p^\mathcal P$ is obtained from $p^{\mathcal P'}$
by a finite sequence of merges, each of which (by Lemma~\ref{lem:merge-shannon}) cannot increase
$H_{\mathrm{Sh}}$. Thus $H_{\mathrm{Sh}}(p^\mathcal P)\le H_{\mathrm{Sh}}(p^{\mathcal P'})$.
Finally, the finest partition (singletons) yields $p^{\mathcal P}=p$, so $H_{\mathrm{Sh}}(p^\mathcal P)\le H_{\mathrm{Sh}}(p)$.
\end{proof}

\begin{corollary}[Existence of a refinement limit for lumped Shannon entropy]
\label{cor:S-limit-shannon}
Let $\{\mathcal{P}_n\}_{n\ge 1}$ be a refining sequence ($\mathcal P_{n+1}\succeq\mathcal P_n$) and
$S_n\coloneqq H_{\mathrm{Sh}}(p^{\mathcal{P}_n})$. Then $(S_n)$ is monotone nondecreasing and bounded
above by $H_{\mathrm{Sh}}(p)$, hence converges:
\[
  \lim_{n\to\infty} S_n = S_\infty \le H_{\mathrm{Sh}}(p).
\]
\end{corollary}

\subsubsection{Internal energy and temperature under lumping (energy-space setting)}

We now specialise to the setting of this paper, where the coarse-graining acts on the \emph{energy}
axis. Let $C=\{c_j\}_{j=1}^n$ be the discretised energy support and let $p_E\in\Delta_n^\circ$ be a
distribution on $C$ (e.g.\ an empirical energy histogram or its Stage-I MaxEnt projection).
A partition $\mathcal P=\{B_k\}_{k=1}^K$ groups energy bins into macro-bins. Define the macro
probabilities $p_E^{\mathcal P}=C_{\mathcal P}p_E$ as above.

To define internal energy after coarse-graining, one must choose a representative energy level
for each macro-bin. A natural choice is the conditional mean energy within each macro-bin:
\begin{equation}
  \bar c_k \coloneqq \mathbb E[c\mid c\in B_k]
  = \frac{\sum_{j\in B_k} c_j\,p_{E,j}}{\sum_{j\in B_k} p_{E,j}}
  = \frac{\sum_{j\in B_k} c_j\,p_{E,j}}{(p_E^{\mathcal P})_k},
  \label{eq:cg-mean-energy}
\end{equation}
with the convention that $\bar c_k$ is arbitrary if $(p_E^{\mathcal P})_k=0$ (which does not affect
expectations). The coarse-grained internal energy is then
\begin{equation}
  U^{\mathcal P}\coloneqq \sum_{k=1}^K (p_E^{\mathcal P})_k\,\bar c_k.
  \label{eq:cg-U}
\end{equation}

\begin{proposition}[Energy is preserved under energy-axis lumping]
\label{prop:cg-U-preserved}
With $\bar c_k$ defined by \eqref{eq:cg-mean-energy}, one has
\[
  U^{\mathcal P} = \sum_{j=1}^n p_{E,j}c_j \eqqcolon U,
\]
i.e.\ coarse-graining on the energy axis preserves internal energy.
\end{proposition}

\begin{proof}
Substitute \eqref{eq:cg-mean-energy} into \eqref{eq:cg-U}:
\[
U^{\mathcal P}
=\sum_{k=1}^K (p_E^{\mathcal P})_k \frac{\sum_{j\in B_k} c_j p_{E,j}}{(p_E^{\mathcal P})_k}
=\sum_{k=1}^K\sum_{j\in B_k} c_j p_{E,j}
=\sum_{j=1}^n c_j p_{E,j} = U.
\]
\end{proof}

For temperature, recall that in our framework $T^{-1}$ is defined from the thermodynamic relation
$T^{-1}=dS/dU$ along a family of states. Since $U$ is preserved by energy-axis lumping, temperature
changes under coarse-graining only through the entropy functional evaluated on the lumped
distribution. In particular, when using Shannon as an illustrative baseline, refinement increases
the lumped entropy and therefore (for fixed $U$) decreases $T^{-1}$ along the coarse-grained family.
For general learned $G_{\mathrm{gen}}$, the corresponding statement depends on the specific generator
and on whether one evaluates $S$ on $p_E$ or on its Stage-I projection; we therefore use this section
primarily to emphasise energy preservation and the need to specify the entropy notion.

\subsubsection{Remark on alternative ``coarse-grained entropies''}

Many statements in statistical physics that ``coarse-graining increases entropy'' refer not to the
macro-entropy $H_{\mathrm{Sh}}(p^{\mathcal P})$ of the block label, but to a \emph{renormalised} or
\emph{conditional} coarse-grained entropy that adds the within-block uncertainty. For Shannon, if
$Y=C_{\mathcal P}(X)$ denotes the block label, then
\[
  H_{\mathrm{Sh}}(X)=H_{\mathrm{Sh}}(Y)+H_{\mathrm{Sh}}(X\mid Y),
\]
so one can define a coarse-grained entropy as $H_{\mathrm{Sh}}(Y)+H_{\mathrm{Sh}}(X\mid Y)$, which is
\emph{invariant} under the partition and coincides with the microstate entropy. More general
coarse-graining conventions (including renormalisation procedures) are discussed in, e.g.,
K.~Lindgren, \emph{Entropy} \textbf{17}, 3332 (2015). In this paper we focus on the lumped macro
distribution because it is the natural object on the energy axis, where macro-bins represent
aggregated energy levels and internal energy is preserved by choosing conditional means
\eqref{eq:cg-mean-energy}.

\subsubsection{Internal energy under energy coarse-graining}

Now suppose that coarse-graining is performed along the energy axis. For simplicity, assume
$E_{\mathrm{gen}}$ takes real values in a compact interval $[E_{\min},E_{\max}]$ and that we
coarse-grain by partitioning this interval.

\begin{definition}[Energy partition and approximate internal energy]
Let $\mathcal{I} = \{I_1,\dots,I_K\}$ be a partition of $[E_{\min},E_{\max}]$ into intervals
$I_k$, and choose a representative energy $E_k\in I_k$ for each bin. Define the coarse-grained
probabilities
\[
  p_k \coloneqq \sum_{x : E_{\mathrm{gen}}(x)\in I_k} p(x),
\]
and the approximate internal energy
\[
  U^{\mathcal{I}} \coloneqq \sum_{k=1}^K E_k p_k.
\]
\end{definition}

\begin{proposition}[Uniform bound on internal-energy error]
\label{prop:U-coarse}
Let $\delta(\mathcal{I}) \coloneqq \max_k \operatorname{diam}(I_k)$ be the mesh of the energy
partition. Then
\begin{equation}
  |U^{\mathcal{I}} - U| \le \delta(\mathcal{I}).
  \label{eq:U-error}
\end{equation}
In particular, as $\delta(\mathcal{I})\to 0$, $U^{\mathcal{I}}\to U$.
\end{proposition}

\begin{proof}
Write
\[
  U = \sum_{x} E_{\mathrm{gen}}(x)p(x)
    = \sum_k \sum_{x : E_{\mathrm{gen}}(x)\in I_k} E_{\mathrm{gen}}(x)\,p(x).
\]
Then
\begin{align*}
  U^{\mathcal{I}} - U
  &= \sum_k E_k p_k - \sum_k \sum_{x : E_{\mathrm{gen}}(x)\in I_k} E_{\mathrm{gen}}(x)p(x)\\
  &= \sum_k \sum_{x : E_{\mathrm{gen}}(x)\in I_k} \big(E_k - E_{\mathrm{gen}}(x)\big) p(x).
\end{align*}
Taking absolute values and using the triangle inequality,
\[
  |U^{\mathcal{I}} - U|
  \le \sum_k \sum_{x : E_{\mathrm{gen}}(x)\in I_k}
       \big|E_k - E_{\mathrm{gen}}(x)\big| p(x).
\]
For any $x$ with $E_{\mathrm{gen}}(x)\in I_k$, we have
$\big|E_k - E_{\mathrm{gen}}(x)\big| \le \operatorname{diam}(I_k)\le\delta(\mathcal{I})$. Thus
\[
  |U^{\mathcal{I}} - U|
  \le \delta(\mathcal{I}) \sum_x p(x)
  = \delta(\mathcal{I}).
\]
\end{proof}

\begin{remark}
This shows that internal energy computed from a coarse-grained energy histogram converges
uniformly to the true $U$ as the maximal bin width shrinks. No assumption on $G$ is needed.
\end{remark}

\subsubsection{Temperature from coarse-grained $(S,U)$}

Finally, we consider how temperature defined via $1/T = dS/dU$ behaves when both $S$ and $U$
are approximated by coarse-grained quantities.

Let $\{\mathcal{P}_n\}$ be a sequence of partitions refining both the state space and the energy
axis in a consistent way (each energy bin defining a block of states), and define
\[
  S_n(\alpha) \coloneqq H_G(p_\alpha^{\mathcal{P}_n}),\qquad
  U_n(\alpha) \coloneqq U^{\mathcal{I}_n}(\alpha)
\]
along a smooth path $\alpha\mapsto p_\alpha$.

\begin{assumption}[Smoothness and uniform convergence]
\label{ass:smooth-uniform}
Assume:
\begin{enumerate}
  \item $S(\alpha)$ and $U(\alpha)$ are $C^1$ on a compact interval $I\subset\mathbb{R}$.
  \item $S_n(\alpha)\uparrow S(\alpha)$ pointwise on $I$ and the convergence is uniform.
  \item $U_n(\alpha)\to U(\alpha)$ uniformly on $I$ as $n\to\infty$.
\end{enumerate}
\end{assumption}

Define approximate temperatures $T_n(\alpha)$ via
\[
  \frac{1}{T_n(\alpha)}
  \coloneqq \frac{dS_n}{dU_n}(\alpha)
  = \frac{dS_n/d\alpha}{dU_n/d\alpha},
\]
where derivatives are taken with respect to $\alpha$ and the ratio is understood where
$dU_n/d\alpha\neq 0$.

\begin{theorem}[Convergence of coarse-grained temperatures]
\label{thm:T-coarse}
Under Assumption~\ref{ass:smooth-uniform}, and assuming $dU/d\alpha$ does not vanish on $I$,
we have
\[
  T_n(\alpha) \to T(\alpha)
\]
uniformly on $I$ as $n\to\infty$, where $T(\alpha)$ is defined by
\[
  \frac{1}{T(\alpha)} = \frac{dS}{dU}(\alpha).
\]
\end{theorem}

\begin{proof}[Proof sketch]
By uniform convergence of $S_n$ to $S$ and standard results on differentiation of uniformly
convergent sequences of $C^1$ functions (or by the mean value theorem and equicontinuity), we
can ensure that $dS_n/d\alpha\to dS/d\alpha$ uniformly on $I$; similarly, from uniform convergence
of $U_n$ to $U$ and smoothness of $U$, we obtain uniform convergence of $dU_n/d\alpha$ to
$dU/d\alpha$. Since $dU/d\alpha$ is bounded away from zero on $I$, for sufficiently large $n$
also $dU_n/d\alpha$ is bounded away from zero, and
\[
  \frac{dS_n/d\alpha}{dU_n/d\alpha} \to
  \frac{dS/d\alpha}{dU/d\alpha}
\]
uniformly. Thus $1/T_n(\alpha)\to 1/T(\alpha)$ uniformly on $I$, and hence $T_n(\alpha)\to T(\alpha)$
uniformly as well.
\end{proof}

\begin{remark}
In practice, one often approximates derivatives via finite differences in $\alpha$; the theorem
still applies, with $T_n$ interpreted as the ratio of finite differences of coarse-grained
$S_n$ and $U_n$, provided the finite-difference scheme converges and the uniform convergence
conditions are met. The key point is that the effect of coarse-graining on $S$ and $U$ is
controlled and vanishes as the partitions are refined, so the inferred temperature is
stable in the limit.
\end{remark}

\subsection{Summary}

In this note we have:

\begin{itemize}
  \item Analysed affine \emph{energy gauge transformations}
        $E_{\mathrm{gen}}\mapsto aE_{\mathrm{gen}}+b$ and shown that:
        \begin{itemize}
          \item Internal energy transforms as $U'\!=\!aU\!+\!b$;
          \item Entropy $S$ is gauge invariant (depends only on $p$);
          \item Temperature rescales as $T'=aT$ under energy rescaling and is invariant
                under additive shifts;
          \item Free energies transform affinely, while response quantities such as heat
                capacity $C=dU/dT$ are gauge invariant.
        \end{itemize}
  \item Shown that canonical MaxEnt distributions are \emph{gauge invariant}: an affine
        change in $E_{\mathrm{gen}}$ can be absorbed into a reparametrisation of the
        Lagrange multipliers without changing the probabilities.
  \item Established \emph{coarse-graining stability} of the generative entropy:
        coarse-graining via partitions of the state space (or energy axis) yields entropies
        that are monotone nondecreasing under refinement and converge to an upper limit
        (the microstate-level entropy).
  \item Provided uniform error bounds for internal energy under energy-bin coarse-graining,
        and, under mild smoothness assumptions, shown that temperatures defined via
        $1/T = dS/dU$ computed from coarse-grained $(S,U)$ converge to the true temperature
        as the coarse-graining is refined.
\end{itemize}

These results ensure that the thermodynamic quantities in the generative framework are
\emph{gauge-stable} (insensitive to affine reparametrisations of the energy) and
\emph{coarse-graining-stable} (robust under changes in the resolution at which the energy
distribution is represented).

\section*{Operational temperature validation by thermal contact}
\label{sec:SM_thermal_contact}

\paragraph{Goal.}
Reviewer concerns about ``temperature defined only as a slope'' typically reduce to the question of
whether the inferred temperature predicts an operational outcome.
Here we use the standard equilibrium criterion from thermal contact:
when two subsystems exchange energy under a fixed total energy constraint, equilibrium is achieved at
equal inverse temperatures $\beta=\partial S/\partial U$.

\paragraph{Systems and shared energy axis.}
We consider two synthetic data systems:
Gaussian-mixture observations with separation parameters $\mu_A=2$ and $\mu_B=3$.
A \emph{shared} generative energy map $E_{\mathrm{ref}}(x)$ is learned once from pooled samples
from both systems (by fitting a low-degree even polynomial to $-\log \hat p(x)$).
This shared map ensures that energies and total-energy conservation are defined on a common scale.
Each system is then represented on the same discrete energy support $\{E_j\}$ by empirical histograms
$\hat p_A(E_j)$ and $\hat p_B(E_j)$.

\paragraph{System-specific entropy functionals and microcanonical curves.}
Although the energy axis is shared, the entropy functional is inferred separately for each system
(using the same construction pipeline as in the main text).
For each system $s\in\{A,B\}$ we compute the microcanonical entropy curve
\[
S_s(U)\;=\;\max_{p\in\Delta_n}\;H_s[p]
\quad\text{s.t.}\quad
\sum_{j} p_j E_j = U,
\]
and obtain the inverse temperature along this admissible family as
\[
\beta_s(U)\;=\;\frac{\partial S_s}{\partial U}.
\]

\paragraph{Thermal-contact protocol and equilibrium prediction.}
We define a contact protocol in which the composite system can exchange energy between subsystems
while conserving total energy $U_{\mathrm{tot}}$. To avoid boundary artefacts associated with
truncated feasibility intervals, we choose $U_{\mathrm{tot}}$ in the interior of the common
admissible range (here the midpoint protocol used in the accompanying code).
Under the usual weak-contact assumption (additive entropies and conserved total energy), define
the total entropy as a function of the split,
\[
S_{\mathrm{tot}}(U_A)\coloneqq S_A(U_A)+S_B\big(U_{\mathrm{tot}}-U_A\big).
\]
The predicted equilibrium split is obtained by maximising $S_{\mathrm{tot}}$:
\[
U_A^\star \;=\; \arg\max_{U_A} S_{\mathrm{tot}}(U_A),
\qquad
U_B^\star = U_{\mathrm{tot}}-U_A^\star.
\]
At an interior optimum, the first-order condition yields the operational equalisation rule
\[
\beta_A(U_A^\star)\;=\;\beta_B(U_B^\star),
\]
i.e.\ energy flows until the inferred inverse temperatures match.

\begin{figure}[t]
\centering
\includegraphics[width=\linewidth]{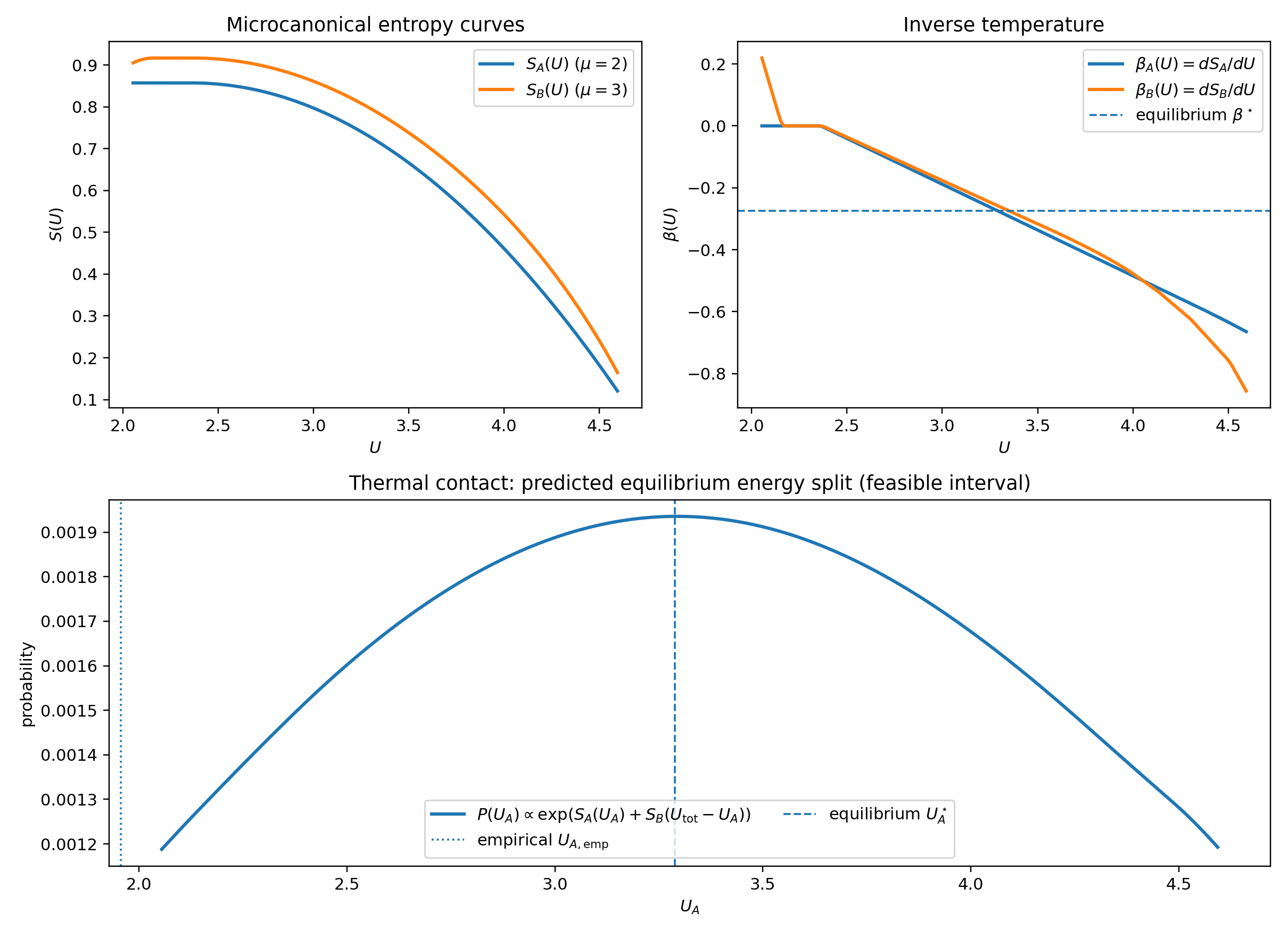}
\caption{\textbf{Operational temperature validation by thermal contact (Gaussian mixtures).}
Two synthetic data systems (Gaussian mixtures with separations $\mu=2$ and $\mu=3$) are mapped to a
\emph{shared} learned energy axis $E_{\mathrm{ref}}$, while their entropy functionals are inferred
separately, producing microcanonical curves $S_A(U)$ and $S_B(U)$ (left).
Inverse temperatures $\beta_s(U)=\partial S_s/\partial U$ (right) determine the predicted equilibrium
energy split under weak thermal contact at fixed total energy $U_{\mathrm{tot}}$.
The equilibrium split is the maximiser of $S_A(U_A)+S_B(U_{\mathrm{tot}}-U_A)$, implying
$\beta_A(U_A^\star)=\beta_B(U_B^\star)$.
For the shown run ($N=2\times 10^5$ per system, seed 1; $U_{\mathrm{tot}}=6.648$),
the predicted equilibrium is $U_A^\star=3.289$ and $U_B^\star=3.360$ with mismatch
$\lvert\Delta\beta\rvert=3.21\times 10^{-3}$ (dashed lines).
The bottom panel shows the induced contact distribution
$P(U_A)\propto \exp(S_A(U_A)+S_B(U_{\mathrm{tot}}-U_A))$ on the feasible interval.}
\label{fig:thermal_contact_mog}
\end{figure}

\paragraph{Result.}
For $\mu_A=2$ and $\mu_B=3$ (seed 1; $N=2\times 10^5$ samples per system),
the protocol selects $U_{\mathrm{tot}}=6.648$ with feasible interval
$U_A\in[2.054,\,4.594]$.
The entropy maximiser occurs at $U_A^\star=3.289$ and $U_B^\star=3.360$,
with inverse-temperature mismatch
$\lvert\beta_A(U_A^\star)-\beta_B(U_B^\star)\rvert=3.21\times 10^{-3}$.
Figure~S\ref{fig:thermal_contact_mog} shows the corresponding microcanonical curves, inverse
temperatures, and the predicted equilibrium split distribution
$P(U_A)\propto \exp(S_A(U_A)+S_B(U_{\mathrm{tot}}-U_A))$.
The accompanying code allows verification across multiple seeds.

